\documentclass[10pt]{article} 
\usepackage{amsmath,amsfonts,amssymb,slashed,makeidx,latexsym,setspace}
\usepackage{graphicx,graphics,floatflt,amssymb,epsf,rotate,subfigure} 
\usepackage{times,axodraw,cite,mathrsfs}
\evensidemargin=\oddsidemargin 
\usepackage{color}

\makeatletter
\@addtoreset{equation}{section}
\makeatother

\def\tlcorner{
  \begin{picture}(1,1)
    \SetWidth{2}
    \Line(-10,15)(50,15)
    \Line(-10,15)(-10,-10)
  \end{picture}
}

\def\blcorner{\newline
  \begin{picture}(50,5)
    \SetWidth{2}
    \Line(-10,0)(50,0)
    \Line(-10,0)(-10,20)
  \end{picture}
}

\newenvironment{bquote}
               {\list{}{\leftmargin=5mm\rightmargin=0mm%
                \listparindent=\parindent\parsep=0pt}%
                \item\tlcorner\relax\footnotesize\em}
               {\blcorner\endlist}

\def\ltap{\raisebox{-.4ex}{\rlap{$\sim$}} \raisebox{.4ex}{$<$}}

\def\H{\hat{H}}
\def\L{\hat{L}}

\def\D{\hat{D}}

\def\O{\cal{O}}

\def\a{\alpha}
\def\b{\beta}
\def\d{\delta}
\def\g{\gamma}
\def\s{\sigma}
\def\G{\Gamma}
\def\D{\Delta}
\def\L{\Lambda}
\def\l{\lambda}
\def\p{\partial}
\def\t{\theta}
\def\T{\Theta}

\def\e{\epsilon}
\def\Lag{{\mathscr{L}}}

\begin{document}


\begin{center}
{\bf \underline{To appear in Reports on Progress in
    Physics}}
\vskip 5pt
  {\Large \bf A Pedagogical Review of Electroweak Symmetry Breaking Scenarios}
  \footnote{\sf Based on a series of lectures given at the Advanced SERC
    School on High Energy Physics at Hyderabad University (2007), the RECAPP
    Workshop at HRI, Allahabad (2008), and the 15th Vietnam
    School of Physics in Dong Hoi (2009). 
} \\
  \vspace*{0.8cm} \renewcommand{\thefootnote}{\fnsymbol{footnote}}
  {\bf Gautam Bhattacharyya} \\
  \vspace{5pt} {\small {\em Saha Institute of Nuclear Physics, 1/AF Bidhan
      Nagar, Kolkata 700064, India}}
    
\normalsize 
\end{center}

\begin{abstract}  
  We review different avenues of electroweak symmetry breaking explored over
  the years. This constitutes a timely exercise as the world's largest and the
  highest energy particle accelerator, namely, the Large Hadron Collider (LHC)
  at CERN near Geneva, has started running whose primary mission is to find
  the Higgs or some phenomena that mimic the effects of the Higgs, i.e. to
  unravel the mysteries of electroweak phase transition.  In the beginning, we
  discuss the Standard Model Higgs mechanism.  After that we review the Higgs
  sector of the Minimal Supersymmetric Standard Model.  Then we take up three
  relatively recent ideas: Little Higgs, Gauge-Higgs Unification, and
  Higgsless scenarios. For the latter three cases, we first present the basic
  ideas and restrict our illustration to some instructive toy models to
  provide an intuitive feel of the underlying dynamics, and then discuss, for
  each of the three cases, how more realistic scenarios are constructed and
  how to decipher their experimental signatures. Wherever possible, we provide
  enough pedagogical details, which the beginners might find useful.
 \end{abstract} 

\setcounter{footnote}{0} 
\renewcommand{\thefootnote}{\arabic{footnote}} 

\setcounter {tocdepth}1
\tableofcontents

\section{Introduction}
The theory of beta decay, which manifests weak interaction, was first
formulated by Fermi. Below we write down the effective Lagrangian of beta
decay. In doing so, we use modern notations and rely on the $(V-A)$ structure
of currents \cite{Sudarshan:1958vf}:
\begin{equation}
\Lag_{\rm eff} = \frac{G_F}{\sqrt{2}} \left(\bar{p} \g_\mu (1+\a \g_5) n\right) 
\left(\bar{e} \g_\mu (1-\g_5) \nu\right) \, .
\end{equation}
Since every fermion field has mass dimension $3/2$ (which follows from power
counting in Dirac Lagrangian), the prefactor $G_F$ has clearly a mass
dimension $-2$. From the neutron decay width and angular distribution, one
obtains $\a \simeq -1.2$ and the `Fermi scale' $G_F^{-1/2} \simeq$ 300
GeV. Particle physics has gone through a dramatic evolution since the time of
Fermi \cite{history}. The success of the Yang-Mills theory revolutionized the
whole scenario \cite{yangmills}. The charged $W^\pm$ boson was eventually
predicted by the Standard Model (SM) \cite{sm} having a mass of around 80 GeV,
which was later experimentally confirmed by direct detection by the UA1
collaboration at CERN. This $W^\pm$ boson induces radioactivity by mediating
the beta decay process. However, a full understanding of the dynamics that
controls the Fermi scale and hence the $W$ boson mass still remains an
enigma. This is the scale of electroweak phase transition, and understanding
the origin of electroweak symmetry breaking (EWSB) constitutes the primary
goal of the Large Hadron Collider (LHC) at CERN.  The readers are strongly
recommended to follow Refs.  \cite{rev_ewsb,Kaul:2008cv,Giudice:2008bi,
  Cheng:2007bu,Rattazzi:2005di,Bhattacharyya:2008ez} to have a broad overview
of different possible EWSB mechanisms.

The SM reigns supreme at the electroweak scale.  But it cannot account for a
few experimentally established facts: neutrino mass, dark matter and the right
amount of baryon asymmetry of the universe.  Any viable scenario beyond the SM
that is expected to trigger EWSB and to answer one or more of the above
questions must pass the strict constraints imposed by the electroweak
precision tests (EWPT) carried out mainly at the Large Electron Positron (LEP)
collider at CERN. Non-abelian gauge theory as the theory for weak interaction
has been established to a very good accuracy: ($i$) the $ZWW$ and $\gamma WW$
vertices have been measured to a per cent accuracy at LEP-2 implying that the
SU(2) $\times$ U(1) gauge theory is unbroken at the vertices, ($ii$) accurate
measurements of the $Z$ and $W$ masses have indicated that gauge symmetry is
broken in masses. Precision measurements at LEP have shown that the
$\rho$-parameter (introduced in section \ref{smhiggs}) is unity to a very good
accuracy. This attests the `SU(2)-doublet' nature of the scalar employed in
the SM for spontaneous electroweak breaking.  Any acceptable new physics
scenario should be in accord with the above observations.  CMS and ATLAS are
the two general purpose detectors of the LHC which are expected to answer a
lot of such questions by hunting not only the Higgs but also any possible
ruler of the terra-electron-volt (TeV) regime.

The primary concern is the following: Is the Higgs mechanism as portrayed in
the SM a complete story?  Bluntly speaking, nobody believes so! Then, what is
the nature of the more fundamental underlying dynamics? A more pointed
question is -- if the Higgs exists, is it {\em elementary} or {\em composite}?
The advantage of working with an elementary Higgs, as in the SM, is that the
two issues of generating gauge boson masses and fermion masses are solved in
one stroke. Also, as it turned out, a theory relying on elementary Higgs is
perfectly comfortable with EWPT. The disadvantage is that the Higgs mass
receives quadratically divergent quantum correction which inevitably calls for
new physics, e.g. supersymmetry, to solve the hierarchy problem by taming the
unruly quantum behavior.  On the other hand, when the Higgs is a composite
object, e.g. in Technicolor, the hierarchy problem is not there any way
because the composite Higgs {\em dissolves} at the scale where new heavy
fermions (e.g. technifermions) condense to break EWSB. But a major
disadvantage of Technicolor is that such models, in general, inflict
unacceptably large flavor changing neutral currents (FCNC) and induce large
contributions to the oblique electroweak parameters $T$ (or $\Delta \rho$) and
$S$.  Although the FCNC problem can be evaded by going to some more
complicated versions of technicolor models, general inconsistency with EWPT in
the post-LEP era have put technicolor far behind supersymmetry in terms of
acceptability.  But the idea of Technicolor was too elegant to die. It simply
went into slumber only to reappear some years later in a different guise
through the AdS/CFT correspondence \cite{Maldacena:1997re} as dual to some
extra dimensional theories.  Many modern nonsupersymmetric ideas, which we
shall discuss in this review, are reminiscent of technicolor, but sufficiently
advanced and equipped over the traditional versions to meet the FCNC and EWPT
challenges.  At this point it is fair to say that supersymmetry and the new
{\em Avatars} of technicolor/compositeness are the two most attractive general
{\em classes of theories} that may dictate the EWSB mechanism and are expected
to be observed at the LHC.  Therefore, before we get going into a systematic
but incremental elaboration of how the idea of EWSB evolved and how the
different concerns at different stages were sorted out, we briefly touch upon
the main features of two most important conceptual pillars on which many
specific models were built, namely, supersymmetry and technicolor.

\subsection{Supersymmetry} 
Supersymmetry is arguably the most favorite extension of physics beyond the
Standard Model. It all started more than 30 years ago from theoretical works
pursued independently by Golfand and Likhtman \cite{Golfand:1971iw}, Volkov
and Akulov \cite{Volkov:1973ix}, and Wess and Zumino \cite{Wess-zumino}.  For
historical developments of the idea of supersymmetry and subsequent model
building and phenomenology, we recommend the text books \cite{susy-books} and
reviews \cite{Rodriguez:2009cd,reviews}. We briefly outline the concept below.

Supersymmetry is a new space-time symmetry interchanging bosons and fermions,
relating states of different spin.  We first recall that Poincar\'e group is a
semi-direct product of translations and the Lorentz transformations (which
involve rotations and boosts), while a super-Poincar\'e group additionally
includes supersymmetry transformations linking bosons and fermions.  More
specifically, the Poincar\'e group is generalized to the super-Poincar\'e
group by adding two anticommuting generators $Q$ and $\bar{Q}$, to the
existing $p$ (linear momentum), $J$ (angular momentum) and $K$ (boost), such
that $\{Q,\bar{Q}\} \sim \g^\mu p_\mu$.  Haag, Lopuszanski and Sohnius
generalized the work of O'Raifeartaigh and by Coleman and Mandula to show that
the most general symmetries of the $S$-matrix are a direct product of
super-Poincar\'e group with the internal symmetry group.  Since the new
symmetry generators linking bosons and fermions are spinors, not scalars,
supersymmetry is not an internal symmetry. Years ago, Dirac postulated a
doubling of states by introducing an antiparticle to every particle in an
attempt to reconcile Special Relativity with Quantum Mechanics.  In
Stern-Gerlach experiment, an atomic beam in an inhomogeneous magnetic field
splits due to doubling of the number of electron states into spin-up and -down
modes indicating a doubling with respect to angular momentum. So it is no
surprise that $Q$ causes a further splitting into particle and superparticle
($f \stackrel{Q}{\rightarrow} f, \widetilde{f}$) \cite{Hall:1996gq}. Since $Q$
is spinorial, the superpartners differ from their SM partners in spin. The
superpartners of fermions are scalars and are called `sfermions', while the
superpartners of gauge bosons are fermions and are called `gauginos'. Put
together, a particle and its superpartner form a supermultiplet. The two
irreducible supermultiplets which are used to construct the supersymmetric
standard model are the `chiral' and the `vector' supermultiplets. The chiral
supermultiplet contains a scalar (e.g. selectron) and a 2-component Weyl
fermion (e.g. left-chiral electron). The vector supermultiplet contains a
gauge field (e.g. photon) and a 2-component Majorana fermion
(e.g. photino). We should remember that ($i$) there is an equal number of
bosonic and fermionic degrees of freedom in a supermultiplet; ($ii$) since
$p^2$ commutes with $Q$, the bosons and fermions in a supermultiplet have the
same mass.

\begin{table}[ht]
\small
\begin{center}
\begin{tabular}{rccc}
\hline
\hline
Particles/superparticles
 & spin 0 & spin 1/2 & 
${\rm SU(3)}_C \times {\rm SU(2)}_L \times {\rm U(1)}_Y$ \\
\hline
\hline
 & & & \\
leptons, sleptons~~ ($L$) & $(\widetilde{\nu}, \widetilde{e}_L)$ &
 $({\nu}, {e}_L)$ & ($1, 2, -1/2$) \\
(in 3 families)~~ ($E^c$) & ${\widetilde{e}_R}^*$ & $e^c_L$ & ($1, 1, 1$)
\\
 & & & \\
\hline
 & & & \\
quarks, squarks~~ ($Q$) & $(\widetilde{u}_L, \widetilde{d}_L)$ &
 $({u}_L, {d}_L)$ & ($3, 2, 1/6$) \\
(in 3 families)~~ ($U^c$) & ${\widetilde{u}_R}^*$ & $u^c_L$ &
($\bar{3}, 1, -2/3$) \\
 ($D^c$) & ${\widetilde{d}_R}^*$ & $d^c_L$ &
($\bar{3}, 1, 1/3$) \\
 & & & \\
\hline
  & & & \\
Higgs, higgsinos~~ ($H_u$) & $(H_u^+, H_u^0)$ &
 $(\widetilde{H}_u^+, \widetilde{H}_u^0)$ & ($1, 2, 1/2$) \\
 (up, down types)~~($H_d$) & $(H_d^0, H_d^-)$ &
 $(\widetilde{H}_d^0, \widetilde{H}_d^-)$ & ($1, 2, -1/2$) \\
 & & & \\
\hline
\hline
Particles/superparticles
& spin $1$ & spin $1/2$ & 
${\rm SU(3)}_C \times {\rm SU(2)}_L \times {\rm U(1)}_Y$ \\
\hline
\hline
 & & & \\
gluon, gluino & $g$ & $\widetilde{g}$ & ($8, 1, 0$) \\
$W$ bosons, winos & $W^\pm, W^0$ & $\widetilde{W}^\pm, \widetilde{W}^0$ &
($1, 3, 0$) \\
$B$ boson, bino & $B^0$ & $\widetilde{B}^0$ &
($1, 1, 0$) \\
 & & & \\
\hline
\hline
\end{tabular}
\caption{{\small \sf The particle content of the minimal supersymmetric 
standard model: superparticles marked by overhead `tilde'.}}
\end{center}
\label{mssm-content}
\end{table}

But, why don't we see the superpartners?  According to supersymmetry every
fermion should have a bosonic partner and {\em vice versa}. Then the
superpartner of electron which is a scalar with the same mass as that of the
electron should have been found. This simply means that supersymmetry is not
only broken but very badly broken and the superpartners are heavy enough to
have escaped detection so far. There are quite a few ideas as to how
supersymmetry is broken. Supersymmetry breaking can be mediated by
supergravity, or by gauge interactions, or superconformal anomaly, and so
on. Although we do not know exactly how it is broken, we know very well how to
parametrize this breaking. Recall that the SM has 18 parameters, but the
minimal supersymmetric standard model contains 106 additional parameters (see
Table \ref{mssm-content} for the particle content). But once we assume a
particular mechanism of supersymmetry breaking many of these parameters will
be related. The next question is how long the superparticles can hide
themselves?  How good is the chance of finding them at the LHC? In other
words, is there a reason of expecting them to appear at the TeV scale. An
interesting observation is that the gauge couplings measured at LEP do not
unify at a high scale when extrapolated using renormalization group (RG)
equations containing beta functions computed with the SM particle content. But
if we use supersymmetric RG equations, i.e. with beta functions computed with
the supersymmetric particle content, the couplings do unify at a high (grand
unification) scale ($M_{\rm GUT}$) provided that the superparticle masses lie
in the 100 GeV $-$ 10 TeV range. Moreover, this GUT scale is somewhat higher
than what is obtained in nonsupersymmetric scenarios which makes the
prediction of proton lifetime more consistent with its non-observation.  A
very attractive property of all supersymmetric models with conserved
$R$-parity is that they all include a stable electrically and color neutral
massive ($\sim$ 100 GeV) particle which could be an excellent candidate of the
observed dark matter of the Universe. In case $R$-parity is violated, even
gravitino could make a reasonable dark matter candidate. Furthermore,
supersymmetry provides a framework to turn on gravity, as when global
supersymmetry is promoted to a local one we get supergravity. Supersymmetric
theories have adapted very well with the LEP data, because they are {\em
  decoupling} theories in the sense that superparticle induced loop
corrections to electroweak observables, in general, rapidly decouple with
increasing superparticle masses.

Supersymmetry provides an important prediction on the Higgs mass. In the
minimal (two-Higgs doublet) supersymmetric standard model the lightest Higgs
cannnot be heavier than about 135 GeV or so provided the superparticles weigh
around a TeV. If we do not find any Higgs within that limit, the minimal
version will be seriously disfavored. We have discussed in detail the
properties of both neutral and charged scalars of the supersymmetric Higgs
sector in section \ref{susyhiggs}.  But in this review we refrain from
discussing their collider search strategies - for detailed search studies see
Djouadi's review in \cite{reviews}.

\subsection{Technicolor} 
Here we present an outline of the main idea behind technicolor theories. For a
detailed survey of the historical development and the evolution of different
concepts of dynamical electroweak symmetry breaking (DWSB) the readers are
recommended to go through the early papers of Susskind \cite{Susskind:1978ms}
and Weinberg \cite{Weinberg:1975gm} and consult the reviews on DWSB breaking
and technicolor \cite{Farhi:1980xs,Kaul:1981uk,King:1994yr}. We also recommend
the readers to subsequently follow two recent articles on Higgs as a
pseudo-Goldstone boson which discuss from a modern perspective as to how the
difficulties of traditional technicolor models are overcome
\cite{Contino:2010rs,Giudice:2007fh}.

QCD provides a strong force that binds the colored quarks. Can it induce EWSB
by creating a bound state of strongly interacting sector which receives a
non-zero expectation value in the vacuum? This is the central theme of
technicolor (TC). Let us for the moment consider only ${\rm SU(3)}_C$
interaction and switch off the electroweak gauge force of the SM. Let us
assume only one generation of massless quark doublet, both left-handed and
right-handed: $Q_L = (u,d)_L^T$ and $Q_R = (u,d)_R^T$. The QCD Lagrangian is
invariant under a global chiral symmetry
$$
{\rm SU(2)}_L \times {\rm SU(2)}_R \, . 
$$  
The symmetry is spontaneously broken down to the diagonal subgroup ${\rm
  SU(2)}_{L+R}$, which corresponds to isospin symmetry, when 
$$
\langle \bar u u\rangle_{\rm vac} = \langle \bar d d\rangle_{\rm vac} \neq 0 \, . 
$$
This chiral symmetry breaking is accompanied by three massless pseudoscalars
which are identified with the pions. These are associated with three axial
currents ($q \equiv (u,d)^T$)
$$
j^\mu_{Aa} = f_\pi \p^\mu \pi_a = \bar q \g^\mu \g^5 \frac{\tau^a}{2} q \, ,
$$ 
where $\tau^a$ are the three Pauli matrices ($a=1,2,3$) and $f_\pi$ is the
pion decay constant. When the electroweak interaction is switched on, the
massless pions are eaten up by the as yet massless gauge bosons to form the
longitudinal components of those gauge bosons which in turn become
massive. The $W$ and $Z$ boson masses are given by 
$$
M_W = g f_{\pi^\pm} / 2 \, , ~~~~~ M_Z = \sqrt{g^2 + g^{\prime 2}} f_{\pi^0}
/2 \, . 
$$
Isospin symmetry guarantees that $f_\pi \equiv f_{\pi^\pm} = f_{\pi^0}$.  This
picture is not phenomenologically acceptable as by putting $f_\pi \sim 93$
MeV, we obtain $M_W \sim 30$ MeV, while in reality $M_W \sim 80$ GeV. So the
QCD force of the SM is not strong enough to generate the correct EWSB
scale. TC does precisely this job. It is a scaled-up version of QCD, where
$f_\pi \to F_\pi \sim v \approx 246$ GeV. So the $W$ and $Z$ bosons do not eat
up the ordinary pions but the technipions. The beauty of this theory is that
the hierarchy problem is solved by dimensional transmutation. Recalling that
the QCD beta function is negative ($\beta < 0$), the electroweak scale ($v$)
is dynamically generated when the TC gauge coupling $g_{TC}$ diverges in the
infrared, in complete analogy with the dynamical generation of $\L_{\rm QCD}$:
$$
\frac{d g_{\rm TC}(\mu)}{d\ln \mu} = \frac{\beta}{16\pi^2} g^3_{\rm TC}(\mu)
\Rightarrow v = M_{\rm Pl} ~{\rm exp} \left(\frac{8\pi^2/\b}
{g_{TC}^2 (M_{\rm Pl})} \right) \, .
$$
The next important question is how fermion masses are generated
\cite{etc}. Let us consider an example by enlarging the TC group ${\rm
  G}_{TC}$ to an {\em extended technicolor} (ETC) group ${\rm G}_{ETC}$ in
which both ${\rm SU(3)}_C$ and ${\rm G}_{TC}$ are embedded:
$$
{\rm G}_{ETC} \supset {\rm SU(3)}_C \times {\rm G}_{TC} \, . 
$$
 It
is assumed that ${\rm G}_{ETC}$ is spontaneously broken at a scale
$\L_{ETC}$. The gauge bosons corresponding to broken ETC generators would
connect ordinary quarks ($q$) which transform under ${\rm SU(3)}_C$ to the TC
quarks ($\Psi_{TC}$) which transform under ${\rm G}_{TC}$, and would generate
effective four-fermion operators (after appropriate Fierz transformations)
$$
\frac{g^2_{ETC}}{\L^2_{ETC}} \left(\bar q  q\right) \left(\bar \Psi_{TC} 
\Psi_{TC}\right) \, .  
$$
At a lower scale $\L_{TC}$, a condensation takes place: $\langle \bar
\Psi_{TC} \Psi_{TC} \rangle \sim \L^3_{TC} \sim F_\pi^3 \sim v^3$.  This
immediately generates the ordinary quark mass
$$
m_q \sim \L_{TC} \left(\frac{\L_{TC}}{\L_{ETC}}\right)^2 \, . 
$$
To generate the mass hierarchy among ordinary quarks, one has to first put all
those ordinary quarks in a single ETC multiplet and arrange to break the
multiplet through different cascades, thus generating different scales. But
the exchanges of the same ETC gauge fields also generate operators with four
ordinary quarks, namely, $(\bar q q)^2/\L^2_{ETC}$, which severely violate
flavor and CP particularly because all those SM quarks are in the same
multiplet. Data on $K$ and $B$ mixing as well as rare meson decays put a very
strong constraint: $\L_{ETC} > 10^{3-5}$ TeV, which is at least two to four
orders of magnitude larger than the value of $\L_{ETC}$ required to predict
the correct strange quark mass. How to resolve this tension between large
enough quark mass {\em vis-\`a-vis} too large FCNC rates? Here comes the
r\^ole of {\em walking technicolor} \cite{wtc}. Without going into the
detailed discussion, we just point out that the dimension of the operator
$\left(\bar q q\right) \left(\bar \Psi_{TC} \Psi_{TC}\right)$ could be
$(6+\g)$, instead of the classical value 6, where $\g$ is the anomalous
dimension generated by the TC group. The TC coupling $g_{TC}$ may have a large
fixed point value at $\mu \sim \L_{TC}$ and its evolution above $\L_{TC}$ may
be slow (hence, `walking', instead of `running'). The formula for the ordinary
quark mass is then modified to
$$
m_q \sim \L_{TC} \left(\frac{\L_{TC}}{\L_{ETC}}\right)^{(2+\g)} \, . 
$$
If $\g$ is large and negative, then for a given $m_q$, one can accommodate a
larger $\L_{ETC}$ than when $\g=0$, i.e. one can have a large $\L_{ETC}$
without suppressing the quark mass. On the other hand, the suppression of FCNC
rates still goes as $1/\L_{ETC}^2$ since the SM color group cannot generate
any large anomalous dimension. This way the quark mass vs FCNC tension is
considerably ameliorated in the walking technicolor scenario. 

We conclude our discussion on technicolor by just mentioning the idea of top
quark condensates.  Although Nambu first postulated it, Bardeen, Hill and
Lindner formulated the theory of dynamical breaking of electroweak theory in
the SM by a top quark condensate \cite{Bardeen:1989ds}. Here the Higgs boson
is a $\bar t t$ bound state. Essentially, one implements the BCS or
Nambu$-$Jona-Lasinio mechanism in which a new interaction at a high scale $\L$
triggers a low energy condensate $\langle t \bar t\rangle$. Generically, top
quark mass turns out to be somewhat larger than the presently known value.
This minimal scheme was further extended by Hill in a specific {\em topcolor}
scheme \cite{Hill:1991at}. In a subsequent development it was shown that in an
ETC theory, where it is hard to generate a large top quark mass without
adversely affecting the $\rho$ parameter, a substantial part of the top quark
mass may be generated by additionally incorporating the topcolor dynamics
\cite{Hill:1994hp}.

\subsection{Plan of the review}
We shall start our discussion with a brief recapitulation of the idea of gauge
invariance.  In the subsequent sections, we shall briefly review the essential
structure of the electroweak part of the SM, illustrate the Higgs mechanism
and raise the issue of the quantum instability of the scalar potential. We
shall then demonstrate how the quadratic divergence is tamed in a toy scenario
reminiscent of a supersymmetric model. Then we go on to explore different
avenues through which one can successfully realize electroweak phase
transition. In the process, we shall discuss minimal supersymmetry (only the
Higgs sector), and some relatively recent ideas like little Higgs, gauge-Higgs
unification and Higgsless scenarios. The latter two scenarios explicitly rely
on the existence of extra dimension with a TeV-size inverse radius of
compactification.  It should be noted that many of these non-supersymmetric
scenarios are often reminiscent of the technicolor models from the standpoint
of AdS/CFT correspondence, which we shall just mention in passing without
actually going into the details of this correspondence.  For each of these
modern nonsupersymmetric scenarios, we shall first illustrate the basic
concepts using simple toy models, and then discuss, without going into
calculational details, their phenomenological features and the strategies for
detecting their signatures at the LHC. Finally, we shall conclude with a brief
stock-taking of different aspects that the model-builders should keep in mind,
followed by a short discussion on how to distinguish apart the different EWSB
models at the LHC.

\section{A short recap of the idea of gauge invariance}
This is a brief survey of the idea of gauge invariance required to formulate
the basic structure of the SM. Let us first
consider QED, which is governed by a U(1) gauge symmetry. We start with 
the Lagrangian of the electron field $\psi(x)$ with a mass $m$,  
\begin{eqnarray}
\label{qedL}
\Lag & = & \bar{\psi} (i\gamma^\a \p_{\a}-m) \psi ,  
\end{eqnarray}
where $\p_{\a} \equiv \frac{\partial}{\partial x^\alpha}$.  Observe that for
$\psi(x)\to\psi^\prime(x)=e^{i\Lambda}\psi(x)$, where $\Lambda=$ real
constant, the Lagrangian remains unaltered: $L(\psi)=L(\psi')$.  Various
transformations of the group U(1) commute. Such groups are called
`abelian'. Since $\L$ is a constant, the group is also called `global'.

Now suppose that the group is still U(1), but `local', i.e., $\Lambda
\equiv \Lambda(x)$. Then $\psi^{\prime}(x) = e^{i\L(x)}\psi(x) \equiv
U(x)\psi(x)$.  Let us see how the derivative $\p_\a\psi(x)$ transform.
\begin{eqnarray}
\p_{\a}\psi(x) & = &\p_\a U^{-1}(x)\psi^\prime(x)  =
U^{-1}(x)\underbrace{U(x)\p_{\a} U^{-1}(x)} \psi^\prime(x) \nonumber \\
&=& U^{-1}(\p_{\a}-i\p_{\a}\L(x))\psi^\prime(x) \, .~~~~~~~~~\left\{
\begin{array}{l}(1+i\L)\p_{\a}(1-i\L) \psi'(x)\\ =
(\p_\a-i\p_\a\L(x))\psi^\prime(x)\end{array}\right.
\end{eqnarray}
Although for illustration we used infinitesimal transformation, it is actually
{\em not a necessary condition}.  Note that in the first term in RHS the
derivative acts on everything on its right, but in the end where we obtain
$(\p_\a - \p_\a \L(x)) \psi'(x)$, the second $\p_\a$ acts only on $\L(x)$ and
not on $\psi'(x)$. The message is the following: although
$\psi(x)=U^{-1}(x)\psi^{\prime}(x)$, $\p_{\a}\psi(x)\ne
U^{-1}(x)\p_\a\psi^{\prime}(x)$, i.e., the field and its derivative do not
transform the same way under a local transformation.  For the global case, if
we recall, they did transform in the same way, and the Lagrangian remained
invariant. But now for the local case, $\Lag(\psi)\ne \Lag(\psi^\prime)$.

Now, we write the Lagrangian in Eq.~(\ref{qedL}) with $D_{\a} \equiv
\p_\a-ieA_\a(x)$ instead of $\p_\a$, where $e$ is a coupling
constant. $D_{\a}$ is called the {\em covariant derivative}. We now observe
the following:
\begin{eqnarray}
[\p_\a-ieA_\a(x)]\psi(x) &=&
U^{-1}U(x)[\p_\a-ieA_\a(x)]U^{-1}(x)\psi^\prime(x) \nonumber \\
&=&U^{-1}(x)[{U(x)\p_\a U^{-1}(x)}-ie
U(x)A_\a(x)U^{-1}(x)]\psi^\prime(x) \nonumber \\
&=&U^{-1}(x)[\p_\a-i\p_\a\L(x)-ieA_\a(x)]\psi^\prime(x)\\
&=&U^{-1}(x)[\p_\a-ieA_\a^\prime(x)]\psi^\prime(x), \nonumber 
\end{eqnarray}
where
\begin{equation}
A^\prime_\a \equiv A_\a(x)+ \frac{1}{e}\p_\a\L(x). 
\end{equation}
We observe that the covariant derivative transforms like the field itself:
$D_\a\psi(x)=U^{-1}(x)D^\prime_\a\psi^\prime(x)$, where $D'_{\a} \equiv
\p_\a-ieA'_\a(x)$.  This ensures that $\Lag$ of Eq.~(\ref{qedL}), after
replacing $\p_\a$ by $D_\a$, is invariant under the gauge transformation.

The gauge field strength tensor is defined as $F_{\a\b} \equiv \p_\a
A_\b-\p_\b A_\a$.  Under gauge transformation
\begin{equation}
F^\prime_{\a\b}=\p_\a(A_\b+\frac{1}{e}\p_\b \L)-\p_\b(A_\a+
\frac{1}{e}\p_\a\L)=F_{\a\b}. 
\end{equation}

The kinetic term of gauge field is given by $L_{\rm kin}=- \frac{1}{4}
F_{\a\b}F^{\a\b}$.  One can also write $F_{\a\b}=\frac{1}{e}[D_\a,D_\b]$.  It
is instructive to check that in terms of the electric and magnetic field
components,
\begin{equation}
F^{\mu\nu} = \left(\begin{array}{cccc} 0& -E_1 & -E_2 & -E_3\\
E_1 & 0 & -B_3 & B_2\\
E_2 & B_3 & 0 & -B_1\\
E_3 & -B_2 & B_1 & 0\end{array}\right).
\end{equation}
It follows immediately that $-{1\over 4}F_{\mu\nu}F^{\mu\nu} =
{1\over2}(\vec{E}^2-\vec{B}^2)$, which is the kinetic term.

Let us now concentrate on the non-abelian group SU(2).  Consider a fermion
field $\psi(x)$ which transforms as a doublet under SU(2):
$\psi(x)=\left(\begin{array}{c}\psi^1(x)\\
    \psi^2(x)\end{array}\right)$.  Let us follow its local SU(2)
transformation : $\psi \to \psi^\prime = e^{i\frac{\tau_a}{2}\L_a(x)}\psi(x)$,
and $\bar{\psi} \to \bar{\psi}^\prime
=\bar{\psi}(x)e^{-i\frac{\tau_a}{2}\L_a(x)}$, where $\tau_a(a=1,2,3)$ are the
Pauli matrices which satisfy $[\tau_a,\tau_b]=2i\epsilon_{abc}\tau_c$.  It is
easy to check that $\p_\a\psi(x)=
U^{-1}(x)[\p_\a-i\frac{\tau_a}{2}\p_\a\L_a(x)]\psi^\prime(x)$, i.e. $\psi(x)$
and $\p_\a \psi(x)$ do not transform identically, and hence the Lagrangian is
not invariant under SU(2) transformation. To ensure gauge invariance we must
start with the covariant derivative $D_\a\psi(x)\equiv
\left[\p_\a-ig\frac{\tau_a}{2}A_\a^a(x)\right]\psi(x)$, where $g$ is the
coupling constant (like the symbol $e$ used for U(1)). We obtain
\begin{eqnarray}
  D_\a\psi(x)&=U^{-1} \underbrace{U(x)\left(\p_\a-ig\frac{\tau_a}{2}
      A_\a^a(x)\right)
    U^{-1}(x)}\psi^\prime(x) & = U^{-1}D^\prime_{\a}\psi^\prime(x), \nonumber \\
  &=\p_\a - ig\frac{\tau_a}{2} A_\a^{\prime a} \equiv D^\prime_{\a}
\end{eqnarray}
where
\begin{equation}
A_\a^{\prime a} = A_\a^a + {1\over g} \p_\a\L^a + \epsilon^{abc}A_{\a b}\L_c
\, .
\end{equation}
If we do a straightforward generalization of the abelian case and construct
$G^a_{\a\b} = \p_\a A_\b^a-\p_\b A_\a^a$, the product 
$G^a_{\a\b} G_a^{\a\b}$ is not gauge invariant. We must redefine field the 
strength tensor in the non-abelian case as 
\begin{equation}
\label{fmunu_nonab}
  F^a_{\a\b}
  \equiv (\p_\a A_\b^a-\p_\b A_\a^a)+g\epsilon^{abc}A_{\a b} A_{\b c}.
\end{equation}
It is instructive to use the transformation properties of the gauge fields,
discussed above, to check that $F^a_{\a\b} F_a^{\a\b}$ remains invariant under
gauge transformation, and constitutes the gauge boson kinetic term in the
Lagrangian.

\section{The Standard Model  Higgs mechanism}
Now we will discuss the idea and implementation of Spontaneous Symmetry
Breaking (SSB).  Whenever a system does not show {\em all} the symmetries by
which it is governed, we say that the symmetry is `spontaneously' broken. More
explicitly, when there is a solution which does not exhibit a given symmetry
which is encoded and respected in the Lagrangian, or Hamiltonian, or the
equations of motion, the symmetry is said to be spontaneously broken. In the
context of the SM, the SSB idea is used to generate gauge boson masses without
spoiling the calculability (which we technically call {\em renormalisability})
of the theory. To gain insight into different aspects of SSB, we will consider
different cases one by one.

\subsection{SSB of discrete symmetry}
Consider a real scalar field $\varphi(x)$. The Hamiltonian is given by
\begin{eqnarray}
H &=& {1\over2} {\dot{\varphi}}^2 + 
{1\over2} \left(\vec{\triangledown} \varphi \right)^2 + 
V(\varphi),
~~~~\text{where}~~~
V(\varphi) ={1\over2} m^2 \varphi^2 + {1\over4} \l^2 \varphi^4.
\end{eqnarray}
Above, we have assumed a $\varphi \leftrightarrow - \varphi$ discrete symmetry
which prohibits odd powers of $\varphi$.  Clearly, the minimum of
$V(\varphi)$ is at $\varphi = 0$.  Now, as the next step, consider
\begin{eqnarray}
\label{potnordis}
V(\varphi)& = - {1\over2}m^2\varphi^2+{1\over4}\l^2\varphi^4, 
~~~\text{where}~~~ m^2 > 0.
\end{eqnarray}%
\begin{floatingfigure}[r]{0.3\textwidth} 
\epsfxsize=4cm
\centering{\epsfbox{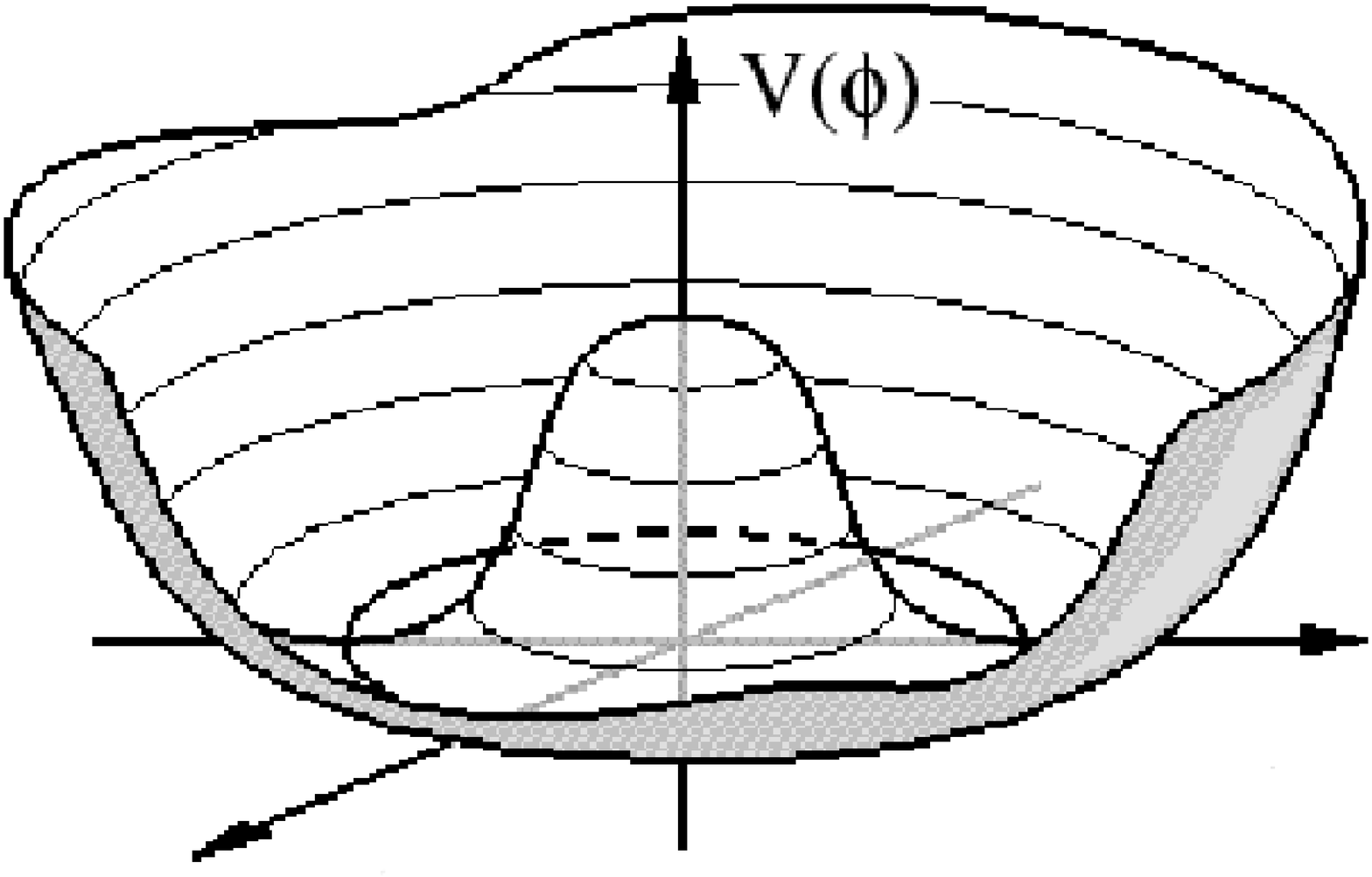}}
  \caption{\small \sf The Mexican hat potential.} 
\label{mexican} 
\end{floatingfigure}%
\noindent 
Since $V^\prime(\varphi)|_{\varphi=0}=0$, it follows that $\varphi = 0$ is
indeed an extremum. Moreover, $V^{\prime\prime} (\varphi)|_{\varphi=0} = -m^2$
implies that $\varphi = 0$ is rather a {\em maximum} and not a minimum.
Stable minima occur at two points $\varphi=\pm{m\over\l}$, where
$V\left({m\over \l}\right) = -{m^4\over4\l^2} $.  Recall, we can always add a
constant term in $V(\varphi)$, which does not change physics. Using this idea,
we can write the potential as a complete square as
\begin{eqnarray}
V(\varphi) &=& {1\over4}\l^2(\varphi^2-v^2)^2, 
\end{eqnarray}
where $v={m\over\l}$. With this redefined potential, the system can be at
either of the two minima ($\pm v$).  Once one solution is chosen, the symmetry
breaks spontaneously. Note, the potential $V(\varphi)$ attains its minimum
value zero for a nonzero value of $\varphi$. The zero energy state,
characterized by $V(\varphi)=0$, is called the ground state or the minimum
energy state, while $v \equiv \langle 0|\varphi|0\rangle$ is called the
`vacuum expectation value (vev)'.

We should remember two points: 
\begin{itemize}
\item When we consider the vev of a field, this field has to be a `classical'
  field. Remember, a quantum field can always be expanded in terms of creation
  and annihilation operators whose vacuum expectation would always vanish.

\item When we write $v \equiv \langle 0|\varphi(x)|0\rangle$, a na\"ive
  question comes to mind as to how the LHS is independent of $x$ while the RHS
  is a function of $x$. It happens because the translational invariance of the
  vacuum can be used to write:
  $$\langle 0|\varphi(x)|0 \rangle =
  \langle 0|e^{ipx}\varphi(0)e^{-ipx}|0\rangle = \langle 0|\varphi|0\rangle =
  v. $$
\end{itemize}

\subsection{ SSB of  global  U(1) symmetry}
For U(1) symmetry, we must start with a {\em complex} scalar field $\varphi$.
The scalar potential is given by
\begin{eqnarray}
V(|\varphi|) & = & {1\over2}\l^2(|\varphi|^2-{1\over2}v^2)^2~~.
\end{eqnarray}
This potential has a global U(1) symmetry: $\varphi \to \varphi' = e^{i\alpha}
\varphi$, where $\a$ is any real constant. The potential is minimum (which is
zero) at all points on the orbit of radius $|\varphi| = v/\sqrt{2}$, different
points corresponding to different values of Arg($\varphi$).  The shape of the
potential takes the form of a `Mexican hat' -- see Fig.~\ref{mexican}.  We
write
\begin{eqnarray}
\varphi(x) = 
{1\over\sqrt{2}} \left(v+ \chi(x)+i\psi(x)\right), 
\end{eqnarray}
where $\chi(x)$ and $\psi(x)$ are the two components of the complex quantum
field around the stable minima.  The Lagrangian in terms of the $\chi$ and
$\psi$ fields can be expressed as
\begin{eqnarray}
\label{lag_global_u1}
\Lag  =  {\cal{K}} - 
{1\over2}\l^2\left[{1\over2}|v+\chi+i\psi|^2-{1\over2}v^2\right]^2 =
{\cal{K}} -{1\over8}\l^2\left[\chi^2+2v\chi+\psi^2\right]^2 ,  
\end{eqnarray}
where ${\cal{K}} = \frac{1}{2} (\p_\mu\chi)^2 + \frac{1}{2} (\p_\mu\psi)^2$ is
the kinetic term.  It is clear from Eq.~(\ref{lag_global_u1}) that the
component of excitation along the $v$-direction ($\chi$) acquires a mass
$(m_\chi=\l v)$, while the component ($\psi$), which is in a direction
tangential to the orbit, remains massless ($m_\psi=0$). That $\psi$ is
massless is not surprising as traversing along the orbit does not cost any
energy. What is important is that as a result of a spontaneous breaking of a
continuous global symmetry, a massless scalar has been generated. Such a
massless scalar field is called the `Nambu-Goldstone boson', or often the
`Goldstone boson'.

\subsection {SSB of global SU(2) symmetry}
Here, the complex scalar field is a doublet of SU(2), given by
$\Phi= \left(\begin{array}{c} \varphi_+\\
    \varphi_0\end{array}\right)$. The Lagrangian is given by
\begin{equation}
\Lag = (\p_\mu{\Phi})^\dagger(\p_\mu\Phi) - V({\Phi}^\dagger\Phi)~~,
\end{equation}
where ${\Phi}^\dagger \Phi =\varphi^*_+\varphi_++\varphi^*_0\varphi_0$.  Here
$\varphi_+$ and $\varphi_0$ have 2 real components each, i.e. there are in
total 4 degrees of freedom (d.o.f.).  At this level, the subscripts $+$ and
$0$ are simply labels. We will see later on that these labels would correspond
to electric charge $+1$ and $0$, respectively.  After SSB, 3 d.o.f. remain
massless, one becomes massive.  It can be proved that the number of Goldstone
bosons is the number of broken generators. To appreciate this from a geometric
point of view, note that a 4d sphere has 3 tangential directions, and clearly,
quantum oscillations along these directions yield massless modes.

Of course, the next question is -- what happens when a global symmetry is
gauged?

\subsection{SSB with local U(1) symmetry}
Now we deal with a local U(1) symmetry. The Lagrangian can be written as
\begin{equation}
\Lag=|D_\mu\Phi|^2 - {1\over2} \l^2 \left(|\Phi|^2-{v^2\over2}\right)^2 - 
{1\over4} F_{\mu\nu}F^{\mu\nu},
\end{equation}
Here we have used a slightly different notation compared to the global U(1)
case.  The complex scalar will be denoted by $\Phi$, which can be written as
$\Phi(x) = \varphi(x) e^{i\theta(x)}$ where $\varphi(x)$ and $\theta(x)$ are
the two real d.o.f. Recall that the covariant derivative and the gauge field
strength tensors are given by
$$
D_\mu = \p_\mu-ie A_\mu~~,~~F_{\mu\nu}=\p_\mu A_\nu - \p_\nu A_\mu \, . 
$$
Now, under gauge transformation $\Phi \to \Phi^\prime = e^{i\alpha(x)} \Phi$
and the Lagrangian still remains invariant.  This phase $\alpha(x)$ is
different at different space-time points, but it is not a physical parameter
and at each and every such point one has the liberty to choose it in such a
way that it precisely cancels the $\t(x)$ at that point. This choice of gauge
is called {\em unitary gauge}.  In other words, $\Phi(x)$ can be chosen to be
real everywhere, and can be written, without any loss of generality, as
\begin{eqnarray}
\Phi(x) =  \varphi(x)= {1\over\sqrt{2}}(v+\chi(x)).
\end{eqnarray}
The Lagrangian takes the following form: 
\begin{eqnarray}
\Lag & = & |(\p_\mu-ieA_\mu)\varphi(x)|^2 -
{\l^2\over2} \left[{1\over2} (v+\chi(x))^2-{1\over2}v^2\right]^2 
-{1\over4}F_{\mu\nu}F^{\mu\nu} \nonumber \\
&=& {1\over2}(\p_\mu \chi(x))^2 + {e^2\over2}A_\mu A_\nu \left(v^2+2v
\chi(x)+\chi^2(x)\right) - {\l^2\over8} \left(2v+\chi(x)\right)^2 \chi^2(x)
-{1\over4}F_{\mu\nu} F^{\mu\nu} . 
\end{eqnarray}
This describes a real scalar field $\chi(x)$ with mass $\l v$ and a massive
vector field $A_\mu$ with a mass $ev$. Note that SSB resulted in a
redistribution of fields: one of the two real fields forming the complex
scalar has been {\em gauged away} but it has reappeared in the form of
longitudinal component of the vector field $A_\mu$. The total number of
d.o.f. thus remains unaltered: 2+2=3+1. The Goldstone boson is {\em eaten up}
by the gauge boson. This is called the {\em Higgs mechanism} and $\chi(x)$ is
called the {\em Higgs field}\footnote{The basic idea of Higgs mechanism was
  borrowed from condensed matter physics. Similar things happen in the BCS
  theory of superconductivity. Electromagnetic gauge invariance is
  spontaneously broken and photon becomes massive inside a superconductor from
  where magnetic fields are repelled due to Meissner effect. For historical
  reasons, the mechanism is also known as Anderson-Higgs mechanism, and,
  Higgs-Brout-Englert-Guralnik-Hagen-Kibble mechanism.}.

\subsection{SSB with local SU(2) symmetry}
Denoting the complex scalar doublet as $\Phi$, the Lagrangian can be written
as
\begin{eqnarray}
L=|D_\mu\Phi|^2-{1\over2}\l^2\left(|\Phi|^2-{1\over2}v^2\right)^2-
{1\over4}F^a_{\mu\nu}F^a_{\mu\nu} \, ,  
\end{eqnarray}
where
\begin{eqnarray}
\Phi & = & \left(\begin{array}{c} \varphi_+\\
\varphi_0\end{array}\right) =
{1\over\sqrt{2}}\left(\begin{array}{l}
\varphi_1+i\varphi_2\\ \varphi_3+i\varphi_4\end{array}\right) \, ,  \\
D_\mu\Phi &=& \left(\p_\mu-{ig\over2}\tau^aW^a_\mu\right)\Phi~~,~~
F^a_{\mu\nu} = \p_\mu W^a_\nu - \p_\nu W^a_\mu + g\epsilon_{abc}W^b_\mu W^c_\nu
~~(a,b=1,2,3). 
\end{eqnarray}
Note that the definition of the field strength tensor $F^a_{\mu\nu}$ follows
from Eq.~(\ref{fmunu_nonab}). Here, $|\Phi|^2 = \varphi^*_+\varphi_+ +
\varphi_0^*\varphi_0 =
{1\over2}(\varphi_1^2+\varphi_2^2+\varphi_3^2+\varphi_4^2)$.  The potential is
minimum when $\sum_i \varphi_i^2 = v^2$, where $v$ is the radius of the orbit.
Without any loss of generality we can assume that the entire vev is in the
$\varphi_3$ direction, i.e.,
$\langle\Phi\rangle=\Phi_0={1\over\sqrt{2}}\left(\begin{array}{c}0\\
    v\end{array}\right)$.  The Higgs field $h(x)$ is the real excitation
around the vev. Thus, in the unitary gauge where the scalar field has only the
real component, $\Phi(x) = {1\over\sqrt{2}} \left(\begin{array}{c} 0 \\
    v+h(x)\end{array}\right)$.

The gauge boson masses arise from the expansion of $|D_\mu\Phi|^2$-piece of
the Lagrangian. This gives
$$
\frac{g^2}{8}
\left|\tau^a W^a_\mu\left(\begin{array}{c}0\\v\end{array}\right)\right|^2 
= \frac{g^2}{8} \left|\left(\begin{array}{cc}
W^3_\mu & W^1_\mu-iW^2_\mu\\ W^1_\mu+iW^2_\mu &
-W^3_\mu\end{array}
\right) \left(\begin{array}{c}0\\v\end{array}\right)\right|^2 
= \left({gv\over 2}\right)^2 W^+_\mu W^-_\mu
+{1\over2}\left({gv\over2}\right)^2 W^3_\mu W^3_\mu , 
$$
where $W^\pm \equiv (W_\mu^1 \mp i W_\mu^2)\sqrt{2}$.  This means that all the
three gauge bosons have the same mass ($gv/2$).  The equality of $W^\pm$ and
$W^3$ masses {\em does not follow from gauge symmetry but results from a
  global `custodial' symmetry}. What is this custodial symmetry?  Looking at
the orbit structure, $\varphi^2_1+\varphi^2_2 + \varphi^2_3 +
\varphi^2_4=v^2$, we note that before the SSB the potential had a SO(4)
symmetry, which is reduced to SO(3) once one direction is fixed for the
vev. The group SO(3) is isomorphic to SU(2). This SU(2) is global and should
not be confused with the SU(2) we gauged. It is {\em this} SU(2) that we call
the custodial SU(2). This remains unbroken even after the vev is generated,
and this unbroken symmetry enforces the equality of the gauge boson masses.
The bottom line is that all the three Goldstone bosons related to the global
SU(2) have now disappeared, and three massive (but degenerate) gauge bosons
have emerged.

\subsection{ SSB with local SU(2) $\times$ U(1) symmetry 
(the Electroweak part of the SM)}
\subsubsection{Why SU(2) $\times$ U(1)?}
Obviously we need two gauge bosons to meet the observations already
made. There has to be a massive charged gauge boson which would mediate beta
decay. The smallest unitary group which provides an off-diagonal generator
(corresponding to the charged gauge boson) is SU(2). The relevant generators
are $\tau^1$ and $\tau^2$.  We further need a massless gauge boson. Any
association of photon with the neutral generator $\tau^3$ would lead to
contradiction with respect to the charge assignment of particles. The gauge
charges of fermions in a doublet coupling to $W^3$ are $\pm {1\over 2}$,
clearly different from the electric charges. Moreover, $W^3$ couples to
neutrino, but photon does not. All in all, just with SU(2) gauge theory we
cannot explain both weak and electromagnetic interactions. The next simplest
construction is to avoid taking a simple group, but consider SU(2) $\times$
U(1).

The covariant derivative will now contain gauge bosons of both SU(2) and U(1):
\begin{eqnarray}
D_\mu &=& \p_\mu - ig \frac{\tau^a}{2} W^a_\mu - 
ig^\prime \frac{Y}{2} B_\mu , 
\end{eqnarray}
where the quantum number $Y$ is the `hypercharge' of the particle on 
which $D_\mu$ acts.

The SM contains 5 representations of fermions (quarks and leptons) for each
generation -- 2 doublets and 3 singlets.
$$
L\equiv \left(\begin{array}{c} \nu\\ e\end{array}\right)_L, ~~e_R\, ,~~
Q \equiv \left(\begin{array}{c} u\\ d\end{array}\right)_L, ~~u_R \, , ~~d_R
$$
$\Psi_L$ and $\Psi_R$ are left- and right-chiral states of a fermion field
$\Psi$, such that $\g_5\Psi_L=-\Psi_L$ and $\g_5\Psi_R=\Psi_R$.

\subsubsection{Notion of hypercharge}
\begin{align*}
\begin{array}{ccccc}_{\left(\begin{array}{c} \nu\\
e\end{array}\right)_L} &\swarrow t_3={1\over2}, & Q=0 &
\therefore Q-t_3=-{1\over2}\\
&\nwarrow t_3=-{1\over2}, & Q=-1 & \therefore Q-t_3=-{1\over2}\end{array}\\
\\
\begin{array}{ccccc}_{\left(\begin{array}{c} u\\
d\end{array}\right)_L} &\swarrow t_3={1\over2}, & Q={2\over3} & \therefore
Q-t_3={1\over6}\\
&\nwarrow t_3=-{1\over2}, & Q=-{1\over3} & \therefore Q-t_3={1\over6}\end{array}
\end{align*}
Note that the $(Q-t_3)$ assignments are same for all members inside a given
multiplet, i.e., the generator corresponding to $(Q-t_3)$ commutes with all
the SU(2) generator $\tau_a$.  Hence, either $(Q-t_3)$ or some multiple of it
can serve as the hypercharge quantum number of $U(1)_Y$. We follow the
convention
\begin{equation}
2(Q-t_3)=Y ~~~~~~~~~\Longrightarrow ~~~{Q = t_3+ {Y\over2}}. 
\end{equation}
It is instructive to check that the currents satisfy $J_\mu^Q = J_\mu^{3} +
{1\over2} J_\mu^Y$.

\subsubsection{How is the symmetry broken?}
If a generator $\hat{O}$ is such that the corresponding operator
$e^{i\hat{O}}$ acting on the vacuum $|0\rangle$ cannot change it, i.e.
$e^{i\hat{O}} |0\rangle = |0\rangle$, then obviously the operation corresponds
to a symmetry of the vacuum and the corresponding generator kills the vacuum,
i.e.  $\hat{O} |0\rangle = 0$. In the context of gauge theory, when the vacuum
is left unbroken by a generator, the gauge boson corresponding to that
generator would remain massless.  Let us now check how the neutral (diagonal)
generators of SU(2) and U(1) act on the scalar vev.
\begin{eqnarray}
t_3\Phi_0 &= {1\over2\sqrt{2}}\left(\begin{array}{cc}1 & 0\\
0&-1\end{array}\right)
\left(\begin{array}{c}0 \\
v\end{array}\right) = {1\over2\sqrt{2}} \left(\begin{array}{c} 0\\
-v\end{array}\right) \ne 0 \, , \nonumber \\
{Y\over2}\Phi_0 &= {1\over2\sqrt{2}}\left(\begin{array}{cc}1 & 0\\
0&1\end{array}\right)
\left(\begin{array}{c}0 \\
v\end{array}\right) = {1\over2\sqrt{2}} \left(\begin{array}{c} 0\\
v\end{array}\right) \ne 0 \, , 
\end{eqnarray}
but $(t_3+{Y\over2})\Phi_0=0$. This means that $Q_{\rm em}= t_3 + {Y\over2}$
is indeed the electromagnetic charge generator and consequently photon is
massless.  This is the {\em only} combination that yields a massless gauge
boson, and the massless state is neither a SU(2) nor a U(1) state, but a mixed
state. In other words, the masslessness of photon is a consequence of the
vacuum being invariant under the operation by $e^{iQ_{\rm em}}$.

\begin{list}{$\bullet$}{\rightmargin=1cm\small\rm}
\item An electrically charged field does not acquire any vev, as otherwise
  charge will be spontaneously broken in the following way: If $\varphi^+$ is
  the charge ($+$) field, then one can write $[Q,\varphi^+] =+\varphi^+$. This
  means that if $\langle 0|\varphi^+|0\rangle = v \ne 0$, then using the
  commutator relation one can show that $Q|0\rangle \ne 0$, i.e. electric
  charge is spontaneously broken!
\end{list}

\subsubsection{Masses of the gauge bosons}
\label{smhiggs}
There are four gauge bosons. One of them is the massless photon, but the
other three are massive. Here we calculate their masses. To do this we look 
into the kinetic term with the covariant derivative.
\begin{eqnarray}
&&|D_\mu\Phi|^2 \Longrightarrow \left|\left(-{ig\over2}\tau^aW^a_\mu -
{ig^\prime\over2}B_\mu\right)\Phi_0\right|^2
= {1\over8}\left|\left(\begin{array}{ll} gW^3_\mu+g^\prime B_\mu & 
\sqrt{2} gW^-_\mu\\
\sqrt{2} gW^+_\mu & -gW^3_\mu+g^\prime B_\mu\end{array}\right)
\left(
\begin{array}{c}0\\ v\end{array}\right)\right|^2 \nonumber \\
&&=\left({1\over2}gv\right)^2W^+_\mu W^-_\mu + {1\over8}v^2(W^3_\mu
~~B_\mu)\left(\begin{array}{ll}g^2 & -gg^\prime\\ -gg^\prime &
g^{\prime2}\end{array}\right) \left(\begin{array}{c}W_\mu^3 \\ B_\mu
\end{array}\right). 
\end{eqnarray}
Clearly, the charged $W^\pm$ gauge boson mass is given by $M_W = gv/2$.
Recall, $W_\mu^\pm$ has been constructed out of $W_\mu^1$ and $W_\mu^2$, the
gauge bosons corresponding to the off-diagonal generators $\tau^1$ and
$\tau^2$.

We now look into the neutral part. The mass matrix in the $(W_\mu^3, B_\mu)$
basis has zero determinant. This is not unexpected as one of the states has to
be the massless photon ($A$). The other eigenstate is the $Z$ boson.  Thus the
orthogonal neutral states and their masses are:
\begin{eqnarray}
A_\mu &=&  \frac{g B_\mu + g^\prime W_\mu^3}{\sqrt{g^2+g^{\prime2}}}~:
~M_A=0 \nonumber \\
Z_\mu &=& \frac{g W_\mu^3 - g^\prime B_\mu}{\sqrt{g^2+g^{\prime2}}}~:
~M_Z= \frac{v}{2} \sqrt{g^2+g'^2}
\end{eqnarray}
Introducing $\cos\t_W \equiv {g\over\sqrt{g^2+g^{\prime2}}}$, $\sin\t_W \equiv
{g^\prime\over\sqrt{g^2+g^{\prime2}}}$, where $\theta_W$ is called the weak
angle, one can express
\begin{eqnarray}
\label{photonZ}
A_\mu &=& \cos\t_W B_\mu + \sin\t_W W^3_\mu , \nonumber \\
Z_\mu &=& \cos\t_W W^3_\mu - \sin\t_W B_\mu , \\
{M_W\over M_Z} &=& {{1\over2}gv\over{v\over2}\sqrt{g^2+g^{\prime2}}}
= \cos\t_W \nonumber.
\end{eqnarray}
Observe that $M_Z > M_W$, i.e. the custodial symmetry associated with the
SU(2) gauge group is broken, and it has been broken by hypercharge mixing,
i.e. by expanding the gauge group to SU(2) $\times$ U(1). One can easily check
that in the $g'\to 0$ limit, one recovers the custodial
symmetry. Experimentally, $M_Z = 91.1875 \pm 0.0021$ GeV and $M_W = 80.399 \pm
0.025$ GeV, which are almost the same values as predicted by the SM. The weak
mixing angle is given by $\sin^2\t_W \simeq 0.23$.

We will here define an important parameter:
\begin{equation}
\rho \equiv {M_W^2 \over M_Z^2 \cos^2\t_W}. 
\end{equation}
With the SU(2) doublet scalar representation (and at tree level), one can
easily check from the above relations that $\rho = 1$.  Experimental
measurements on the $Z$ pole at LEP also indicate $\rho$ to be very close to
unity within a {\em per mille} precision.

\begin{list}{$\bullet$}{\rightmargin=1cm\small\rm}
\item If there are several representations of scalars whose
    electrically neutral members acquire vev's $v_i$, then 
\begin{equation}
\rho \equiv {M_W^2\over M_Z^2\cos^2\t_W} =
{\displaystyle{\sum_{i=1}^N} v_i^2[T_i(T_i+1)-{1\over4}Y_i^2]\over
{\displaystyle\sum_{i=1}^N} {1\over2} v_i^2Y^2_i} \, , 
\end{equation}
where $T_i$ and $Y_i$ are the weak isospin and hypercharge of the $i$-th
multiplet.  It is easy to check that only those scalars are allowed to acquire
vevs which satisfy $(2T+1)^2 - 3Y^2 =1$, as otherwise $\rho = 1$ will not be
satisfied at the tree level.  The simplest choice is to have a scalar with
$T=\frac{1}{2}$ and $Y=1$, which corresponds to the SM doublet $\Phi$. More
complicated scalar multiplets, e.g. one with $T=3$ and $Y=4$, also satisfy
this relation.
\end{list}

\subsubsection{Couplings of photon, $Z$ and $W^{\pm}$ with fermions}
The interaction of the gauge bosons with the fermions arise from $i \bar{\Psi}
\gamma^\mu D_\mu \Psi$, where $D_\mu = \p_\mu - ig \frac{\tau^a}{2} W^a_\mu -
ig^\prime \frac{Y}{2} B_\mu$.  In the SM, a generic fermion field ($\Psi$) has
a left-chiral SU(2)-doublet representation: $\Psi_L =
\left(\begin{array}{c}\psi_1\\ \psi_2\end{array}\right)_L$, and right-chiral
SU(2)-singlet representations: $\psi_{1R}$ and $\psi_{2R}$.

Now we look into the charged-current interaction. We write the relevant part
of the Lagrangian as
\begin{eqnarray}
\Lag_{CC} &=& \frac{g}{2}(J^1_\mu W^1_\mu + J^2_\mu W^2_\mu), ~~~{\rm where}~~~
J^{1,2}_\mu = \bar{\Psi}\g_\mu P_L\tau^{1,2}\Psi 
~~~({\rm using}~~P_{L,R} \equiv (1 \mp \g_5)/2) \, . 
\end{eqnarray}
Expressing $W_\mu^{\pm} = (W_\mu^1 \mp iW_\mu^2)/\sqrt{2}$, we rewrite the
charged-current Lagrangian as
\begin{eqnarray}
\Lag_{CC} & =& \frac{g}{\sqrt{2}} \left[\bar{\psi}_1\g_\mu
P_L\psi_2W^+_\mu+\bar{\psi}_2\g_\mu P_L\psi_1W^-_\mu\right]. 
\end{eqnarray}
Now we come to the neutral-current part. We can express the Lagrangian as 
\begin{eqnarray}
\Lag_{NC} & = & \frac{g}{2} J^3_\mu W^3_\mu + \frac{g^\prime}{2} J_\mu^Y B_\mu, 
~~~~{\rm where} \nonumber \\
J_\mu^3 &=& \bar{\Psi} \g_\mu P_L\tau^3 \Psi~~, 
J_\mu^Y = \Psi \g_\mu P_LY_L \Psi + \bar{\psi}_1 \g_\mu P_R Y^1_R \psi_1
+ \bar{\psi}_2 \g_\mu P_R Y^2_R \psi_2 , 
\end{eqnarray}
where $Y_L$ is the hypercharge of the left-handed doublet while $Y_R^1$ and
$Y_R^2$ are hypercharges of the two right-handed singlets.  Now rewriting
$W^3$ and $B$ in terms of the photon ($A$) and the $Z$ boson, as $W_\mu^3 =
\cos\t_WZ_\mu + \sin\t_WA_\mu$ and $B_\mu = -\sin\t_WZ_\mu + \cos\t_WA_\mu$,
one can write the neutral current Lagrangian in the $(A,Z)$ basis as
\begin{eqnarray} 
\Lag_{NC} &=& J^Q_\mu A_\mu + J^Z_\mu Z_\mu , ~~~{\rm where} \nonumber \\
J_\mu^Q &=& e Q_i \bar{\psi}_i \g_\mu \psi_i , 
~~\text{with} ~ e \equiv g\sin\t_W , ~\text{and sum over $i$ implied,}\\
J^Z_\mu &=& {g\over\cos\t_W} \left[a_L^i \bar{\psi}_i \g_\mu
P_L \psi_i + a_R^i \bar{\psi}_i \g_\mu P_R \psi_i\right], ~\text{with}~
a_L^i \equiv t_3^i - Q_i \sin^2 \t_W, ~~ a_R^i \equiv  - Q_i \sin^2 \t_W .  
\end{eqnarray}
As we observe, the $Z$ boson couples to the left- and right-handed fermions
with different strengths. Quite often, the $Z$ boson's interaction with
fermions are expressed in terms of vector and axial-vector couplings, which
are simply linear combinations of $a_L$ and $a_R$. Thus, for a given fermion
$f$, the $Zf\bar f$ vertex is given by,
\begin{equation}
{g \over \cos\t_W} \g_\mu \left(a^f_LP_L + a^f_RP_R\right) \equiv 
{g\over2\cos \t_W} \g_\mu \left(v^f -a^f \g_5\right), ~\text{where}~~
v^f \equiv t^f_3-2Q_f\sin^2\t_W, ~ a^f \equiv t^f_3 \, ,
\end{equation}
are the tree level couplings of the $Z$ boson to the fermion $f$.

\subsubsection{The decay width of the $Z$ boson}
The $Z$ boson decays into all $f\bar{f}$ pair, except the $t\bar{t}$ because
$m_t \simeq 173$ GeV, while $M_Z \simeq 91$ GeV. The expression of the decay
width of the $Z$ boson in the $f\bar{f}$ channel is given by ({\em the
  derivation can be found in text books})
\begin{eqnarray}
\G_f &=& {G_F\over6\pi\sqrt{2}} M_Z^3 \left(v^2_f + a_f^2\right) 
f\left({m_f\over M_Z}\right), ~~\text{where}~~ \\
f(x)&=& (1-4x^2)^{1/2}\left(1-x^2+3x^2 {v_f^2-a_f^2\over v^2_f+a_f^2}\right). 
\nonumber 
\end{eqnarray}
One can easily verify some of the SM predictions of the $Z$ boson properties:
total decay width $\G_Z \simeq 2.5$ GeV, hadronic decay width $\G_{\rm had}
\simeq 1.74$ GeV, charged lepton decay width (average of $e,\mu,\tau$)
$\G_\ell \simeq 84.0$ MeV, invisible decay width (into all neutrinos) $\G_{\rm
  inv} \simeq 499.0$ MeV, hadronic cross section (peak) $\s_{\rm had} \simeq
41.5$ nano-bern \cite{pdg}.  

While doing the algebraic manipulation it will be useful to remember that the
Fermi coupling $G_F$ can be expressed in many ways:
\begin{equation}
{G_F\over\sqrt{2}} = {g^2\over 8M_W^2} = \frac{1}{2 v^2} = 
{g^2\over 8M_Z^2 \cos^2\t_W} = {e^2\over 8M_Z^2 \sin^2 \t_W \cos^2\t_W} \, .
\end{equation}

\section{The LEP Legacy}
\subsection{Cross section and decay width}
Let us consider the total cross section of $e^+e^- \to \mu^+\mu^-$ mediated by
the photon and the $Z$ boson. It is given by ($\sqrt{s} =$ c.m. energy)
\begin{eqnarray}
\label{sigma-a1}
\s &=& {4\pi\a^2\over3 s}(1 + a_1) , ~~\text{where}\\
a_1 &=& 2v_\ell^2f_Z + (v_\ell^2 + a_\ell^2)^2 f_Z^2 , ~~\text{with}~~
f_Z = {s\over s - M_Z^2} \left({1\over \sin^2 2\t_W}\right) . \nonumber
\end{eqnarray}
Note that the effect of the $Z$ mediation is encoded in $a_1$, whereas setting
$a_1=0$ we get the contribution of the photon.  For the leptons
$\ell=e,\mu,\tau$, $v_\ell \propto (1-4\sin^2 \t_W) \sim$ zeroish.  Therefore,
\begin{equation}
\s(e^+e^-\longrightarrow^{^{\hspace{-0.5cm}\g,Z}}
~\mu^+\mu^-) \simeq {4\pi\a^2\over3s} \left[1+{1\over16\sin^42\t_W}
{s^2\over(s-M_Z^2)^2}\right].
\end{equation}
Thus, in the vicinity of $\sqrt{s}=M_Z$, we would expect a sharp increase of
cross section. This is the sign of a resonance of the $Z$ boson mediation.
But, in reality the cross section does not diverge at $s=M_Z^2$. The reason is
that the $Z$ boson has a decay width $\G_Z$, which would lead to the following
modification:
$$
{s^2\over(s-M^2_Z)^2} \to
{s^2\over\left[s-|M_Z-{i\over 2}\G_Z|^2\right]^2}
$$
The factor $\frac{1}{2}$ in front of $\G_Z$ comes from the definition of the
width as half width at the maximum. Consequently, 
\begin{eqnarray}
\s_{\max} &\simeq& {4\pi\a^2\over3M_Z^2}
\left[1+{1\over 16 \sin^42\t_W} {M_Z^2\over\G^2_Z}\right]
\simeq {4\over27} {\pi \a^2\over\G_Z^2}.
\end{eqnarray}
In general, for $e^+e^-\to f\bar{f}$, $(v_\ell^2+a_\ell^2)^2$ should be
replaced by $(v^2_e+a^2_e)(v^2_f+a_f^2)$, i.e., $f$ is not necessarily
$\mu$. Therefore,
\begin{eqnarray}
\s(e^+e^-\to f\bar{f})|_{s=M_Z^2-{\G_Z^2\over4}} \simeq
{4\pi\a^2\over3M_Z^2}\left[1+{(v^2_e+a_e^2)(v_f^2+a_f^2)\over
\sin^42\t_W} {M_Z^2\over\G_Z^2}\right].
\end{eqnarray}
Substituting 
$\G_f=\a M_Z(v_f^2+a_f^2)/\left(3\sin^2 2\t_W\right)$, we obtain  
\begin{equation}
\s_{\max}(e^+e^-\to
f\bar{f}) \simeq {4\pi\a^2\over3M_Z^2}\left(1+{9\over\a^2}
{\G_e\G_f\over\G_Z^2}\right).
\end{equation}
Numerically, $9 \G_e \G_f \gg \a^2 \G_Z^2$. Thus we arrive at the {\em Master
  Formula}:
\begin{equation}
\s_{\max}^f \simeq {12\pi\over M_Z^2} {\G_e\G_f\over\G_Z^2}.
\end{equation}

Now, we make some important observations. 
\begin{enumerate}
\item From the peak position of the Breit-Wigner resonance, we can measure 
$M_Z$ for any final state $f$. 

\item The half-width at the maximum gives us the {\em total} width $\G_Z$ for
  any final state $f$.

\item By measuring Bhabha scattering cross section ($\s^e$) at the $Z$ pole,
  we can calculate $\G_e$.

\item By measuring the peak cross section for any other final state
  ($f=e,\mu,\tau, \text{hadron}$), we can calculate the corresponding $\G_f$. 

\item Since neutrinos are invisible, we cannot directly measure the neutrino
  decay width. But the total invisible decay width $\G_{\rm inv} = \G_Z -
  \G_{\rm visible} = \G_Z - \G_e - \G_\mu - \G_\tau - \G_{\rm had}$.
  
\item The number of light neutrinos is $N_\nu = \G_{\rm inv}/ \G_\nu^{\rm SM}
  = 2.984 \pm 0.008$, which for all practical purposes is 3.
\end{enumerate}

\subsection{Forward-backward asymmetry}
The differential cross section in the $\ell^+ \ell^-$ channel
($\ell=\mu,\tau$) is given by
\begin{eqnarray} 
  \frac{d\s}{d\Omega} (e^+e^-\xrightarrow{\g,Z}
  \ell^+ \ell^-)&=&\frac{e^4}{64 \pi^2 s} 
\left[(1+a_1)(1+\cos^2\t)+a_2\cos\t\right],
\end{eqnarray}
where, $a_1$ has been defined in Eq.~(\ref{sigma-a1}), and
$a_2 = 8v_\ell^2 a_\ell^2 f_Z^2 + 4a_\ell^2 f_Z$ . 

The $a_1$ contribution has the same angular dependence - $(1+\cos^2\t)$ - as
in QED. The $a_2$ contribution makes a vital qualitative and quantitative
difference by introducing a term proportional to $\cos\t$. This term arises
due to interference between vector and axial-vector couplings. This gives rise
to the forward-backward asymmetry, which is defined as
\begin{equation}
A_{\rm FB}^l =  {\int_0^{\pi/2} d\t \sin\t {d\s\over d\Omega} -
\int_{\pi/2}^\pi d\t \sin\t {d\s\over d\Omega}\over \int_0^{\pi/2}
d\t \sin\t {d\s\over d\Omega} + \int_{\pi/2}^\pi d\t \sin\t
{d\s\over d\Omega}} = {3\over8} \left({a_2\over1+a_1}\right). 
\end{equation}

\begin{list}{$\bullet$}{\rightmargin=1cm\small\rm}
\item Even though top quark could not be produced at LEP due to kinematic
  reason, its existence was inferred from the measurement of $\G_b \equiv \G
  (Z \to b\bar b)$ and the forward-backward asymmetry $A_{\rm FB}^b$ in the
  following way. Note
\begin{eqnarray}
\G^{\rm SM}_b &=& \frac{G_FM_Z^3}{3\pi\sqrt{2}} 
\left[\left(a_L^b\right)^2 + \left(a_R^b\right)^2\right] 
= \frac{G_FM_Z^3}{3\pi\sqrt{2}} 
\left[\left(t_3^b - Q_b \sin^2\t_W\right)^2 + 
\left(- Q_b \sin^2\t_W\right)^2  \right] \nonumber \\
&=& \frac{1.166 \cdot 10^{-5} ~{\rm GeV}^{-2} \times (91.2 ~{\rm GeV})^3}
{3 \pi \sqrt{2}} \left[\left(-\frac{1}{2} + \frac{1}{3} \times 0.23 
\right)^2 + \left(\frac{1}{3} \times 0.23\right)^2  \right]
\simeq 376 ~{\rm MeV} \, .
\end{eqnarray}
If the top quark did not exist, i.e. the bottom quark were a SU(2) singlet,
its isospin would have been zero. In that situation, by putting $t_3^b = 0$ in
the above formula, we would get $\G_b \simeq 23.5$ MeV. Even though the
lighter quarks could not be well discriminated from one another, bottom
tagging was quite efficient thanks to the micro-vertex detector at LEP. As a
result, $\G_b$ could be measured with good accuracy and the measurement was
very close to the SM value. The discrepancy (between 376 MeV and 23.5 MeV) was
too much to be {\em blamed} to radiative corrections! The immediate conclusion
was that the bottom quark should have a partner, the top quark. But is the
bottom an isospin `minus half' or a `plus half' quark? The measured decay
width is consistent with $t_3^b = -\frac{1}{2}$. One could reach the same
conclusion from the measurement of $A_{\rm FB}^b$. If the bottom quark were an
SU(2) singlet, its coupling to the $Z$ boson would have been vector-like and
$A_{\rm FB}^b$ would have been identically zero. But LEP measured a
statistically significant non-vanishing asymmetry. Moreover, $A_{\rm FB}^b$ is
sensitive to $a_b = t_3^b$ ({\em not} $a_b^2$). This way too it was settled
that $t_3^b = -\frac{1}{2}$. Thus even before the top quark was discovered,
not only its existence was confirmed but also all its gauge quantum numbers
were comprehensibly established by studying how the $Z$ boson couples to the
bottom quark. Measurements of electroweak radiative effects at LEP further
provided some hint of what would be the expected value of the top mass. This
will be discussed in the context of the quantum corrections to the tree level
value of the $\rho$ parameter.
\end{list}

\subsection{Main radiative corrections}
The main radiative corrections relevant at the $Z$-pole originate from one
particle irreducible gauge boson two-point functions.  A generic
fermion-induced two point correlation function with gauge bosons in the two
external lines has the following structure ($\l$ and $\l^\prime$ can be $+1$
or $-1$):
\begin{eqnarray}
&&X^{\mu\nu} (m_1,m_2,\l,\l') = (-) \int {d^4k\over(2\pi)^4}
\frac{{\rm Tr} \left\{\g^\mu {1-\l\g_5\over2}(\slashed{q} + \slashed{k} +m_1)
\g^\nu {1-\l^\prime\g_5\over2} (\slashed{k} +m_2)\right\}} 
{\{(q+k)^2-m_1^2\}(k^2 -m_2^2)} \nonumber \\ 
 &&= \frac{i}{16\pi^2} \int_0^1dx \left[\D-\ln
 \left\{\frac{-q^2 x(1-x)+m_1^2 x + m_2^2(1-x)}{\mu^2}\right\}\right]
 \left[2(1+\l\l^\prime)x(1-x)(q_\mu q_\nu-q^2g_{\mu\nu}) \right. \nonumber \\
&&\left. ~~~~+ (1+\l\l^\prime)(m_1^2x+m^2_2(1-x))g_{\mu\nu}-(1-\l\l^\prime)m_1m_2
 g_{\mu\nu}\right] \, . 
\end{eqnarray}
Above, $m_1$ and $m_2$ are the masses of the fermions inside the loop, and
$\Delta (\equiv 2/(4-d) - \g + \ln 4\pi)$ is a measure of divergence in the
dimensional regularization scheme.  The terms of our interest are proportional
to $g_{\mu\nu}$, which we will call $F$. Below, we will write the
$\Pi$-functions, which are defined as $\Pi (q^2, m_1, m_2) = -i X(q^2, m_1,
m_2)$. By putting $\l=1$ and $\l'=1$, we will get the left-left (LL)
$\Pi$-function, given by
\begin{eqnarray}
\Pi_{LL}(q^2,m_1^2,m_2^2)&=& -{1\over 4\pi^2}\int_0^1 dx \left[\D+\ln
{\mu^2\over -q^2x(1-x)+ M^2(x)}\right] \left[q^2x(1-x)-{1\over2}M^2(x)\right],
\\ 
&& \text{where}~~~ M^2(x)=m_1^2x+m_2^2(1-x) \nonumber.
\end{eqnarray}
As before, we denote the SU(2) currents by $J_\mu^i$. Then 
\begin{eqnarray}
\Pi_{33}(q^2) &=& \left<J_\mu^3,J_\mu^3\right> = 
t^2_{3L} \Pi_{LL}(q^2,m^2,m^2) \, ,  \\
\Pi_{11}(q^2) &=& \left<J_\mu^+,J_\mu^-\right> = 
\frac{1}{2} \Pi_{LL}(q^2,m_1^2,m_2^2) \, .
\end{eqnarray}
Now, supposing $m_1$ and $m_2$ are the masses of the two fermion states
appearing in a SU(2) doublet, it immediately follows that 
\begin{eqnarray}
\Pi_{33}(q^2) &=& \frac{1}{4} \left[\Pi_{LL}(q^2,m_1^2,m_1^2)
  + \Pi_{LL}(q^2,m_2^2,m_2^2)\right] \, ,  \nonumber \\
\Pi_{11}(q^2) &=& \frac{1}{2} \Pi_{LL}(q^2,m_1^2,m_2^2) \, .
\end{eqnarray}
The $\rho$ parameter, which is unity at tree level (discussed earlier),
receives one-loop radiative correction due to the mass splitting $m_1 \ne
m_2$. This is a consequence of the breaking of custodial SU(2) due to weak
isospin violation. The effect is captured by
\begin{equation}  
\D\rho \equiv \a T = \a {4\pi\over
\sin^2 \t_W \cos^2 \t_W M_Z^2}\left[\Pi_{11}(0)-\Pi_{33}(0)\right]. 
\end{equation}
The dominant effect of isospin violation indeed comes from top-bottom mass
splitting, given by 
\begin{equation}
\label{rho-1}
\D\rho^{t-b} = \a{4\pi\over \sin^2 \t_W \cos^2 \t_W M_Z^2} {N_c\over 32\pi^2}
\left[{m_t^2+m_b^2\over2} - {m_t^2m_b^2\over m_t^2-m_b^2} \ln
{m_t^2\over m_b^2}\right] \simeq {\a\over\pi} {m_t^2\over M_Z^2} . 
\end{equation}
The last step follows from the approximation that the ratio $(m_b^2/m_t^2)$ is
very small.  The dependence on the fermion mass is quadratic because the
longitudinal gauge bosons are equivalent to the Goldstones whose coupling to
fermions are proportional to the fermion mass. Also note that in the limit
$m_t = m_b$, the contribution to $\D \rho$ vanishes, as expected.

The Higgs contribution is milder in the sense that the dependence on the Higgs
mass is logarithmic. The contribution arises from $ZZh$ and $W^+W^-h$
interactions. It turns out that
\begin{equation}
\label{rho-2}
\D\rho^h = - \frac{3G_F}{8\pi^2\sqrt{2}} (M_Z^2 - M_W^2) 
\ln\left(\frac{m_h^2}{M_Z^2}\right) \simeq
-{\a\over2\pi}\ln{m_h\over M_Z} .
\end{equation}
The Higgs contribution to $\D\rho$ follows from custodial SU(2) violation due
to hypercharge mixing, i.e. the fact that the gauge group is not just SU(2)
but SU(2) $\times$ U(1).  Besides $T (\equiv \Delta \rho/\a)$, two more
parameters $S$ (isospin preserving) and $U$ (isospin violating but less
important than $T$) capture the radiative effects. The $S$ parameter is
particularly sensitive to non-decoupled type of physics (see definition
below).  The Higgs contribution to the $S$ parameter is again logarithmic:
\begin{equation}
\label{s}
  S \equiv \frac{16\pi} {M_Z^2}\left[\Pi_{3Y}(0) - \Pi_{3Y}(M_Z^2)\right]
  \longrightarrow^{^{\hspace{-0.5cm}\text{Higgs}}}
\frac{1}{6\pi} \ln\left(\frac{m_h}{M_Z} \right) .
\end{equation} 

\begin{list}{$\bullet$}{\rightmargin=1cm\small\rm}
\item Note that Bose symmetry does not admit $Zhh$ coupling.  The $Z$ boson is
  a spin-$1$ particle. If it has to decay into two scalars, then the system of
  two scalars would be in an antisymmetric $l=1$ state and there is no other
  quantum number to symmetrize the system of two identical Bose particles.
  One can also argue as follows: The $Z$ boson couples in a gauge
  invariant manner through the corresponding $F_{\mu\nu}$, but $\p_\mu h
  \p_\nu h$ being symmetric in ($\mu,\nu$) would not couple to $F_{\mu\nu}$.

  \item {\em $S,T,U$: Why just three?} There are four two-point functions:
    $\Pi_{\g\g}(q^2), \Pi_{\g Z}(q^2), \Pi_{ZZ}(q^2),
    \Pi_{WW}(q^2)$. Measurements have been made at two energy scales: $q^2=0,
    M_Z^2$. So there are eight two-point correlators (four types at two
    different scales). Of these eight, $\Pi_{\g\g}(0)=\Pi_{\g Z}(0) =0$ due to
    QED Ward identity. Of the remaining six, three linear combinations are
    absorbed in the redefinition of the experimental inputs: $\alpha$, $G_\mu$
    (Fermi coupling extracted from muon decay) and $M_Z$. The remaining three
    independent combinations are $S$, $T$ and $U$. The parameters $T$ and $U$
    capture the effects of custodial and weak isospin violation, while $S$ is
    custodially symmetric but weak isospin breaking \cite{stu}\footnote{A
      generalization of the number of such parameters required to cover {\em
        all} electroweak results was done in \cite{Barbieri:2004qk}.}.

  \item Through the total and partial $Z$ decay width measurements, LEP
    settled the number of light families to be just 3. What about heavier
    ($>M_Z/2$) families, which cannot be produced at LEP due to kinematic
    inaccessibility? If the heavier generations are {\em chiral}, i.e. receive
    mass through Higgs mechanism, then no matter how heavy they are, there is
    a (non-decoupled) contribution to the $S$ parameter ($S = 2/3\pi$ for each
    degenerate chiral family) \cite{stu}. After maintaining consistency with
    precision electroweak data, a heavy {\em fourth} chiral family can be
    barely accommodated. This has a lot of interesting consequences, e.g. it
    broadens the allowed range of the Higgs mass \cite{Kribs:2007nz}.

\end{list}

\subsection{Measurements of the radiative effects}
The $\rho$ parameter is essentially the wavefunction renormalization of the
external $Z$ boson line. Therefore, it is of paramount importance in the
context of LEP physics. There are three places where radiative corrections
enter in a sizable fashion: ($i$) the vector ($v_f$) and axial vector ($a_f$)
couplings receive an overall $\sqrt{\rho}$ multiplication, ($ii$) the weak
angle $\t_W$ is modified to effective $\bar{\t}_W$, and ($iii$) the
$Zb\bar{b}$ vertex receives a large ($m_t^2$-dependent) radiative
correction. We will not talk about the $Zb\bar{b}$ vertex any more.  The
other corrections are called `oblique' corrections which are lumped inside the
following parametrization:
\begin{eqnarray}
v_f = \sqrt{\rho} \left(t_3^f - 2 Q_f \sin^2\bar{\t}_W\right) , ~~~
a_f = \sqrt{\rho} t_3^f .
\end{eqnarray}
Now note that the width $\G_f \propto (v_f^2 + a_f^2)$, while the
forward-backward asymmetry $A_{\rm FB}^f$ is a function of $v_f/a_f$. So,
through a combined measurement of $\G_f$ and $A_{\rm FB}^f$, one can measure
$v_f$ and $a_f$. It is then straightforward to compare the measured $v_f$ and
$a_f$ with their radiatively corrected SM expectations. Noting,
$\sin^2\bar{\t}_W \simeq \sin^2\t_W - \frac{3}{8} \D\rho$, it is intuitively
clear that one can make a prediction on the Higgs mass, as the top quark mass
is now known to a pretty good accuracy.

\begin{list}{$\bullet$}{\rightmargin=1cm\small\rm}
\item To appreciate why radiative corrections became necessary not long after
  LEP started running, let us look back into the situation of summer 1992
  \cite{Novikov:1992rj}: the measured $v_\ell^{\rm exp} =
  -0.0362^{+0.0035}_{-0.0032}$, when compared with its tree level SM
  prediction $v_\ell^{(\rm SM, tree)} = - 0.5 + 2 \sin^2\theta_W = -0.076$
  ($\sin^2\theta_W$ obtained from the muon decay data: $G_\mu =
  \pi\alpha(0)/\sqrt{2}M_Z^2\sin^2\theta_W\cos^2\theta_W$), showed a
  $13\sigma$ discrepancy, inevitably calling for the necessity of dressing the
  Born-level prediction with radiative corrections. However, just the
  consideration of running of the electromagnetic coupling $\alpha(0)
  \rightarrow \alpha(m_Z)$ and extracting $\sin^2\theta$ (to replace
  $\sin^2\theta_W$ in the expression of $v_\ell$) from $\cos^2\theta
  \sin^2\theta = \pi\alpha(M_Z)/\sqrt{2}G_\mu M_Z^2$, enabled one to obtain
  $v_\ell = -0.037$, i.e. within $1\sigma$ of its experimental value at that
  period. The essential point is that it was possible to establish a
  significant consistency between data and predictions just by considering the
  running of $\alpha$ and it was only much later, with a significantly more
  data, that the weak loop effects (${\cal{O}} \left(G_F m_t^2\right)$) were
  felt. In fact, before the discovery of the top quark at Fermilab (1995), the
  main indirect information on the top quark mass used to come from $\D\rho$.
\end{list}

\section{Constraints on the Higgs mass}
\subsection{Electroweak fit}
As emphasized in the previous section, the Higgs mass enters electroweak
precision tests through $\Delta \rho$ and $S$.  The quantum corrections, as we
noticed in Eqs.~(\ref{rho-1},\ref{rho-2},\ref{s}), exhibit a logarithmic
sensitivity to the Higgs mass:
\begin{eqnarray}
\label{st}
\Delta \rho^{\rm SM}  \simeq {\a\over\pi} {m_t^2\over M_Z^2} -
{\a\over2\pi}\ln\left(m_h\over M_Z\right) ,~~  
 S^{h({\rm SM})} \simeq \frac{1}{6\pi} \ln\left(\frac{m_h}{M_Z} \right) .
\end{eqnarray}
At present, the CDF and D0 combined estimate is $m_t = 173.3 \pm 1.1$ GeV
(updated July 2010 \cite{lepewwg}). This translates into an upper limit on the
Higgs mass: $m_h < 186$ GeV at 95\% C.L.  The lower limit $m_h > 114.4$ GeV on
the Higgs mass is obtained from non-observation of the Higgs by direct search
at LEP-2 via the Bjorken process $e^+e^- \to Zh$ \cite{lepewwg}. Why the limit
is so is not difficult to understand: simple kinematics tells us that the
limit should roughly be $\sqrt{s} - M_Z \simeq 205 - 91 = 114$ GeV, where
$\sqrt{s}$ is the maximum c.m.~energy at LEP-2.

\begin{figure*}
\begin{minipage}[t]{0.42\textwidth}
\epsfxsize=6cm
\centering{\epsfbox
{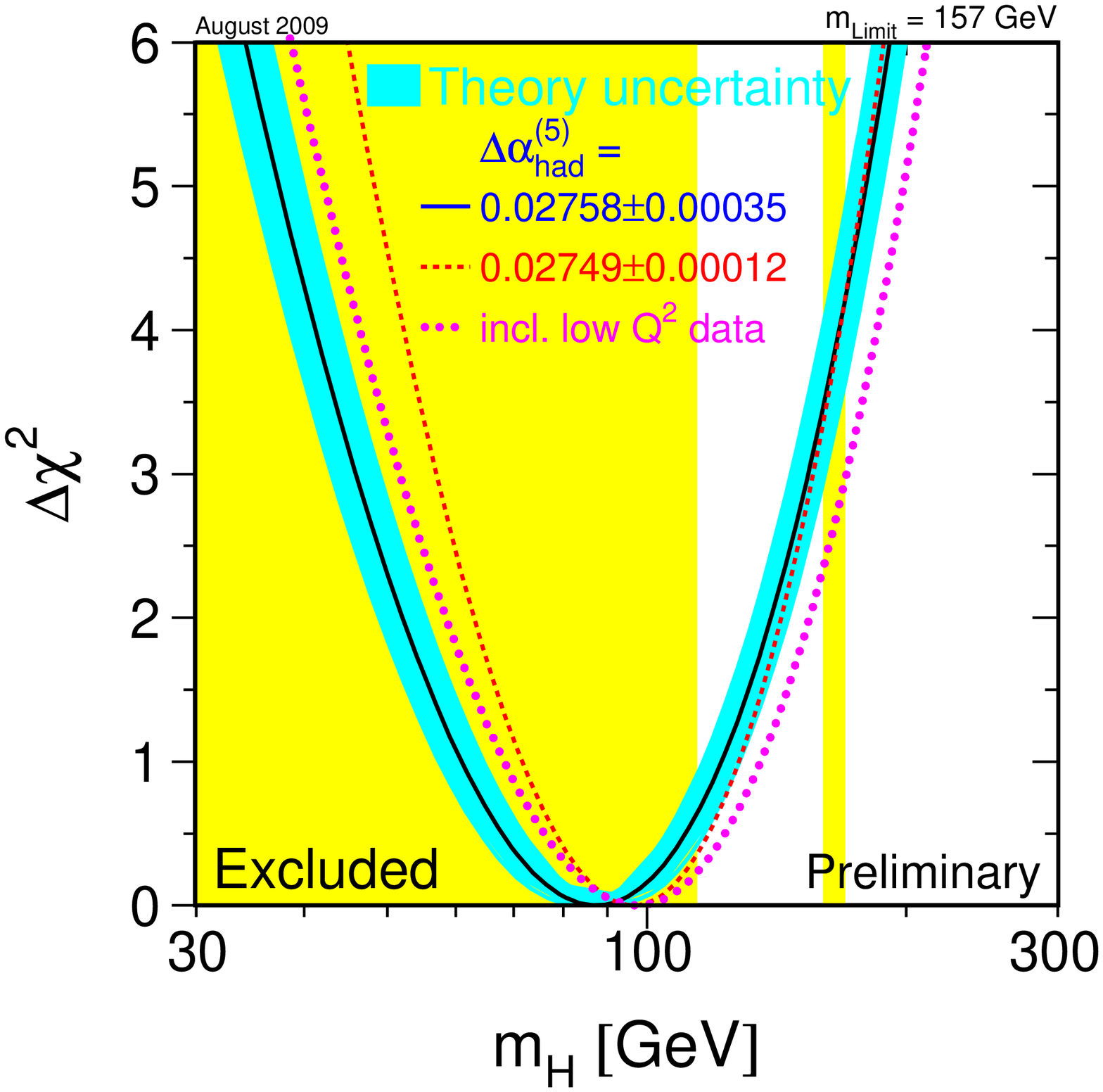}}
\end{minipage}
\hfill
\begin{minipage}[t]{0.42\textwidth}
\epsfxsize=6cm
\centering{\epsfbox
{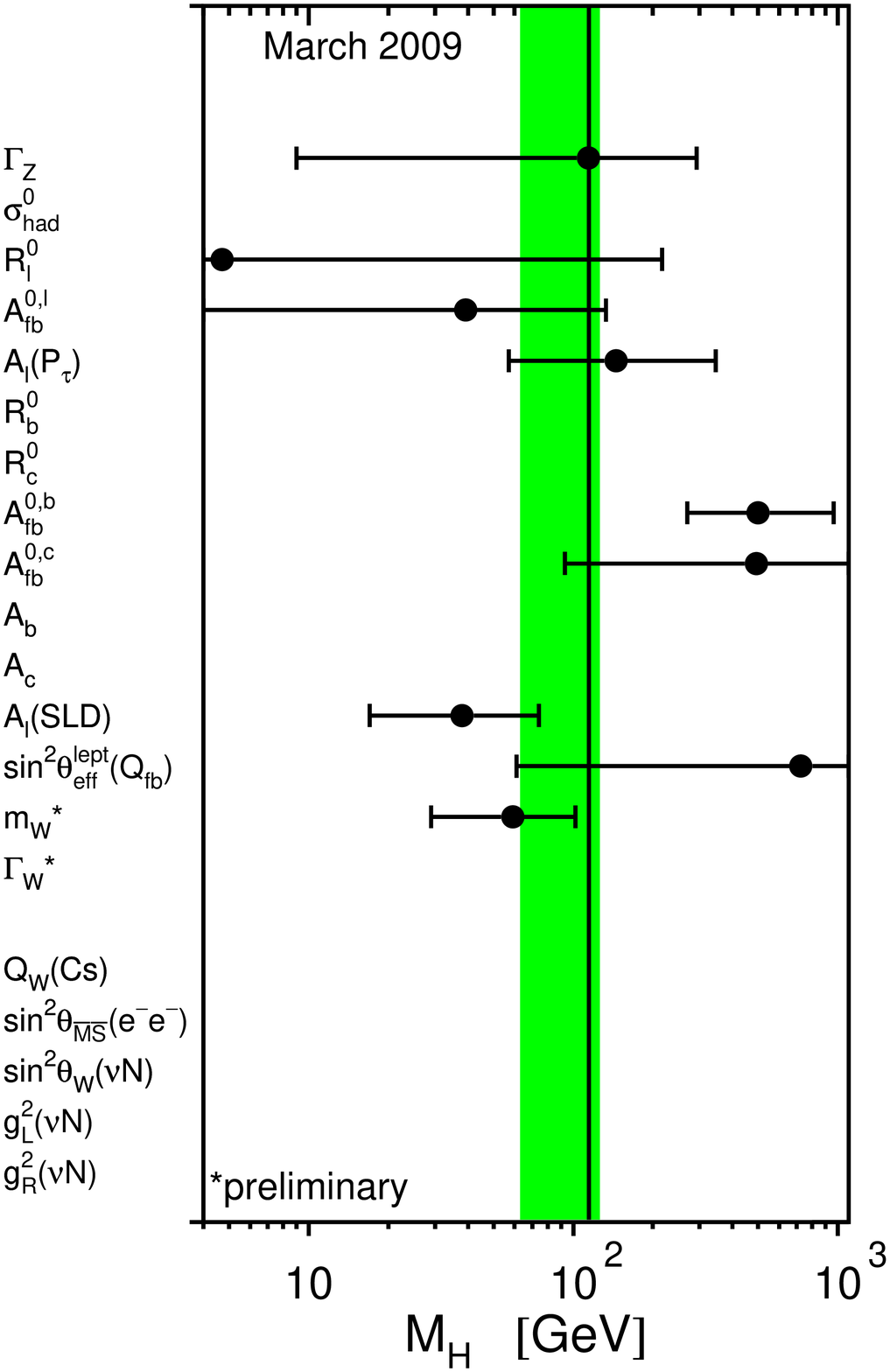}}
\end{minipage}
\vspace{5pt}
\caption[]{\small{\sf (a) Left panel: The blue-band plot showing the Higgs
    mass upper limit \cite{lepewwg}. (b) The upper limits on the Higgs mass
    from different measurements. The central band corresponds to the `average'
    \cite{lepewwg}.}}
\label{higgs}
\end{figure*}

Fig.~\ref{higgs}a is the famous blueband plot (August 2009 update shown) which
is generated using electroweak data obtained from LEP and by SLD, CDF and D0,
as a function of the Higgs mass, assuming that Nature is completely described
by the SM.  The preferred value for the Higgs mass, corresponding to the
minimum of the curve, is 87 GeV, with an experimental uncertainty of $+$35 and
$-$26 GeV (68\% C.L. which corresponds to $\Delta \chi^2 = 1$).  This serves
as a guideline in our attempt to find the Higgs boson. The 95\% C.L. upper
limit (corresponding to $\Delta \chi^2 = 2.7$) on the Higgs mass is 157 GeV,
which is pushed up to 186 GeV when the LEP-2 direct search limit of 114 GeV is
taken as a constraint in the fit. In a recent development, the Tevatron
experiments CDF and D0 have excluded the Higgs mass in the range 160 to 170
GeV at 95\% C.L. In Fig.~\ref{higgs}b we see that the extraction of the Higgs
mass from individual measurements indicates different ranges, though all are
consistent within errors.

\subsection{Theoretical limits} 
\subsubsection{Perturbative unitarity} {\em Unitarity} \cite{Lee:1977eg}
places an upper bound on $m_h$ beyond which the theory becomes
non-perturbative. Here, we shall call it a `tree level unitarity' as we would
require that the tree level contribution of the first partial wave in the
expansion of different scattering amplitudes does not saturate unitarity (in
other words, some probability should not exceed unity). The scattering
amplitudes involving gauge bosons and Higgs can be decomposed into partial
waves, using the `equivalence theorem', as ($\theta$ is the scattering
angle)
\begin{equation}
A = \sum_{J=0}^\infty (2J+1) P_J (\cos\theta) a_J,  
\end{equation}
where $a_J$ is the $J$th partial wave and $P_J$ is the $J$th Legendre
polynomial (where $P_0(x) = 1, P_1(x) = x, P_2(x) = 3x^2/2 -1/2, \cdots $).
Using the orthogonality of the Legendre polynomials, the cross section can be
written as
\begin{eqnarray}
\label{optical}
\sigma = \frac{16\pi}{s} \sum_{J=0}^\infty (2J+1)|a_J|^2 
= \frac{16\pi}{s} \sum_{J=0}^\infty (2J+1)~{\rm Im}~a_J.
\end{eqnarray}
The second equality in Eq.~(\ref{optical}) is obtained using optical
theorem. Therefore,
\begin{equation}
{|a_J|}^2 = {\rm Re}~ (a_J)^2 + {\rm Im}~ (a_J)^2 = {\rm Im}~ a_J . 
\end{equation}
This translates to the bound 
\begin{equation}
|{\rm Re}(a_J)| \leq \frac{1}{2}.
\end{equation}
For the channel $W_L^+ W_L^- \to W_L^+ W_L^-$, and for $s \gg m_h^2$, the $J=0$
mode is given by (at tree level)
\begin{equation}
a_0 = -\frac{m_h^2}{8\pi v^2}.
\end{equation}
The requirement that $|a_0| \leq 0.5$ thus sets an upper limit $m_h
<2\sqrt{\pi} v = 870$ GeV.  The {\em most} divergent scattering amplitude
arises from $2 W_L^+ W_L^- + Z_L Z_L$ channel leading to $a_0 = -5 m_h^2/64
\pi v^2$, which yields $m_h < 780$ GeV.

\subsubsection{Triviality} 
The {\em triviality} argument provides an upper limit on the Higgs mass
\cite{Altarelli:1994rb,Kolda:2000wi}.  First, recall that the SM scalar
potential has the following form ({\em Be alert that the normalizations are
  different from those in Eq.~(\ref{potnordis})}):
\begin{equation}
V(\Phi) = -|\mu^2|(\Phi^\dagger\Phi)+\l(\Phi^\dagger\Phi)^2 \, ,
\end{equation}
where 
\begin{eqnarray}
\Phi = \left(\begin{array}{c} \varphi_+\\
\varphi_0\end{array}\right) =
{1\over\sqrt{2}}\left(\begin{array}{l}
\varphi_1+i\varphi_2\\ \varphi_3+i\varphi_4\end{array}\right) 
\stackrel{\rm unitary ~ gauge}{\Longrightarrow} 
{1\over\sqrt{2}} \left(\begin{array}{c} 0 \\ v+h(x)\end{array}\right) \, .
 \nonumber
\end{eqnarray}
Now consider only the scalar sector of the theory. The scalar quartic
coupling evolves as 
\begin{equation}
{d\l\over dt} = {3\l^2 \over 4\pi^2}, ~~~~ \text{where}~~~
t = \ln \left({Q^2\over Q^2_0}\right) .
\end{equation}
Here $Q_0$ is some reference scale, which could as well be the vev $v$. 
The solution of the above equation is  
\begin{eqnarray}
 \l(Q) &=& {\l(Q_0)\over 
1-{3\l(Q_0)\over4\pi^2}\ln\left({Q^2\over Q_0^2}\right)} .
\end{eqnarray}
This means there is a pole at $Q_c =Q_0 e^{4\pi^2/3\l(Q_0)}$, which is called
the `Landau pole'. This pole has to be avoided during the course of RG
running.  The general `triviality' argument states that in order to remain
perturbative at all scales one needs to have $\l=0$ (which means Higgs remains
massless), thus rendering the theory to be `trivial', i.e non-interacting.
However, one can have an alternative view: use the RG of quartic coupling $\l$
to establish the energy domain in which the SM is valid, i.e.  find out the
energy cutoff $Q_c$ below which $\l$ remains finite.  If we denote the cutoff
by $\L$, then
\begin{eqnarray} 
{1\over\l(\L)}  =  {1\over\l(v)} - {3\over4\pi^2} \ln
  {\L^2\over v^2} > 0 . 
\end{eqnarray} 
The above inequality follows from the requirement $\l(\L)<\infty \Rightarrow
{1\over\l(\L)}>0$. This immediately leads to 
\begin{eqnarray}
\l(v) \le \frac {4\pi^2}{3\ln\left(\frac{\L^2}{v^2}\right)} 
 \Longrightarrow m_h^2 = 2\l v^2
 < \frac{8\pi^2v^2}{3\ln\left(\frac{\L^2}{v^2}\right)} \, . 
 \end{eqnarray}
Putting numbers, $m_h < 160$ GeV, for a choice of the cutoff close to the 
typical GUT scale $\L=10^{16}$ GeV.

Now let us include the full structure of fermions and gauge bosons in RG
equations.
\begin{eqnarray}
\label{full_lambda}
{d\l\over dt} \simeq {1\over 16\pi^2} \left[12\l^2+12\l h^2_t -
12h_t^4-{3\over2}\l(3g_2^2+g_1^2) +
{3\over16}\left\{2g_2^4+(g_2^2+g_1^2)^2\right\}\right], 
\end{eqnarray}
where $h_t=\sqrt{2}m_t/ v$ is the top quark Yukawa coupling.  For a rather
large $\l > h_t, g_1,g_2$, i.e., for a `heavy' Higgs boson, the dominant
contribution to running is
\begin{equation}
{d\l\over dt} \simeq {1\over16\pi^2}\left[12\l^2+12\l
h_t^2-{3\over2}\l(3g_2^2+g_1^2)\right]. 
\end{equation}
Note that whenever the quartic coupling $\l$, calculated at the weak scale
$v$, is equal to $\l_c \equiv {1\over8}(3g_2^2+g_1^2)-h_t^2$, which corresponds
to the vanishing right-hand side of the above RG equation, the coupling
reaches a critical limit. If one starts the evolution with a $\l(v)> \l_c(v)$,
i.e. for $m_h > m_h^c \equiv \sqrt{2\l_c}\,v$, then during the course of RG
running the quartic coupling hits the Landau pole, i.e. becomes infinite, at
some scale and the theory ceases to be perturbative. From this requirement,
one obtains an upper limit (at two-loop level)
\begin{equation}
m_h <m_h^c = 170~\text{GeV} ~\text{for}~\L=10^{16}~ \text{GeV}. 
\end{equation}
The limits for other choices of $\L$ can be read off from
Figs.~\ref{stability}a and \ref{stability}b.

\subsubsection{Vacuum stability} 
\begin{figure*}
\begin{minipage}[t]{0.42\textwidth}
\epsfxsize=6cm
\centering{\epsfbox
{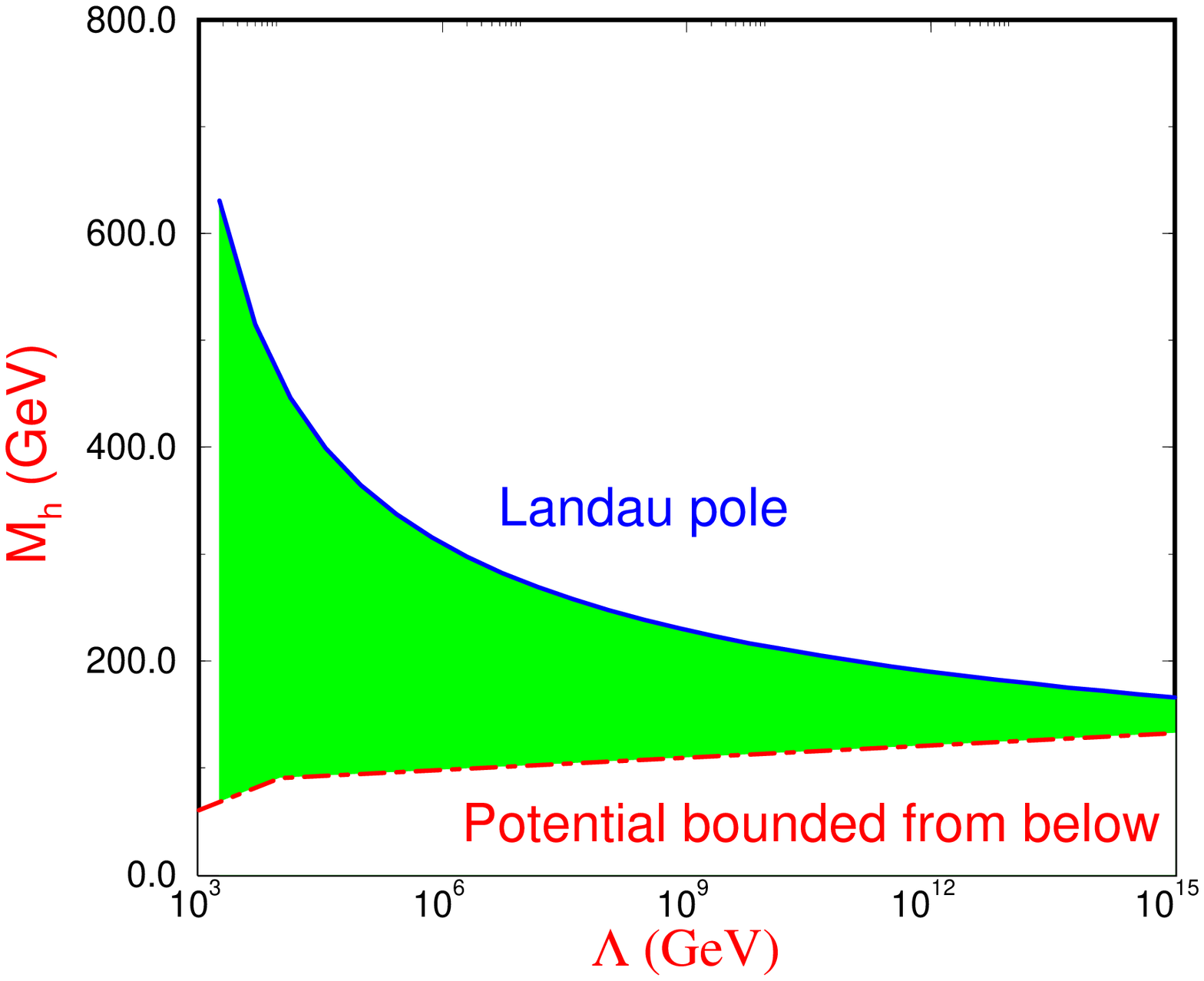}}
\end{minipage}
\hfill
\begin{minipage}[t]{0.42\textwidth}
\epsfxsize=6cm
\centering{\epsfbox
{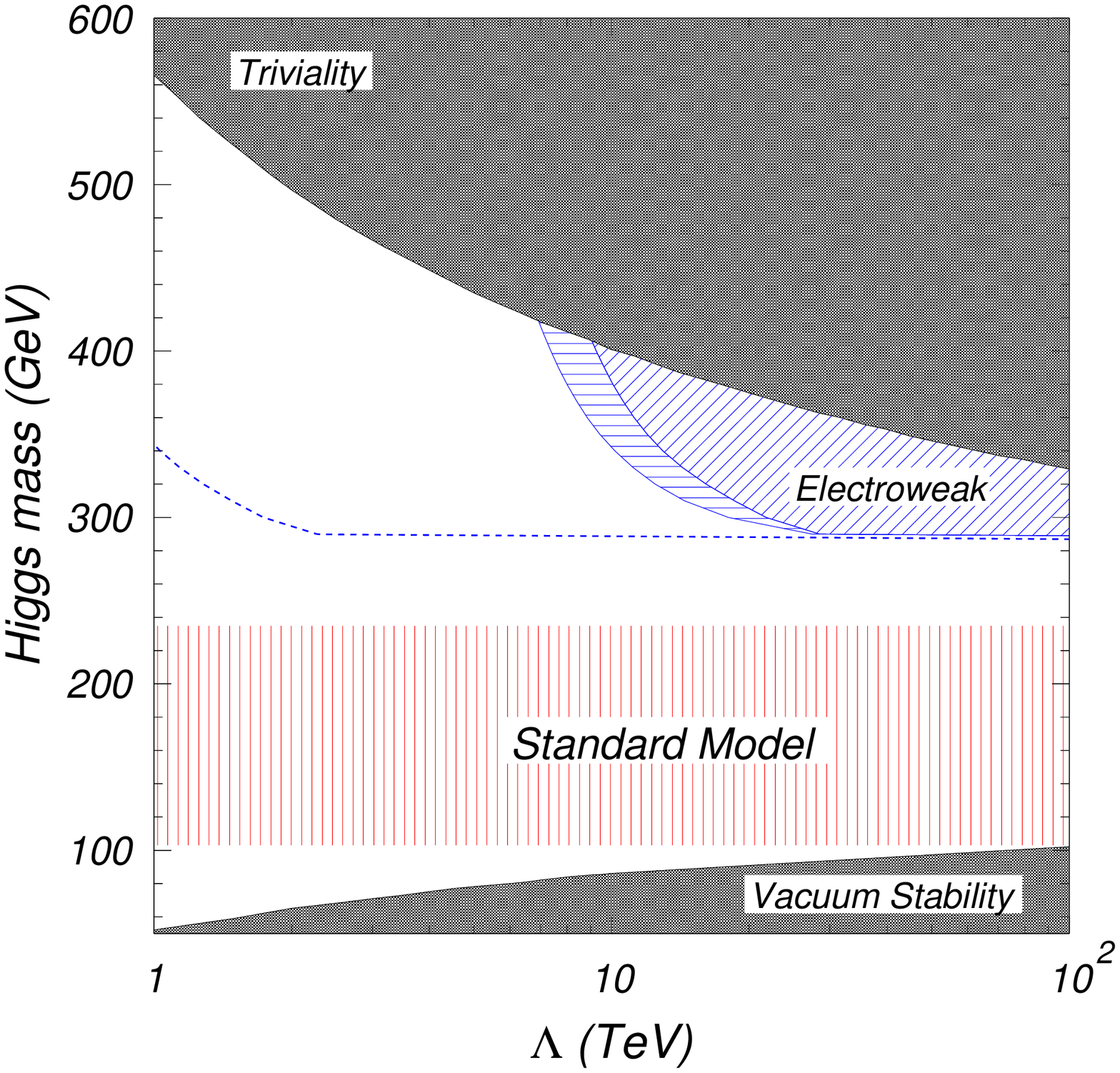}}
\end{minipage}
\vspace{5pt}
\caption[]{\small{\sf (a) Left panel: The triviality and vacuum stability
    limits (taken from \cite{Dawson:1998yi}). (b) Right panel: The region
    upto $\Lambda = 100$ TeV is zoomed.  The hatched region `Electroweak' is
    ruled out by electroweak precision data. Only the `Standard Model' region
    is allowed at 95\% CL \cite{Kolda:2000wi}.}}
\label{stability}
\end{figure*}

The argument of {\em vacuum stability} is based on the requirement that the
potential is always bounded from below. This means $\l(Q)$ has to remain
positive throughout the history of RG running. This gives rise to a lower
bound on the Higgs mass \cite{Altarelli:1994rb,Kolda:2000wi,Casas:1996aq}.  If
the Higgs mass is too small, i.e., $\l$ is very small, then the top quark
contribution dominates which can drive $\l$ to a negative value. If it happens
then the vacuum is not stable as it has no minimum.  For small $\l$,
Eq.~(\ref{full_lambda}) becomes
\begin{eqnarray}
{d\l\over dt} \simeq 
{1\over16\pi^2}\left[-12h_t^4+{3\over16}\{2g_2^4+(g_1^2+g_2^2)^2\}\right].
\end{eqnarray}
To provide intuitive understanding through easy analytic implementation, we
perform a one-step integration and obtain
\begin{eqnarray}
\l(\L) = \l(v) +
{1\over16\pi^2}\left[-12h_t^4+{3\over16}\{2g_2^4+(g_1^2+g_2^2)^2\}\right]\ln
\left({\L^2\over v^2}\right).
\end{eqnarray} 
To ensure that $\l(\L)$ remains positive, the Higgs mass must satisfy
\begin{equation}
  m_h^2 > \frac{v^2}{8\pi^2}
  \left[12h_t^4-{3\over16}\{2g_2^4+(g_1^2+g_2^2)^2\} \right]\ln
  \left({\L^2\over v^2}\right).
\end{equation}
Clearly, the above steps are very simple-minded, yet provides the rationale
behind the lower limit. By actually solving the RG equation at 2-loop level,
one obtains 
\begin{equation}
m_h > 134~\text{GeV} ~\text{for}~\L=10^{16}~ \text{GeV}. 
\end{equation}
If the cutoff $\L = 1$ TeV, then \cite{Casas:1996aq}
$$
m_h > 50.8 + 0.64~(m_t - 173.1 ~{\rm GeV}) \, ,
$$
which indicates that such a low cutoff is clearly disfavored by LEP (see also
C. Quigg's article in \cite{rev_ewsb}). Again, the limits for other choices
of the cutoff can be read off from Figs.~\ref{stability}a and
\ref{stability}b.

\section{Gauge hierarchy problem}
\subsection{Quadratic divergence}
Let us illustrate the problem of quadratic divergence in the Higgs sector
through an explicit calculation.  Recall that in the unitary gauge the doublet
$\Phi(x) = \left(\begin{array}{c} 0 \\ \varphi(x) \end{array}\right) =
{1\over\sqrt{2}} \left(\begin{array}{c} 0 \\ v+h(x)\end{array}\right)$.  We
write the Yukawa interaction Lagrangian as
\begin{eqnarray*}
\Lag =-h_f \varphi  \bar{f}_Lf_R + ~\text{h.c.},
\end{eqnarray*}
where $f_{L,R}$ are the left and right chiral projection of the fermion $f$.
After spontaneous symmetry breaking, 
\begin{equation}
\label{fermioncpl}
\Lag = -\frac{h_f}{\sqrt{2}} h \bar f_L f_R 
 -\frac{h_f}{\sqrt{2}} v  \bar{f}_L f_R + ~\text{h.c.} 
\end{equation}
The fermion mass is therefore given by $m_f=h_f {v\over{\sqrt{2}}}$. 

\begin{figure}
\begin{center}
\begin{picture}(338,38)(-38,-1)
\Text(-38,0)[c]{$\Delta(m_{h}^2)\> =\, $}
\SetWidth{0.7}
\DashLine(0,0)(25,0){5}
\DashLine(65,0)(90,0){5}
\SetWidth{1.3}
\CArc(45,0)(20,0,360)
\Text(3,8)[c]{$h$}
\Text(45,28)[c]{$f$}
\Text(45,-30)[c]{[a]}
\SetWidth{0.7}
\DashLine(120,0)(145,0){4.5}
\DashLine(185,0)(210,0){4.5}
\SetWidth{1.3}
\DashCArc(165,0)(20,180,360){4}
\DashCArc(165,0)(20,0,180){4}
\Text(124,8)[c]{$h$}
\Text(165,28)[c]{$\tilde f_{L,R}$}
\Text(165,-30)[c]{[b]}
\SetWidth{0.7}
\DashLine(240,-5)(275,-5){4.5}
\DashLine(310,-5)(275,-5){4.5}
\SetWidth{1.3}
\DashCArc(275,11)(16,-90,270){4}
\Text(243,3)[c]{$h$}
\Text(275,35)[c]{$\tilde f_{L,R}$}
\Text(275,-30)[c]{[c]}
\end{picture}
\end{center}
\vspace{1cm}
\caption{\small{\sf One-loop quantum corrections to the Higgs mass, due to a
    Dirac fermion $f$ [a], and scalars $\tilde f_{L,R}$ ([b] \&
    [c]).\label{fig:higgscorr1}}}
\end{figure}
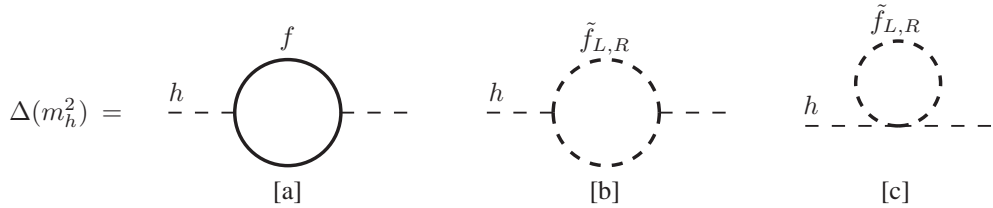

Let us compute the two-point function with the zero momentum Higgs as the two
external lines and fermions inside the loop.  The corresponding diagram is in
figure (\ref{fig:higgscorr1}[a]).
\begin{eqnarray}
\label{pi_f}
 i \Pi^f_{hh}(0)&=&(-)\int \frac{d^4k}{(2\pi)^4} {\rm Tr}~\left[ \left(-i 
\frac{h_f}{\sqrt{2}}\right)
\frac{i}{\slashed{k} - m_f} \left(-i\frac{h_f}{\sqrt{2}}\right)
\frac{i}{\slashed{k} - m_f}    \right] \nonumber\\
 &=& -2h_f^2\int\frac{d^4k}{(2\pi)^4} 
\left[ \frac{1}{k^2 - m_f^2} + \frac{2m_f^2}{(k^2-m_f^2)^2}\right] .
\end{eqnarray}
The correction $\D m_h^2$ is proportional to $\Pi^f_{hh}(0)$.  The first term
in the RHS is quadratically divergent. The divergent correction to $m_h^2$
looks like 
\begin{equation}
\Delta m_h^2 (f) = \frac{\L^2}{16\pi^2} (-2h_f^2) \, . 
\end{equation}
Another divergent piece will appear from quartic Higgs vertex.  The
corresponding diagram is similar to what is displayed in figure
(\ref{fig:higgscorr1}[c]), except that the internal line is also $h$.  The
divergent contribution to $m_h^2$ is
\begin{equation}
\Delta m_h^2 (h) = \frac{\L^2}{16\pi^2} (\l) \, .
\end{equation}
For the sake of simplicity, we neglect the gauge boson contributions to the
quadratic divergence. Combining the above two divergent pieces, we obtain
\begin{equation}
\Delta m_h^2 = \frac{\L^2}{16\pi^2} \left(-2h_f^2 + \l\right) \, . 
\end{equation}
Now, we contemplate on the following issues: 

\noindent {$(i)$}~~The Yukawa coupling $h_f$ and the quartic scalar coupling
$\l$ are totally unrelated. Suppose, we set $\l = 2h_f^2$. First of all, this
is a huge fine-tuning. Second, at higher loops, this relation will not be able
to prevent the appearance of divergence. {\em It is also amusing to note that
  if we set $\l = h_f^2$, then we would require two scalars to cancel the
  quadratic divergence caused by one fermion}.

\noindent {$(ii)$}~~Suppose we do not attempt to relate $\l$ and $h_f$ for
canceling the quadratic divergence. Now, remember that we have a tree level
bilinear mass term, which is the bare mass. We can absorb the quadratic
divergent in a redefinition of the bare mass. Still, there is a residual
finite part to the mass correction, given by $\sim \frac{h_f^2 m_f^2}{8\pi^2}$
[see Eq.~(\ref{pi_f})]. What is the value of the loop mass $m_f$?  If SM gives
way to some GUT theory at high scale we can have fermions where $m_f \sim
M_{\rm GUT} \sim 10^{16}$ GeV. In that case, even after removing the quadratic
cutoff dependence, the leading contribution to $\Delta m_h^2$ would be order
$M_{\rm GUT}^2/(8\pi^2)$. One would then have to do an unnatural fine-tuning
($1 \div 10^{26}$) between the bare term $m_{h_0}^2$ and the correction term
$\Delta m_h^2$ to maintain the renormalized mass ($m_h^2 = m_{h_0}^2 + \Delta
m_h^2$) at around 100 GeV.  Furthermore, this fine-tuning has to be done
order-by-order in perturbation theory to prevent the Higgs mass from shooting
up to the highest mass scale of the theory.  This constitutes what is
technically called the {\em gauge hierarchy problem} \cite{hierarchy}.

\noindent {$(iii)$}~~The primary problem is that the correction is independent
of $m_h$. Setting $m_h=0$ does not increase the symmetry of the theory.  In
QED, in the limit of vanishing electron mass we have exact {\em chiral
  symmetry}, and since the photon mass is zero we have exact {\em gauge
  symmetry}. But there is no symmetry that protects the Higgs mass.


One of the biggest challenges in the SM is to stabilize the scalar potential,
i.e. to protect it from a run-away quantum behavior.  Although we said that it
is the Higgs mass which is not stable but, more precisely, it is the
electroweak vev ($v$) which is unstable. {\em Since $v$ feeds into all masses
  in the SM through SSB, none of them which is proportional to $v$ is stable
  either}. In fact, the argument of protection from gauge and chiral symmetry
applicable to QED is strictly not applicable for the SM because all the SM
particle masses are proportional to $v$.

\subsection{Cancellation of quadratic divergence in a toy 
supersymmetric scenario}
Supersymmetry, a theory with an intrinsic fermion $\leftrightarrow$ boson
symmetry, unambiguously solves the gauge hierarchy problem and restores
naturalness.  For an early study of supersymmetric model building and
demonstration of quadratic divergence cancellation, we refer to
\cite{intro-susy,romesh}.  The content of this subsection is adapted from the
text book by Drees, Godbole and Roy in \cite{susy-books}.

We consider a toy model which contains $\varphi(x) = \frac{1}{\sqrt{2}}
(v+h(x))$ plus two additional complex scalar fields $\tilde
f_{L,R}(x)$. Suppose the interaction is encoded in the following effective
Lagrangian:
\begin{eqnarray}
 \mathcal{L}_{\tilde{f}\tilde{f}\varphi} &=& 
-\tilde{\l}_f |\varphi|^2\left(|\tilde f_{L}|^2 +|\tilde f_{R}|^2  \right) +
\left({h}_f A_f \varphi \tilde f_{L} \tilde f_{R}^* + {\rm h.c.} \right) 
\nonumber \\
&=& -\frac{1}{2}\tilde{\l}_f h^2\left(|\tilde f_{L}|^2 +|\tilde f_{R}|^2  
\right) -\tilde{\l}_f h v\left(|\tilde f_{L}|^2 +|\tilde f_{R}|^2
\right) 
+ \frac{{h}_f}{\sqrt{2}} A_f \left(h \tilde f_{L} \tilde f_{R}^* +
  {\rm h.c.} \right) + \cdots
\end{eqnarray}
Above, the dots correspond to Higgs independent terms which need not be spelt
out. $A_f$ has the dimension of mass and it measures the strength of triple
scalar vertex. The Yukawa coupling $h_f$ is multiplied to it by convention.
The fermion loop, described before, is shown in
Fig.~(\ref{fig:higgscorr1}[a]).  The new loops involving scalars are displayed
in Figs.~\ref{fig:higgscorr1}[b] and \ref{fig:higgscorr1}[c]. 
The contribution of the scalar loops are given by 
\begin{eqnarray}
\label{pi_S}
i \Pi^{\tilde f}_{hh}(0)&=& \tilde{\l}_f\int\frac{d^4k}{(2\pi)^4} 
\left[ \frac{1}{k^2 - m_{{\tilde f}_L}^2} + \frac{1}{k^2-m_{{\tilde
        f}_R}^2}\right] ~~~~~~~
{\Longleftarrow \rm {\bf Fig.~\ref{fig:higgscorr1}[c]}} 
\nonumber \\
&+ & (\tilde{\l}_f v)^2\int\frac{d^4k}{(2\pi)^4}
\left[ \frac{1}{(k^2 - m_{{\tilde f}_L}^2)^2} + 
\frac{1}{(k^2-m_{{\tilde f}_R}^2)^2}\right] 
~~~~~~~{\Longleftarrow \rm {\bf Fig.~\ref{fig:higgscorr1}[b]}}  
\nonumber \\
&+ &  |{h}_f A_f|^2\int\frac{d^4k}{(2\pi)^4}
\left[ \frac{1}{k^2 -m_{{\tilde f}_L}^2} \frac{1}{k^2-m_{{\tilde
        f}_R}^2}\right]\, . 
~~~~~~~{\Longleftarrow \rm {\bf Fig.~\ref{fig:higgscorr1}[b]}} 
\end{eqnarray}

Combining Eqs.~(\ref{pi_f}) [fermion loop] and (\ref{pi_S}) [scalar loops] we
make the following observations:

\begin{itemize}
\item The fermion loop contribution (\ref{fig:higgscorr1}[a]) and the scalar
  loop contribution (\ref{fig:higgscorr1}[c]) give quadratic
  divergence. However, if one computes the net contribution to the two-point
  function, given by $\Pi^{f}_{hh}(0) + \Pi^{\tilde f}_{hh}(0)$, the quadratic
  divergence exactly cancels if one sets ${\tilde{\lambda}_f = {h}_f^2}$. This
  cancellation of quadratic divergence occurs regardless of the magnitude of
  any mass dimensional parameter, namely, $m_{{\tilde f}_{L,R}}$ or $A_f$.

\item A log sensitivity to the cutoff (or, the unknown scalar mass) still
  remains.  If we assume $m_{{\tilde f}_L} =m_{{\tilde f}_R} = m_{{\tilde
      f}}$, then
\begin{eqnarray}
   \Pi^{f}_{hh}(0) + \Pi^{\tilde f}_{hh}(0) &=& 
\frac{{h}_f^2}{16\pi^2}\left[
    -2m_f^2\left\{ 1-\ln\left( \frac{m_f^2}{\mu^2}\right)\right\} + 
4m_f^2\ln\left( \frac{m_f^2}{\mu^2}\right) \right. \nonumber \\
  &+ &\left.  2m_{\tilde f}^2
\left\{1-\ln \left( \frac{m_{\tilde f}^2}{\mu^2}\right)\right\} - 
4m_f^2 \ln\left( \frac{m_{\tilde f}^2}{\mu^2}\right) - 
|A_f|^2 \ln\left( \frac{m_{\tilde f}^2}{\mu^2}\right) \right]. 
\end{eqnarray}
 
\item Now, if we further assume that ($i$) $m_f=m_{\tilde f}$ and ($ii$) $A_f
  =0$, then we have $\left(\Pi^{f}_{hh}(0) + \Pi^{\tilde f}_{hh}(0) \right) =
  0$ i.e. even the finite contribution vanishes.
\end{itemize}

All these points are shared by supersymmetric extension of the standard
model. Quadratic divergence cancels due to the equality of two types of
dimensionless couplings. If supersymmetry is broken in masses, e.g. $m_f \ne
m_{\tilde{f}}$, i.e., gives rise to the `soft' terms (mass dimension $< 4$) of
the Lagrangian, the quadratic divergence still cancels. Also, in the limit of
exact supersymmetry, i.e., ($i$) $m_f=m_{\tilde f}$ and ($ii$) $A_f =0$, the
correction to the Higgs mass exactly vanishes. This toy scenario is
reminiscent of supersymmetric models.

\subsection{The Higgs bosons of the Minimal Supersymmetric 
Standard Model (MSSM)}
\label{susyhiggs}
We need two complex scalar doublets of opposite hypercharge to ensure
electroweak symmetry breaking. 
\begin{equation}
H_1=\left(\begin{array}{c}h^0_1\\
h^-_1\end{array}\right)_{Y=-1}, ~~~
H_2=\left(\begin{array}{c}h^+_2\\
h^0_2\end{array}\right)_{Y=1} .
\end{equation}
There are three reasons behind the need for at least two doublets. 
\begin{itemize}
\item Chiral or ABJ (Adler-Bardeen-Jackiw) anomaly cancellation requires $\sum
  Y_f=0 = \sum Q_f$, where the sum is on fermions only.  If we use only one
  Higgs doublet, its spin-1/2 (Higgsino) components will spoil the
  cancellation. We therefore need two Higgs doublets with opposite
  hypercharge. ({\em This anomaly arises from triangular fermionic loops
    involving axial vector couplings. The theory ceases to be renormalizable
    if it has ABJ anomaly}).

\item Recall that in the SM we use the scalar doublet $\Phi$ and
  $\tilde{\Phi}=i\tau_2\Phi^*$ for giving masses to up- and down-type
  fermions. In supersymmetry, $\Phi$ is a chiral superfield, and we cannot use
  a chiral superfield and its complex conjugate in the same
  superpotential. Therefore, we need two chiral superfields.

\item Unless we introduce both $H_1$ and $H_2$, we cannot provide the right
  number of degrees of freedom necessary to make the charginos massive.  In
  this sense, introducing at least two complex doublets is an experimental
  compulsion.
\end{itemize}
In the MSSM, the scalar potential $V_H$ receives contributions from
three sources. 
\begin{itemize}
\item[{(a)}] {\em The $D$ term}:~ 
$V_D=\displaystyle{
{1\over2}\sum_{a=1}^3\left(\sum_i g_a S^*_iT^aS_i\right)^2 :
\begin{array}{l}a~\text{runs over groups}\\ i~\text{runs over
particles}\end{array}}$
~~~($S_i$ is a generic scalar)

Keeping only the Higgs contributions, i.e. neglecting slepton/squark
contributions, we obtain, 

For ${\rm U(1)}_{Y} : V_D^{(1)} =
\frac{1}{2}\left[\frac{g_1}{2}\left(|H_2|^2-|H_1|^2\right)\right]^2$ \, , 

For ${\rm SU(2)}_{L} : V_D^{(2)} =
\frac{1}{2} \left[\frac{g_2}{2}\left(H_1^{i*} \tau_{ij}^a H_1^j +
H_2^{i*}\tau_{ij}^a H_2^j \right)\right]^2$. Here, $g_1 \equiv g'$ and $g_2
\equiv g$. 

Using $\tau_{ij}^a\tau^a_{kl} = 2\d_{il}\d_{jk}-\d_{ij}\d_{kl}$,
one obtains, 

$V_D  = V_D^{(1)} + V_D^{(2)} = 
\frac{g_2^2}{8}\left[4|H_1^\dagger H_2|^2 - 2|H_1|^2 |H_2|^2 + |H_1|^4 + |H_2|^4
\right] + {g_1^2\over8}(|H_2|^2-|H_1|^2)^2$. 

\item[{(b)}] {\em The $F$ term}:~ 
$\displaystyle{ V_F = \sum_i\left|{\p
W(\varphi_j)\over\partial\varphi_i}\right|^2}$. The superpotential
$W=\mu\hat{H}_1\hat{H}_2$ (`hat' denotes superfields) leads to: \\
$
V_F = \mu^2\left(|H_1|^2+|H_2|^2\right). 
$

\item[{(c)}] {\em The soft supersymmetry breaking terms}:~ $V_{\rm soft} =
  m^2_{H_1}|H_1|^2 + m_{H_2}^2 |H_2|^2 + (B_\mu H_2 H_1 + {\rm h.c.})$.
\end{itemize}
We now introduce the notations: ~$\bar{m}_1^2 \equiv
|\mu|^2+m_{H_1}^2$, ~ $\bar{m}_2^2 \equiv |\mu|^2+m^2_{H_2}$, ~ $\bar{m}_3^2
\equiv B_\mu$.  Using the charged and neutral components of the doublet
scalars, we can write the full scalar potential as
\begin{eqnarray}
V_H =&
\bar{m}_1^2(|h^0_1|^2+|h_1^-|^2)+\bar{m}_2^2(|h_2^0|^2+|h_2^+|^2)+
\bar{m}_3^2(h_1^-h^+_2-h_1^0h_2^0+ {\rm h.c.}) \nonumber\\
&+
\left(\frac{g_2^2+g_1^2}{8}\right)(|h_1^0|^2+|h_1^-|^2-|h_2^0|^2-|h_2^+|^2)^2
+ \frac{g_2^2}{2}|h_1^{-*}h_1^0+h_2^{0*}h_2^+|^2 .
\end{eqnarray}
We then require that the minimum of $V_H$ breaks ${\rm SU(2)}_L\times {\rm
  U(1)}_Y$ to ${\rm U(1)}_{Q}$. One can always choose $\langle h_1^-\rangle=
\langle h^+_2\rangle = 0$ to avoid breakdown of electromagnetism without any
loss of generality.  Note two important features at this stage:
\begin{itemize}
\item Only $B_\mu$ can be complex. However, the phase can be absorbed into the
  phases of $H_1$ and $H_2$. Hence, the MSSM tree level scalar potential has
  no source of CP violation.

\item The quartic scalar couplings are fixed in terms of the SU(2) and U(1)
  gauge couplings.
\end{itemize}
Note that it is sufficient to write the potential keeping only the (neutral)
fields which can acquire vevs.
\begin{equation}
V_H^0 ={1\over8}(g_1^2+g_2^2)(|h_1^0|^2-|h_2^0|^2)^2 +
\bar{m}_1^2|h_1^0|^2 + \bar{m}_2^2|h_2^0|^2 -\bar{m}_3^2(h_1^0 h_2^0 
+ {\rm h.c.})
\end{equation}
Again, note the following points:
\begin{itemize}
\item $V_H^0$ will be bounded from below if $\bar{m}_1^2 + \bar{m}_2^2 >
  2\bar{m}_3^2$.  This relation has to be valid at all scales.  (Note, there
  is no quartic term in the direction $|h_1^0|=|h_2^0|$).
 
\item $V_H^0 ~
({\rm quadratic~part})= (h_1^{0*} h_2^0)\left(\begin{array}{cc} \bar{m}_1^2 &
-\bar{m}_3^2\\ -\bar{m}_3^2 &
\bar{m}_2^2\end{array}\right) \left(\begin{array}{c} h_1^0 \\
h_2^{0*}\end{array}\right)$\\
SSB requires $\bar{m}_3^4>\bar{m}_1^2\bar{m}_2^2$. This has to be necessarily
valid at the weak scale where SSB occurs. 

\item The above two conditions cannot be satisfied simultaneously if
  $\bar{m}_1^2=\bar{m}_2^2$. Hence, $\bar{m}_1^2\ne \bar{m}_2^2 \Rightarrow
  m^2_{H_1}\ne m^2_{H_2}$, which indicates to a connection between
  supersymmetry breaking and electroweak symmetry breaking.
\end{itemize}
Putting $\langle h_1^0\rangle = \frac{v_1}{\sqrt{2}}$ and $\langle
h_2^0\rangle = \frac{v_2}{\sqrt{2}}$,
\begin{equation}
V_H^0 (\min) ={1\over{32}}(g_1^2+g_2^2)(v_1^2-v_2^2)^2+{1\over2}
\bar{m}_1^2v_1^2+{1\over2}\bar{m}_2^2 v_2^2 - \bar{m}_3^2v_1v_2
\end{equation}
The minimization conditions ${\p V_H^0(\min)\over\p v_i} =0$, for $i=1,2$ yield
\begin{align}
\bar{m}_1^2 =  \bar{m}_3^2{v_2\over v_1} -
{1\over8}(g_1^2+g^2_2)(v_1^2-v_2^2) \, , ~~~~\text{and}~~
\bar{m}_2^2 = \bar{m}_3^2{v_1\over v_2} +
{1\over8}(g_1^2+g^2_2)(v_1^2-v_2^2) \, . 
\end{align}
Now using the above equations and putting back $\bar{m}_1^2 \equiv
m_{H_1}^2+|\mu|^2$, $\bar{m}_2^2 \equiv m^2_{H_2}+|\mu|^2$, we obtain the two
conditions of electroweak symmetry breaking:
\begin{eqnarray}
{1\over2} M_Z^2 &=& \left({m^2_{H_1}-m^2_{H_2}\tan^2\b\over\tan^2\b-1}\right) -
|\mu|^2 , ~~{\rm where}~~ \tan\b \equiv \frac{v_2}{v_1} \, , \\
- 2B_\mu &=& (m^2_{H_1}-m^2_{H_2})\tan2\b + M^2_Z\sin 2\b .
\end{eqnarray}
Our next task is to extract the different masses from the quadratic part of
the potential: $V_H^{\rm quad} = {1\over2} m^2_{ij}\varphi_i\varphi_j$.

\subsubsection {Charged Higgs and Goldstone}
The mass matrix is given by 
\begin{eqnarray}
V_{h^{\pm}} & = &
\left({\bar{m}_3^2\over v_1v_2} + {1\over4}g_2^2\right)
(h_1^+ ~h_2^+) \left(\begin{array}{cc}
v_2^2&v_1v_2\\ v_1v_2 & v_1^2\end{array}\right)
\left(\begin{array}{c}h^-_1\\ h^-_2\end{array}\right). 
\end{eqnarray}
Note that the determinant of the mass matrix is zero, which is a consequence
of the masslessness of the Goldstones ($m^2_{G^\pm}=0$). 
The charged Higgs mass is given by 
\begin{equation}
m^2_{h^\pm} = \left({\bar{m}_3^2\over v_1v_2}+{1\over4}g^2_2\right)
(v_1^2+v_2^2) \, . 
\end{equation}
The mass eigenstates are given by 
\begin{eqnarray}
H^{\pm}&=&\sin\b ~ h_1^{\pm}+\cos\b ~ h_2^\pm  \, , ~~~~
G^{\pm}=-\cos\b ~ h_1^{\pm}+\sin\b ~ h_2^\pm \, .
\end{eqnarray}

\subsubsection{Neutral CP-odd Higgs and Goldstone}
The Goldstone is massless, while the mass of the CP odd scalar depends on
$\bar{m}_3^2 = B_\mu$: 
\begin{equation}
  m^2_{G^0}=0, ~~~m^2_A = {2\bar{m}_3^2\over\sin2\b} .
\end{equation}
The physical states are given by
\begin{eqnarray}
{A\over\sqrt{2}} &=& \sin\b~ {\rm Im}~ h_1^0  + \cos\b~ {\rm Im}~ h_2^0  \, , 
~~~~
{G^0\over\sqrt{2}} = - \cos\b~ {\rm Im}~ h_1^0  + \sin\b~ {\rm Im}~ h_2^0  \, . 
\end{eqnarray}

\subsubsection{Neutral CP-even Higgses}
The 2$\times$2 mass-squared matrix for the neutral CP-even sector in the
(${\rm Re}~ h_1^0, {\rm Re}~ h_2^0$) basis is given by
\begin{align}
M^2_{{\rm Re}~h^0} &= {1\over2} \left(\begin{array}{lr}
2\bar{m}_1^2+{1\over4}(g_2^2+g_1^2)(3v_1^2-v_2^2)&-
2\bar{m}_3^2-{1\over2}(g_1^2+g_2^2) v_1 v_2\\
\\ -2\bar{m}_3^2-{1\over2}(g_1^2+g_2^2) v_1 v_2 &
2\bar{m}_2^2+{1\over4}(g_1^2+g_2^2)(3v_2^2-v_1^2)\end{array}\right) \nonumber \\ 
&= \left(\begin{array}{lr} m_A^2\sin^2\b + M_Z^2\cos^2\b &
-(m_A^2 + M_Z^2)\sin\b\cos\b\\ \\-(m_A^2+M_Z^2)\sin\b\cos\b &
m_A^2\cos^2\b + M_Z^2\sin^2\b\end{array}\right) \, . 
\end{align}

\noindent The mass-squared eigenvalues are then given by ($h$ is lighter, $H$
heavier)
\begin{equation}
m^2_{h,H}= {1\over2}\left[m_A^2 + M_Z^2 \mp
\left\{(m_A^2 + M_Z^2)^2 - 4 M_Z^2 m_A^2 \cos^2 2\b\right\}^{1/2}\right].
\end{equation}

\subsubsection{Important equalities and inequalities}
These are some of the important relations.
\begin{align}
&m_h < \min~ (m_A,M_Z)|\cos2\b|<\min~(m_A,M_Z) ,
& m_h^2 + m^2_H = m_A^2 + M_Z^2 , \nonumber \\
&m_H > \max~(m_A, M_Z) , 
&m^2_{H^{\pm}} = m_A^2+M_W^2 \, . 
\end{align}
The tree level inequality $m_{h}<M_Z$ is an important prediction of the
MSSM. This is a consequence of the fact that the quartic couplings in MSSM are
related to the gauge couplings.

\subsubsection{Radiative correction to the lightest Higgs mass}
The lightest neutral Higgs mass ($m_h$) receives large quantum corrections.
The correction is dominated by the top quark Yukawa coupling ($h_t$) and the
masses of the stop squarks ($\tilde{t}_1$, $\tilde{t}_2$). The corrected Higgs
mass-squared is given by (original references can be found in
\cite{susy-books,reviews})
\begin{equation}
m_h^2 \simeq M_Z^2 \cos^2 2\beta + \frac{3 m_t^4}{2\pi^2 v^2}
\ln \left(\frac{m_{\tilde{t}}^2}{m_t^2}\right) \, , 
\end{equation}
where $m_{\tilde{t}} = \sqrt{m_{\tilde{t}_1} m_{\tilde{t}_2}}$ is an average
stop mass, This is a one-loop expression. Including two-loop calculations
pushes the upper limit on the Higgs mass to around 135 GeV. If a neutral Higgs
is not found at LHC approximately within this limit, the {\em two-Higgs
  doublet} version of supersymmetric model will be strongly disfavored.  In
the next-to-minimal supersymmetric model (NMSSM) \cite{nmssm}, which contains
an additional gauge singlet scalar ($N$) coupled to $H_1$ and $H_2$ through
the superpotential $\lambda N H_1 H_2$, there is an additional tree level
contribution to $m_h^2$. It turns out that \cite{Drees:1988fc}
$
m_h^2 ({\rm tree,~NMSSM}) = M_Z^2 \left[\cos^2 2\b + 2\l^2(g^2+g^{\prime
    2})^{-1} \sin^2 2\b\right].
$
Including radiative corrections, the upper limit on $m_h$ is relaxed to about
150 GeV \cite{limit_nmssm}.

\subsection{Radiative electroweak symmetry breaking in MSSM}
One of the most attractive features of supersymmetry is that the electroweak
symmetry is broken radiatively. Recall that in the SM we had to put a {\em
  negative} sign by hand in front of $\mu^2$ to ensure EWSB, which was {\em ad
  hoc}. In supersymmetry this happens dynamically thanks to the large top
quark Yukawa coupling.  We will demonstrate how one of the Higgs mass-squared,
more precisely $m^2_{H_2}$, starting from a positive value at a high scale is
driven to a negative value at low scale by RG running. To appreciate the
salient features, we will take into consideration only the effect of $h_t$ in
RG evolution and ignore the gauge and other Yukawa couplings' contributions
({\em for details, see text books}). This estimate may be crude, but it brings
out the essential features. First we write down the RG evolution of
$m^2_{H_2}$, $m^2_{\tilde{Q}_3}$ and $m^2_{\tilde{u}_3}$:
\begin{eqnarray} 
\label{indi-eqs}
\frac{dm^2_{H_2}}{dt} = -3 h_t^2 (m^2+A_t^2), ~~
\frac{dm^2_{\tilde{Q}_3}}{dt} = - h_t^2 (m^2+ A_t^2), ~~
\frac{dm^2_{\tilde{u}_3}}{dt} = -2 h_t^2 (m^2 +A_t^2) \, ; 
\end{eqnarray}
where $\tilde{Q}_3$ and $\tilde{u}_3$ are the third generation squark doublet
and singlet respectively, $t \equiv \ln(M_{\rm GUT}^2/Q^2)/16\pi^2$, $h_t$ is
the top quark Yukawa coupling, $A_t$ is the scalar trilinear coupling
involving the top squark, and $m^2 \equiv m^2_{H_2} + m^2_{\tilde{Q}_3} +
m^2_{\tilde{u}_3}$.  Now recall that the Bernoulli's equation
$$
\frac{dy}{dx} + y P(x) = Q(x)
$$
has a solution 
$$
  y \exp\left(\int dx P(x)\right) = \int dx Q(x) \exp\left(\int dx P(x)\right) 
+ ~{\rm constant} \, . 
$$
Therefore, the equation (obtained by summing the individual RG's in
Eq.~(\ref{indi-eqs}))
\begin{equation}
\frac{dm^2}{dt} + 6 h_t^2 m^2 = -6 h_t^2 A_t^2 
\end{equation}
has a solution 
\begin{equation}
  m^2 \exp\left(6 \int_0^t dt' h_{t}^2\right) = 
\int_0^t dt' \left( -6 h_{t}^2 A_{t}^2 \right) 
\exp\left(6 \int_0^{t'} dt'' h_{t}^2\right) + ~{\rm constant} \, . 
\end{equation}
Now, ignore the running of $h_t$ and $A_t$ to avoid complications, i.e. treat
them as fixed values. This eases calculational hassles but preserves the
important features of radiative EWSB. Then 
\begin{equation}
m^2 \exp\left(6 h_{t}^2 t\right) = 
- 6 h_t^2 A_t^2 \int_0^t dt' \exp\left(6 h_t^2 t'\right) + C 
= - A_t^2 \exp\left(6 h_t^2 t\right) + C \, . 
\end{equation}
At $t=0$ (i.e. $Q=M_{\rm GUT}$), assume universal boundary conditions, i.e.
$m_0^2 \equiv m^2_{H_2} = m^2_{\tilde{Q}_3} = m^2_{\tilde{u}_3}$. Therefore,
$m^2 (t=0) = 3 m_0^2$, hence $C = 3m_0^2 + A_0^2$ (where $A_t=A_0$, since we
ignored the running of $A_t$). Using these relations, it is simple to obtain
the solution
\begin{equation}
m^2 = - A_t^2 \left[1-\exp\left(-6h_t^2 t\right) \right] 
+ 3 m_0^2 \exp\left(-6h_t^2 t\right) \, . 
\end{equation}
Now we solve the individual equations in (\ref{indi-eqs}). The mathematical
steps are easy and we do not display them. The solutions are 
\begin{eqnarray} 
m^2_{H_2} &=& \frac{1}{2} (3m_0^2 + A_t^2) \exp\left(-6h_t^2 t\right) - 
\frac{1}{2} m_0^2 - \frac{1}{2} A_t^2 \xrightarrow{t\to \infty, A_t = 0} 
- \frac{1}{2} m_0^2 \,  ,
\nonumber \\
m^2_{\tilde{Q}_3} &=& \frac{1}{6} (3m_0^2 + A_t^2) \exp\left(-6h_t^2 t\right) +
\frac{1}{2} m_0^2 - \frac{1}{6} A_t^2 \xrightarrow{t\to \infty, A_t = 0}  
 \frac{1}{2} m_0^2 \, , \\
m^2_{\tilde{u}_3} &=& \frac{1}{3} (3m_0^2 + A_t^2) \exp\left(-6h_t^2 t\right) 
- \frac{1}{3} A_t^2 \xrightarrow{t\to \infty, A_t = 0} 
0 \, . \nonumber 
\end{eqnarray}
The limit $t \to \infty$ refers to the electroweak scale ($v\simeq$ 246
GeV). We observe that at low energy the up-type Higgs mass-squared is driven
to a negative value due to strong $h_t$-effect. The above assumptions are
indeed too simplistic. Addition of gauge loops yield additional positive
contributions proportional to the gaugino mass-square $(M_i^2)$. Moreover,
running of $h_t$ and $A_t$ should also be considered which make the solutions
more complicated. All in all, RG evolution enforces a sign flip in $m^2_{H_2}$
only at the low scale, thus triggering EWSB.

\section{Little Higgs} {\sf The contents of this section (key ideas and
  illustration) have been developed together with Romesh K. Kaul. See also the
  discussion on little Higgs models in \cite{Kaul:2008cv}.}

We first discuss the basic ideas. Pions are spin-0 objects. Higgs is also a
spin-0 particle. Pions are composite objects. Higgs is perhaps elementary (as
indicated by electroweak precision measurements), but it can very well turn
out to be composite. The important thing is that the pions are light, and
there are reasons.  Can Higgs be light too for similar reasons? The lightness
of the pions owes its origin to their pseudo-Goldstone nature. These are
Goldstone bosons which arise when the chiral symmetry group ${\rm SU(2)}_L
\times {\rm SU(2)}_R$ spontaneously breaks to the isospin group ${\rm
  SU(2)}_I$. The Goldstone scalar $\phi$ has a shift symmetry $\phi \to \phi +
c$, where $c$ is some constant. Therefore, any interaction which couples
$\phi$ not as $\partial_\mu \phi$ breaks the Goldstone symmetry and attributes
mass to the previously massless Goldstone. Quark masses and electromagnetic
interaction {\em explicitly} break the chiral symmetry. Electromagnetism
attributes a mass to $\pi^+$ (more precisely, to the mass difference between
$\pi^+$ and $\pi^0$) of order $m_{\pi^+}^2 \sim (\alpha_{\rm em}
/4\pi)\Lambda_{\rm QCD}^2$. Can we think of the Higgs mass generation in the
same way?  We know that Yukawa interaction has a non-derivative Higgs
coupling, so it must break the Goldstone symmetry. Then, if we replace
$\alpha_{\rm em}$ by $\alpha_t \equiv h_t^2/4\pi$ and $\Lambda_{\rm QCD}$ by
some cutoff $\L$, we obtain
\begin{equation}
\label{mhpion}
m_h^2 \sim \left(\frac{\alpha_t}{4\pi}\right) \L^2 \, . 
\end{equation}
Is this picture phenomenologically acceptable? The answer is a big `no', since
a 100 GeV Higgs would imply $\L \sim 1$ TeV. This is what happens in {\em
  technicolor} models. Such a low cutoff is strongly disfavored by electroweak
precision tests.  Suppose that we arrange the prefactor in front of $\L^2$ to
be not $\left(\frac{\alpha_t}{4\pi}\right)$ but
$\left(\frac{\alpha_t}{4\pi}\right)^2$, i.e. the leading cutoff sensitivity
appears not at one-loop but parametrically at two-loop order, then the problem
might be {\em temporarily} solved. Let us see how. The Higgs mass will then be
given by
\begin{equation}
\label{lhmh1}
m_h^2 \sim 
\left(\frac{\alpha_t}{4\pi}\right)^2 \L^2 \, .    
\end{equation}
For $m_h \sim 100$ GeV, the cutoff would now be $\L \sim$ 10 TeV. In a sense,
this is nothing but a {\em postponement} of the problem as the cutoff of the
theory is now pushed by one order of magnitude.  {\em The idea of `little
  Higgs' is all about achieving this extra prefactor of $(\alpha_t/4\pi)$} --
see reviews \cite{lhreviews} and Refs.~\cite{littlest,simplest}. There are
indeed other concerns, which we will discuss later.
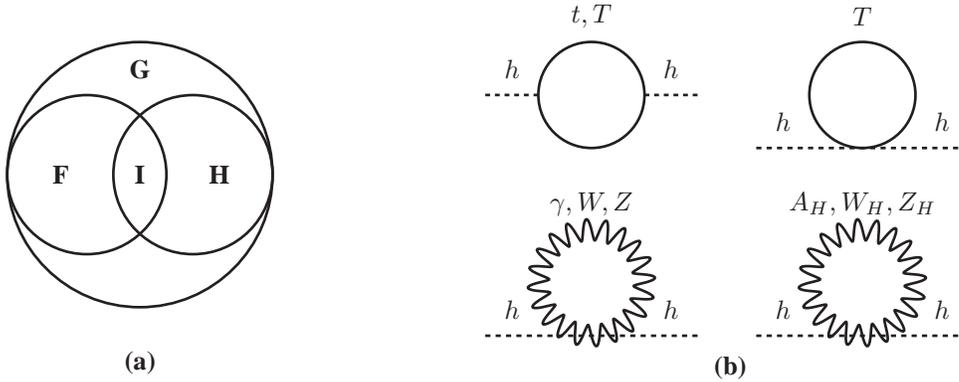
\begin{figure*}
\begin{minipage}[t]{0.42\textwidth}
\begin{center}
\begin{picture}(100,100)(-50,-10)
\SetWidth{1}
\CArc(0,0)(50,0,360)
\CArc(-20,0)(30,0,360)
\CArc(20,0)(30,0,360)
\Text(0,40)[]{\bf G}
\Text(-30,0)[]{\bf F}
\Text(0,0)[]{\bf I}
\Text(30,0)[]{\bf H}
    \Text(0,-71)[]{\bf (a)}
\end{picture}
\end{center}
\end{minipage}
\hspace{0.5cm}
\begin{minipage}[t]{0.42\textwidth}
\begin{picture}(100,70)(-50,0)
\SetWidth{1}
    \CArc(0,40)(20,0,360)
    \DashLine(-20,40)(-40,40)2 
    \DashLine(20,40)(40,40)2 
    \Text(-30,50)[]{$h$}
    \Text(30,50)[]{$h$}
    \Text(0,70)[]{$t,T$}
\end{picture}
\begin{picture}(100,70)(-50,0)
\SetWidth{1}
    \CArc(0,40)(20,0,360)
    \DashLine(-40,20)(40,20)2 
    \Text(-30,30)[]{$h$}
    \Text(30,30)[]{$h$}
    \Text(0,70)[]{$T$}
\end{picture}
\\
\begin{picture}(100,70)(-50,0)
\SetWidth{1}
    \PhotonArc(0,40)(20,0,360){4}{20}
    \DashLine(-40,20)(40,20)2 
    \Text(-30,30)[]{$h$}
    \Text(30,30)[]{$h$}
    \Text(0,70)[]{$\g,W,Z$}
\end{picture}
\begin{picture}(100,70)(-50,0)
\SetWidth{1}
    \PhotonArc(0,40)(20,0,360){4}{20}
    \DashLine(-40,20)(40,20)2 
    \Text(-30,30)[]{$h$}
    \Text(30,30)[]{$h$}
    \Text(0,70)[]{$A_H,W_H,Z_H$}
    \Text(-50,8)[]{\bf (b)}
\end{picture}
\end{minipage}
\vspace{5pt}
\caption[]{\small{\sf (a) Left panel: Little Higgs cartoon.  (b) Right panel:
    Feynman diagrams among which same statistics cancellation takes
    place. Note that $T$ is a (new) heavy quark, and $A_H, W_H, Z_H$ are
    (new) heavy gauge bosons -- see text.}}
\label{lhfig}
\end{figure*}

To appreciate the little Higgs trick we look into Fig.~\ref{lhfig}a.  A global
group $G$ spontaneously breaks to $H$ at a scale $f$. The origin of this
symmetry breaking is irrelevant below the cutoff scale $\Lambda \sim 4\pi f$.
$H$ must contain SU(2) $\times$ U(1) as a subgroup so that when a part of $G$,
labeled $F$, is weakly gauged the unbroken SM group (more precisely, the
electroweak part of the SM) $I =$ SU(2) $\times$ U(1) comes out.  The Higgs
doublet (under SU(2) of $I$), that would ultimately trigger electroweak
breaking, is a part of the Goldstone multiplet that parametrizes the coset
space $G/H$.  Choosing $G$, $H$ and $F$ is an open game. There are many
choices. We will give some examples in a while. {\em In fact, the little Higgs
  idea would work if the Higgs is a Goldstone boson under two different shift
  symmetries, i.e., $h \to h+c_1$ and $h \to h+c_2$. Both symmetries have to
  be broken. This is the idea of `collective symmetry breaking'}. It is
important to note that the generators of the gauged part of $G$ do not commute
with the generators corresponding to the Higgs, and thus gauge interaction
breaks the Goldstone symmetry. Yukawa interaction also breaks the Goldstone
symmetry.  Thus both gauge and Yukawa interactions induce Higgs mass at
one-loop level (the cutoff dependence would appear parametrically at two-loop
order, as we will see towards the end of this section).

\subsection{A simple example with G$=$SU(3) $\times$ SU(3)}
For the purpose of illustration in this review, let us consider a global group
${\rm SU(3)_V ~\times~ SU(3)_A}$.  Assume that there are two scalars $\Phi_1$
and $\Phi_2$ which transform as ($3,3$) and ($3,\bar 3$) respectively. Now,
imagine that each SU(3) spontaneously breaks to SU(2). So we start with
$8+8=16$ generators from the two SU(3), and end up with $3+3=6$ unbroken
generators corresponding to the two SU(2) groups. This means that $16-6=10$
generators are broken, thus yielding 10 massless Goldstone bosons.

Now, we gauge ${\rm SU(3)_V}$, but keep ${\rm SU(3)_A}$ global.  Hence, 5 out
of 10 broken generators are {\em eaten up} as the gauged ${\rm SU(3)_V}$ is
broken to ${\rm SU(2)}$, but 5 Goldstone bosons still remain.  This happens at
a scale higher than that of electroweak symmetry breaking, i.e. the
corresponding order parameter $f$ is larger than the electroweak vev $v$.
Note that since both $\Phi_1$ and $\Phi_2$ transform as 3 under ${\rm
  SU(3)_V}$, both couple to the same set of gauge bosons with identical
couplings. We can write $\Phi_1$ and $\Phi_2$ as
\begin{align}
\Phi_1 & = e^{i\frac{\t_E}{f}} e^{i\frac{\t_A}{f}} 
\left(\begin{array}{c}0\\ 0\\
f + \rho_1(x)\end{array}\right)~~,~~\Phi_2= e^{i\frac{\t_E}{f}} 
e^{-i\frac{\t_A}{f}}\left(\begin{array}{c}0 \\
0 \\ f + \rho_2(x) \end{array}\right) .
\end{align}
Above, $\rho_1$ and $\rho_2$ are real scalar fields which acquire masses $\sim
f$.  The phase $\t_E$ (where $E$ stands for `eaten') contains the d.o.f which
are eaten up (i.e. gauged away), while $\t_A$ contains 5 Goldstone bosons:
$\t_A = \sum_{a=4}^8 \t_A^a T_a$, where $T_4,..,T_8$ are broken
generators. One can express
\begin{align}
\t_A ={1\over\sqrt{2}}\left(\begin{array}{ccc}0&0&h^+\\ 0&0&h^0\\
h^-&h^{0*}&0\end{array}\right) + \frac{\eta}{4} 
\left(\begin{array}{ccc}1&0&0\\ 0&1&0\\
0&0&-2\end{array}\right) . 
\end{align}
The complex scalar $H = \left(\begin{array}{c}h^+\\ h^0\\ \end{array}\right)$
doublet under the {\em yet} unbroken SU(2) is {\em our} Higgs doublet,
i.e. the one with which we will implement the electroweak SSB. But, until this
point, $H$ (in fact, both the charged and neutral components contained in $H$)
and $\eta$ are both massless. 

Now recall that in the case of pions, the original ${\rm SU(2)}_L \times {\rm
  SU(2)}_R$ symmetry was {\em not} there to start with, as it was explicitly
violated by electromagnetic interaction and quark masses. In the present case,
the gauge and Yukawa interactions {\em explicitly} violate ${\rm
  SU(3)_A}$. This is the reason as to why we will be able to finally write
down a potential involving $H$.

\subsubsection{How does gauge interaction violate ${\rm \mathbf{SU(3)_A}}$ ?}
\begin{figure*}
\begin{minipage}[t]{0.42\textwidth}
  \begin{picture}(80,120)(-50,-70)
    \PhotonArc(0,0)(20,0,360){4}{20}
    \DashLine(0,-20)(-10,-40)2 
    \DashLine(0,-20)(10,-40)2
    \Text(0,35)[]{$A_H$}
    \Text(-12, -42)[]{$\Phi_{1,2}$}
    \Text(12, -42)[]{$\Phi_{1,2}$}
    \Text(0,-60)[]{\bf (a)}
  \end{picture}
\quad \quad 
  \begin{picture}(80,120)(-50,-70)
    \PhotonArc(0,0)(20,0,360){4}{20}
    \DashLine(0,-20)(-10,-40)2 
    \DashLine(0,-20)(10,-40)2
    \DashLine(0,20)(-10,40)2 
    \DashLine(0,20)(10,40)2
    \Text(35,0)[]{$A_H$}
    \Text(-35,0)[]{$A_H$}
    \Text(-12, -42)[]{$\Phi_1$}
    \Text(12, -42)[]{$\Phi_2$}
    \Text(-12, 42)[]{$\Phi_1$}
    \Text(12, 42)[]{$\Phi_2$}
    \Text(0,-60)[]{\bf (b)}
  \end{picture}
\end{minipage}
\hfill
\begin{minipage}[t]{0.48\textwidth}
  \begin{picture}(100,120)(-50,-70)
    \CArc(0,0)(30,0,360)
    \DashLine(-30,0)(-50,0)2 
    \DashLine(30,0)(50,0)2 
    \Text(0,35)[b]{$t_{1,2}$}
    \Text(0,-35)[t]{$Q'$}
    \Text(-50,5)[bl]{$\Phi_{1,2}$}
    \Text(50,5)[br]{$\Phi_{1,2}$}
    \Text(0,-60)[]{\bf (a)}
  \end{picture}
\quad \quad 
  \begin{picture}(100,120)(-50,-70)
    \CArc(0,0)(30,0,360)
    \DashLine(-24,18)(-40,34)2 
    \DashLine(-24,-18)(-40,-34)2 
    \DashLine(24,18)(40,34)2 
    \DashLine(24,-18)(40,-34)2 
    \Text(-42,36)[]{$\Phi_1$}
    \Text(-42,-36)[]{$\Phi_1$}
    \Text(42,36)[]{$\Phi_2$}
    \Text(42,-36)[]{$\Phi_2$}
    \Text(0,35)[b]{$Q'$}
    \Text(0,-35)[t]{$Q'$}
    \Text(-40,0)[]{$t_1$}
    \Text(40,0)[]{$t_2$}
    \Text(0,-60)[]{\bf (b)}
  \end{picture}
\end{minipage}
\vspace{5pt}
\caption[]{\small{\sf (i) Left panel: Heavy gauge boson loops on the SU(3)
    triplet $\Phi$ lines: (a) yields quadratic cutoff dependence which does
    not contribute to the Higgs potential; (b) yields log-divergent
    contribution to the Higgs mass. (ii) Right panel: Heavy fermion loops on
    the SU(3) triplet $\Phi$ lines.  Diagrams (a) yield quadratic cutoff
    sensitivity but does not contribute to the Higgs potential, while diagrams
    (b) contribute to the Higgs potential with a log sensitivity to the
    cutoff.}}
\label{lhfeyn}
\end{figure*}
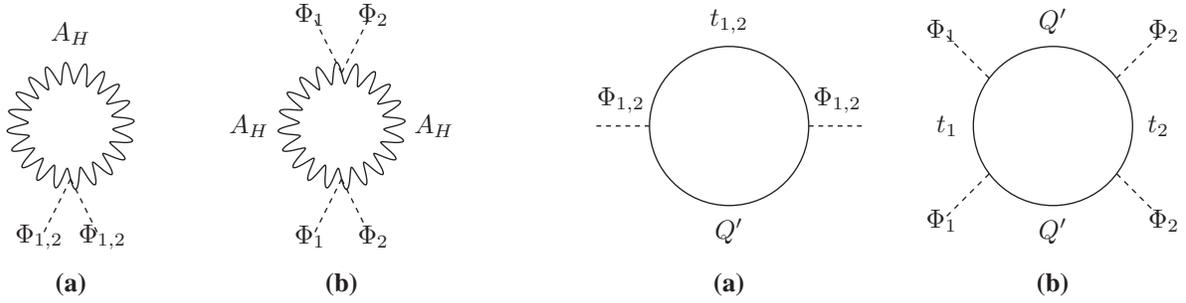

With ${\rm SU(3)_V}$ as the gauge group, the gauge interaction can be
expressed as
\begin{align}
  (D_\mu\Phi_1)^\dagger&(D_\mu\Phi_1) + (D_\mu\Phi_2)^\dagger(D_\mu\Phi_2) ,
  \qquad
  D_\mu=\p_\mu+igA_\mu^aT_a~~(a=1,2,\cdots,8),\\
  {\rm where}~~
  \Phi_1 & = e^{i\frac{\t_A}{f}} \left(\begin{array}{c}0\\ 0\\
      f + \rho_1 \end{array}\right)~~,~~\Phi_2=e^{-i\frac{\t_A}{f}}
\left(\begin{array}{c}0 \\
      0 \\ f + \rho_2\end{array}\right).
\end{align}
After integrating out the heavy ($\sim gf$) gauge bosons -- see
Fig.~\ref{lhfeyn}a (left panel) -- we obtain the following term in the
effective Lagrangian
\begin{align}
\label{g11}
-{g^2\over16\pi^2}\L^2\left(\Phi_1^\dagger\Phi_1+\Phi_2^\dagger\Phi_2 \right)
\, .
\end{align} 
Now, we observe two important things: 
\begin{itemize}
\item $\t_A$-dependence goes away in the above expression. Since
  the Higgs doublet $H={1\over\sqrt{2}}\left(\begin{array}{c}h^+\\
      h^0\end{array}\right)$ is contained inside $\t_A$, it is rotated away in
  $\Phi_i^\dagger \Phi_i$ and is hence insensitive to the quadratic cutoff
  dependence of Eq.~(\ref{g11}). This is not unexpected as the above piece of
  the Lagrangian is ${\rm SU(3)_A}$ invariant, and hence is blind to $\t_A$ or
  $H$.

\item The scalar excitations $\rho_1$ and $\rho_2$ can sense the quadratic
  cutoff, and therefore their masses (na\"ively of order $\sim f$) are not
  protected. This implies that the vev $f$ is also not protected from
  quadratic cutoff dependence\footnote{This is reminiscent of the quadratic
    cutoff sensitivity of the electroweak vev $v$ in the SM. The lack of
    `protection' is identical in the two cases.}.
\end{itemize} 

We reiterate that all the shift symmetries of the Goldstone boson have to be
broken, as any unbroken symmetry would keep the Goldstone massless.  Quadratic
divergence appears in those diagrams which involve only a single coupling
operator, and such an operator cannot sense the breaking of all the
symmetries.  For Higgs mass generation, the responsible pieces of the
Lagrangian involve {\em all} the symmetry breaking operators. Thus, the
relevant Feynman diagrams involve more internal propagators, which is why
there is no quadratic divergence. 

Let us look at the diagram in Fig.~\ref{lhfeyn}b (left panel). After the
heavy gauge bosons are integrated out, one obtains the following piece of the
effective Lagrangian, which breaks the ${\rm SU(3)_A}$ symmetry and hence can
contribute to the Higgs potential. The Lagrangian term has the following form:
\begin{align}
\label{g12}
-{g^4\over16\pi^2} \ln\left({\L^2\over
f^2}\right)|\Phi_1^\dagger\Phi_2|^2 \, . 
\end{align}

\begin{bquote} \small
{
\noindent\underline{\bf Calculation of $|\Phi_1^\dagger\Phi_2|^2$}
\begin{align*}
\Phi_1 &=e^{i\t_A/f}\left(\begin{array}{c}0\\ 0\\
f\end{array}\right) = \left(1+i{\t_A\over f}-{\t_A^2\over2f^2}\right)_{3\times3}
\left(\begin{array}{c}0\\ 0\\ f\end{array}\right)_{3\times1}\\
\t_A&={1\over\sqrt{2}}\left(\begin{array}{ccc}0&0&h^+\\ 0&0&h^0\\
h^-&h^{0*}&0\end{array}\right) ~\therefore \t_A^2={1\over2}
\left(\begin{array}{ccc}0&0&h^+\\0&0&h^0\\h^-&h^{0*}&0\end{array}\right)
\left(\begin{array}{ccc}0&0&h^+\\0&0&h^0\\h^-&h^{0*}&0\end{array}\right)
\end{align*}
\begin{align*}
\therefore \t_A^2|_{\rm 3rd~col} &={1\over2}
\left(\begin{array}{c}0\\0\\h^-h^+ + h^{0*}h^0\end{array}\right) =
\left(\begin{array}{c}0\\ 0\\H^\dagger H\end{array}\right)
\quad\text{where}\quad H={1\over\sqrt{2}}
\left(\begin{array}{c}h^+\\h^0\end{array}\right)\\
\text{Hence,}~~~ \Phi_1&=
\left(\begin{array}{c}0\\0\\f\end{array}\right) + {i\over\sqrt{2}}
\left(\begin{array}{c}h^+\\h^0\\0\end{array}\right) +
\left(\begin{array}{c}0\\0\\{-H^+H\over2f^2}\end{array}\right)f =
\left(\begin{array}{c}iH_{2\times1}\\f
\left(1-{H^\dagger H\over2f^2}\right)_{1\times1}\end{array}\right)\\
\therefore \Phi_1^\dagger&= \left(-iH^\dagger_{1\times2} ~~~~
f\left(1-{H^\dagger H\over2f^2}\right)_{1\times1}\right) \, ~~~~~
\text{Recall,}~~~\Phi_2 =
\left(\begin{array}{c}-iH_{2\times1}\\f
\left(1-{H^\dagger H\over2f^2}\right)_{1\times1}\end{array}\right)
\end{align*}
\begin{align*}
\therefore \Phi_1^\dagger\Phi_2 & =
-(H^\dagger H)+f^2\left(1-{H^\dagger H\over2f^2}\right)^2
= f^2-2(H^\dagger H)+{(H^\dagger H)^2\over4f^2}\\
\therefore |\Phi_1^\dagger \Phi_2|^2
&=-4f^2(H^\dagger H)+{9\over2}(H^\dagger H)^2+\cdots
\end{align*}
}
\end{bquote}
Notice that a potential of $H$ is generated with a bilinear and a quartic
term. Interestingly, the bilinear term has the negative sign required for SSB,
and the sign of the quartic term is positive as required by the stability of
the potential. After SSB, the Higgs mass is given by
\begin{equation}
\label{lh-g-mh}
m_h^2 \simeq {g^4\over16\pi^2}f^2 \ln \left({\L^2\over f^2}\right) \, . 
\end{equation}
It appears somewhat miraculous that unlike in SM, here the one-loop generated
$m_h^2$ is not proportional to $\L^2/16\pi^2$, but $f^2/16\pi^2$. The
cancellation of quadratic divergence takes place between two sets of diagrams,
one that contains the massless SU(2) gauge bosons and the other that contains
the massive gauge bosons (see Fig.~\ref{lhfig}b).  This is an example of {\em
  same statistics cancellation}.

\subsubsection{How does Yukawa interaction violate ${\rm \mathbf{SU(3)_A}}$ ?}
\label{howyukawa}
Consider a left-handed SU(3) triplet $Q'_{L} \equiv
\left(\begin{array}{c}t\\b\\T\end{array}\right)_L$ and three right-handed
singlets $t_R, b_R$ and $T_R$, i.e. the `new' states are $T_{L,R}$.  When the
gauged ${\rm SU(3)}_V$ breaks to SU(2) by the scalar vevs, the part $Q_L
\equiv \left(\begin{array}{c}t\\b\end{array}\right)_L$ inside $Q'_L$
transforms as a doublet under the SU(2).

\noindent Now, start with the following SU(3) invariant Yukawa interaction
Lagrangian:
\begin{eqnarray}
\Lag_Y & ={h_t\over\sqrt{2}}
[t_1^c\Phi_1^\dagger Q'_L+t_2^c\Phi_2^\dagger Q'_L], ~~~ \text{where}~~
h_t\equiv h_t^{(1)}=h_t^{(2)}, ~~
t_{1,2} \equiv {1\over\sqrt{2}}(T_R\pm it_R) \, . 
\end{eqnarray}
\begin{quote} {
\noindent {\em Some algebraic steps:}
\begin{eqnarray*}
\Phi_1^\dagger Q'_L&=& \left(-iH^\dagger_{1\times2} ~~~~~
f\left(1-{H^\dagger H\over2f^2}\right)_{1\times1}\right)
\left(\begin{array}{c}Q_{L(2\times1)}\\ T_{L(1\times1)}\end{array}\right)
=-iH^\dagger Q_L + f\left(1-{H^\dagger H\over2f^2}\right)T_L\\
\text{And,}~~ \Phi_2^\dagger Q'_L&=&
iH^\dagger Q_L + f\left(1-{H^\dagger H\over2f^2}\right)T_L \, .
\end{eqnarray*}
}
\end{quote}
\begin{eqnarray}
\therefore \Lag_Y &=&{h_t \over\sqrt{2}}\left[t_1^c\left\{-iH^\dagger Q_L
+ f\left(1-{H^\dagger H\over2f^2}\right)T_L\right\} +t_2^c\left\{
iH^\dagger Q_L+f\left(1-{H^\dagger H\over2f^2}\right)T_L\right\}\right]
\nonumber \\
&=&h_t\left[{-i\over\sqrt{2}}(t_1^c-t_2^c)Q_LH^\dagger + {f\over\sqrt{2}}
(t_1^c+t_2^c) \left(1-{H^\dagger H\over2f^2}\right)T_L\right] \nonumber \\
&=&h_t\bar{t}_R Q_LH^\dagger + h_t
f\left(1-{H^\dagger H\over2f^2}\right)\bar{T}_R T_L \, . 
\end{eqnarray}
The first term in the above expression contains the SM top quark Yukawa
coupling, and the second term indicates that the $T$ quark is heavy ($\sim
f$). 

Fig.~\ref{lhfeyn}a (right panel) yields an one-loop effective Lagrangian   
\begin{align}
\label{h11}
-{h_t^2\over16\pi^2}\L^2(\Phi_1^\dagger \Phi_1+\Phi_2^\dagger \Phi_2). 
\end{align}
This is exactly similar to Eq.~(\ref{g11}) with $g \leftrightarrow
h_t$. Again, this Lagrangian preserves ${\rm SU(3)_A}$, and hence not relevant
for the Higgs potential. We then turn to Fig.~\ref{lhfeyn}b (right panel),
which yields
\begin{align}
\label{h12}
-{h_t^4\over16\pi^2}\ln\left({\L^2\over
f^2}\right)|\Phi_1^\dagger \Phi_2 + \Phi_2^\dagger \Phi_1|^2. 
\end{align}
This Lagrangian is similar to Eq.~(\ref{g12}) with $g \leftrightarrow
h_t$. This piece of the Lagrangian is of interest to us as it yields the
bilinear and quartic terms involving $H$ with the right sign of the
coefficients.  After SSB the Higgs mass is generated as
\begin{equation}
\label{lh-ht-mh}
m_h^2 \simeq {h_t^4\over16\pi^2}f^2 \ln \left({\L^2\over f^2}\right),
\end{equation}
which is similar to Eq.~(\ref{lh-g-mh}) with $g \leftrightarrow h_t$.  Again,
the apparently miraculous cancellation of quadratic divergence can be
diagrammatically understood by the cancellation occurring between the $t$ and
$T$ loops (see Fig.~\ref{lhfig}b), which is yet another example of {\em same
  statistics cancellation}.

\subsection{Salient features of little Higgs models}
\subsubsection{Quadratic cutoff sensitivity} 
Although {\em same statistics cancellations} enable us to express $m_h^2$ as
proportional to $f^2/16\pi^2$ (i.e. {\em not} as $\L^2/16\pi^2$) with only a
logarithmic cutoff sensitivity at one-loop, as reflected in
Eqs.~(\ref{lh-g-mh}) and (\ref{lh-ht-mh}), the quadratic cutoff sensitivity
comes back parametrically at two-loop order.  To appreciate this, first recall
that in the SM the one-loop correction to the Higgs mass goes as
\begin{equation}
\label{mhsm}
\Delta m_h^2~({\rm SM})~ \sim \frac{\L^2}{16\pi^2} \, .
\end{equation} 
This means that the electroweak vev ($v$) receives a quadratic ($\L^2$)
correction,
\begin{equation}
v^2 \to v^2 + \frac{\L^2}{16\pi^2} \, . 
\end{equation}
Now consider the gauging of ${\rm SU(3)_V}$ as discussed in the previous
subsection. The corresponding order parameter is $f$, but note that $f$ is as
unprotected as the electroweak vev $v$ is in the SM \cite{Kaul:2008cv}. Hence,
    \begin{equation}
       f^2 \to F^2 = f^2 + \frac{a}{16\pi^2} \Lambda^2 = (1+a) f^2
~~~~~({\rm since}~ \L = 4\pi f) \, ,  
     \end{equation}
     where $a \sim {\cal{O}} (1)$.  Then, what did we gain {\em vis-\`a-vis}
     the SM?  For little Higgs models
\begin{equation}
\label{mhlh}
  m_h^2~({\rm LH}) \sim \left(\frac{1}{16\pi^2}\right) 
  F^2 \ln \left(\frac{\L^2}{F^2}\right) ~~ 
  \Longrightarrow ~~\Delta m_h^2~({\rm LH}) 
\sim \left(\frac{1}{16\pi^2}\right)^2 \L^2  \, . 
\end{equation}

\begin{list}{$\bullet$}
\item Note that the quadratic cutoff sensitivity of the Higgs mass-square
  exists not only in the SM but also in the little Higgs models. Then, what
  purpose did little Higgs serve?  In the little Higgs case there is an extra
  loop suppression factor -- compare Eq.~(\ref{mhsm}) with Eq.~(\ref{mhlh}).
  The appearance of the cutoff in the little Higgs models is thus postponed by
  one decade in energy scale compared to the SM. One important thing should be
  kept in mind. A Goldstone boson becoming massive in little Higgs models is
  not a surprise. The global symmetry is {\em explicitly} broken to start with
  by the gauge and Yukawa interactions, and precisely for this reason the
  loosely mentioned Goldstone boson is actually a pseudo-Goldstone boson
  (pGB). Up to this point there is no difference with the theory of pions
  where electromagnetic interaction and quark masses explicitly break the
  Goldstone symmetry. What is new here, i.e. the reason for which we consider
  the little Higgs construction as an important achievement over the SM, is
  the appearance of the quadratic cutoff dependence of the Higgs mass at the
  next order in perturbation theory, i.e. at the two-loop level.
\end{list}

If we want $m_h \sim (f/4\pi) \sim 100$ GeV, it immediately follows that $f
\sim F \sim 1$ TeV, and the cutoff of the theory is $\Lambda \sim 4\pi f \sim$
10 TeV, as against the SM cutoff of $4 \pi v \sim$ 1 TeV.  The ultraviolet
completion beyond 10 TeV in little Higgs models is a detailed model-dependent
issue \cite{Roy:2005hg}. 

\subsubsection{Large quartic coupling}
A clever construction of a little Higgs theory should yield the following
electroweak Higgs potential:
\begin{equation}
  V = - \frac{\left({g~{\rm or}~h_t}\right)^4}{16\pi^2}f^2 
\ln\left(\frac{\L^2}{f^2}\right)
  (H^\dagger H) + \lambda (H^\dagger H)^2 \, , 
\end{equation} 
i.e., the bilinear term should have a one-loop suppression but, crucially, the
quartic interaction should be {\em un-suppressed}, i.e. $\lambda \sim
g^2~({\rm or}~ h_t^2)$. If both quadratic and quartic terms are suppressed, it
is not possible to simultaneously obtain the correct $W$ boson mass and a
phenomenologically acceptable Higgs mass.  In the simple scenario used for our
illustration, both the quadratic and quartic terms are generated by loops, so
the phenomenological problem survives. In more realistic scenarios, as we will
see shortly, this problem can be avoided.  We will discuss only some of these
scenarios below.

\subsection{Realistic little Higgs scenarios - a brief description}  
\subsubsection{Different choices of groups}
The `littlest Higgs' \cite{littlest} construction is based on a choice of a
global group $G = {\rm SU(5)}$ which breaks to $H = {\rm SO(5)}$ by the vev
($\Sigma_0$) of a scalar field, expanded as $\Sigma = e^{2i\Pi/f} \Sigma_0$,
where $\Pi = \Pi^a X_a$ contains the Goldstone bosons, $X_a$ being the broken
generators. The $5 \times 5$ vev matrix is given by $\Sigma_0 =$ anti-diagonal
$\left(1_{2 \times 2}, 1, 1_{2 \times 2}\right)$.  The subgroup of SU(5) that
is gauged is $\left[{\rm SU(2)} \times {\rm U(1)}\right]_1 \times \left[{\rm
    SU(2)} \times {\rm U(1)}\right]_2$ which breaks to ${\rm SU(2)}_D \times
{\rm U(1)}_Y$. Out of the 14 ($=24-10$) pGB's generated during $G \to H$, four
are absorbed as the longitudinal components of the massive gauge bosons $A_H$,
$Z_H$ and $W^\pm_H$ corresponding to the broken ${\rm SU(2)} \times {\rm
  U(1)}$ generators.  The other 10 scalar degrees of freedom arrange
themselves as a complex SU(2) scalar doublet $H$ with the right quantum
numbers required to make a SU(2) Higgs doublet with hypercharge ($= 1/2$) and
a complex scalar SU(2) triplet $\Phi$ with hypercharge ($= 1$). In the limit
when any pair of gauge couplings ($g_1, g_1^\prime$) or ($g_2, g_2^\prime$)
goes to zero, the Higgs field becomes exactly massless. Therefore, any loop
diagram contributing to the Higgs mass must involve a product $g_1 g_2$ (or,
$g_1^\prime g_2^\prime$). Due to this {\em collective} symmetry breaking, all
such diagrams are logarithmically sensitive to the cutoff at one-loop.

The type of little Higgs models discussed earlier for the purpose of
illustration, i.e. where the global group is $G = {\rm SU(3)} \times {\rm
  SU(3)}$ and the gauged subgroup is the simple group SU(3), is called the
`simplest' \cite{simplest}. The difficulty of achieving a large quartic
coupling was overcome by considering $G = [{\rm SU(4)}]^4$ which breaks to $H
= [{\rm SU(3)}]^4$, while the gauged subgroup is ${\rm SU(4)} \times {\rm
  U(1)}$ which breaks down to ${\rm SU(2)} \times {\rm U(1)}$. Out of the 28
pGB's 12 are eaten up by the massive gauge bosons. The 16 degrees of freedom
are distributed as two complex doublets, three complex singlets and two real
singlet scalars. The scalar quartic coupling is generated at tree level.

The authors of \cite{Low:2002ws} have considered $G = {\rm SU(6)}$ and $H =
{\rm Sp(6)}$. The gauged subgroup is $\left[{\rm SU(2)} \times {\rm
    U(1)}\right]^2$ which breaks to ${\rm SU(2)}_D \times {\rm U(1)}_Y$. So,
out of the $35 - 21 = 14$ pGB's four are absorbed by the massive gauge bosons,
and the remaining 10 degrees of freedom are decomposed into two complex
doublet scalars and one complex singlet scalar. A distinct advantage here is
that there is no triplet scalar which could have caused some trouble in
electroweak precision tests (see discussions later).

The moose models are on the other hand based on the concept of {\em
  deconstruction} (a term borrowed from Economics). The electroweak sector is
described by a product global symmetry $G^N$ which is broken by the
condensates transforming as bi-fundamentals under $G_i \times G_j$, where
$i,j$ are the sites.  In Ref.~\cite{ArkaniHamed:2002qx}, the global group
considered is $G^N = [{\rm SU(3)}]^8$, and a subgroup of it is gauged which
eventually breaks to ${\rm SU(2)} \times {\rm U(1)}$. The scalar spectrum
contains two complex SU(2) doublets, a complex SU(2) triplet and a complex
singlet. To ensure custodial SU(2) symmetry, i.e. to maintain consistency with
the oblique $\Delta \rho$ (or, $T$) parameter, the global group was enlarged
in \cite{Chang:2003un} to $[{\rm SO(5)}]^8$ with the gauge group ${\rm SO(5)}
\times {\rm SU(2)} \times {\rm U(1)}$. To further minimize the scalar
contribution to $\Delta \rho$, a coset space ${\rm SO(9)}/[{\rm SO(5)} \times
{\rm SO(4)}]$ was constructed with the gauge symmetry ${\rm SU(2)}_L \times
{\rm SU(2)}_R \times {\rm SU(2)} \times {\rm U(1)}$ \cite{Chang:2003zn}. A
review of these and many other models can be found in \cite{lhreviews}.

\subsubsection{Bounds from electroweak precision tests (EWPT)}  
In an effective field theory description \cite{Skiba:2010xn}, two dimension-6
operators ${\O}_T \propto \left|H^\dagger D_\mu H \right|^2$ and ${\O}_S
\propto H^\dagger \sigma^a H W_{\mu\nu}^a B_{\mu\nu}$ serve as the primary
filters before certifying whether a model passes EWPT or not. Recall that an
SU(2) global custodial symmetry in the SM guarantees the tree level relation
$M_W = M_Z \cos\t_W$. The operator ${\O}_T$ violates that symmetry, which is
not difficult to conceive: when $H$ goes to the vacuum, ${\O}_T \propto Z_\mu
Z^\mu$ but there is no similar contribution for $W_\mu W^\mu$, i.e. there is a
contribution to $M_Z$ but not to $M_W$, and this mismatch violates custodial
symmetry. Similarly, the operator ${\O}_S$ induces kinetic mixing between
$W_\mu^3$ and $B_\mu$. The coefficients of ${\O}_T$ and ${\O}_S$ will,
therefore, indicate the contributions to the $T$ and $S$ parameters,
respectively.

Unless special care is taken, a general class of little Higgs models gives a
large contribution to $T$, and hence receives a strong constraint: $f > (2-5)$
TeV \cite{large_f_ewpt}. A large $f$ means that to obtain the Higgs mass in
the 100 GeV range one has to fine-tune the parameters. The constraints arise
primarily from the tree level mixing of the SM particles with the new
particles. In the littlest Higgs model, the $T$ parameter receives a large
contribution from the custodial symmetry breaking trilinear operator $H^T \Phi
H$, which mixes the doublet $H$ with the triplet $\Phi$.  Also, the $W_L W_H
HH$ term ($W_L$ is the SM gauge boson and $W_H$ is the heavy one) yields a
sizable contribution to $T$. To circumvent these constraints, the authors of
\cite{Low:2004xc} introduced, more in the spirit of $R$-parity in
supersymmetry, what is called $T$-parity under which all ({\em but one}) new
particles are odd and the SM particles are even. It is a discrete $Z_2$
symmetry, which is an automorphism of the gauge groups that exchanges the
gauge bosons of $\left[{\rm SU(2)} \times {\rm U(1)}\right]_1$ and $\left[{\rm
    SU(2)} \times {\rm U(1)}\right]_2$. It also means $g_1 = g_2$ and
$g_1^\prime = g_2^\prime$. Under this symmetry $H \to H$, but $\Phi \to -
\Phi$, so the problematic $H^T \Phi H$ coupling is absent. Contributions to
$T$ and $S$ from heavy particles arise only at the loop level. As a result,
$f$ as low as 500 GeV can be accommodated without facing any inconsistency
with EWPT \cite{Hubisz:2005tx}. It should be noted that there is one new, yet
$T$-even, state in this scenario, the so-called `top partner' which cancels
the standard top induced quadratic divergence to the Higgs mass.  This state
has a positive contribution to the $T$ parameter, and to compensate that one
may need a Higgs mass as large as 800 GeV \cite{Hubisz:2005tx}. Chen's review
in \cite{lhreviews} covers the EWPT and naturalness constraints on quite a few
such scenarios. In a recent development, the authors of
Ref.~\cite{Schmaltz:2010ac} have considered a ${\rm SO(6)} \times {\rm
  SO(6)}/{\rm SO(6)}$ model, called it the `Bestest' little Higgs, and claimed
that quartic coupling can be generated without violating custodial symmetry
($S$ and $T$ vanish at tree level) and at the same time keeping the
fine-tuning within 10\% in the top sector.

\subsubsection{Collider signals of little Higgs models}
Since each little Higgs model involves a $G/H$ coset space and an extended
electroweak gauge sector, there are invariably new weak gauge bosons, new
fermions and new scalars. To confirm little Higgs models, those news particles
have to be looked for in the colliders (see the study made by the ATLAS
collaboration at the LHC \cite{Azuelos:2004dm}).

\noindent{\bf New gauge bosons:}~ In the littlest Higgs model, the couplings
of the heavy gauge bosons $Z_H$ and $W_H$ with the fermions are universal
which, beside a mixing angle factor, depend only on the weak isospin $t_3$ of
the fermions (i.e. purely left-handed) and not on the electric charge $Q$.  It
has been shown that about 30000 $Z_H$ can be produced annually at the LHC with
100${\rm fb}^{-1}$ luminosity. These heavy gauge bosons would decay into the
SM fermions ($V_H \to f \bar f^\prime$), or into the SM gauge bosons ($Z_H \to
W_L^+ W_L^-$, $W_H \to W_L Z_L$, where $V_L \equiv V_{\rm SM}$), or into the
Higgs and SM gauge boson ($V_H \to V_L h$). The branching ratios would follow
a definite pattern, which would serve as the `smoking gun signals'
\cite{Han:2003wu,Burdman:2002ns}.

\noindent{\bf New fermions:}~ Colored vector-like $T$ quark features in almost
all little Higgs models. It may be produced singly by $bW \to T$ at the
LHC. Typically, $\Gamma (T \to th) \approx \Gamma (T \to tZ) \approx
\frac{1}{2} \Gamma (T \to bW)$. This branching ratio relation would constitute
a characteristic signature for $T$ quark discovery
\cite{Han:2003wu,Perelstein:2003wd}. When $T$-parity is conserved, one has a
$T$-odd state $t_{-}$ and a $T$-even state $t_{+}$ (which has been referred
above as the $T$ quark, and which also cancels the SM top induced quadratic
divergence to the Higgs mass), and $m_{t_+} > m_{t_-}$. The QCD production
cross section $\sigma(gg\to t_- t_-) \approx 0.3$ pb for $m_{t_-} = 800$ GeV,
and almost all the time $t_-$ would decay as $t_- \to A_H t$, where $A_H$ is
the lightest $T$-odd gauge boson which, being stable, would escape the
detector carrying missing energy \cite{Hubisz:2005tx}.

\noindent{\bf New scalars:}~ The presence of a doubly charged scalar
$\phi^{++}$, as a component of a complex triplet scalar, is a hallmark
signature of a large class of little Higgs models. Its decay into like-sign
dileptons ($\phi^{++} \to \ell^+ \ell^+$) would lead to an unmistakable signal
with a separable SM background \cite{Han:2003wu}. The other spectacular signal
of the doubly charged scalar would be a resonant enhancement of $W_L W_L \to
W_L W_L$ proceeding via $\phi^{++}$ exchange. An analysis of $M(W^+W^+)$
invariant mass distribution was carried out in \cite{Han:2003wu} with the
claim that with 300${\rm fb}^{-1}$ luminosity at the LHC about 100 events
would pop up over the SM background for $m_{\phi^{++}} = 1.5$ TeV, assuming a
triplet to doublet vev ratio $v^\prime/v = 0.05$. One can do a little further
by employing the triplet scalar in generating neutrino mass via type-II
see-saw. The maximal mixing in the $\mu-\tau$ sector would predict equal
branching ratios of $\phi^{++}$ in the $\mu^+\mu^+$, $\mu^+\tau^+$ and
$\tau^+\tau^+$ channels, which can be tested at the LHC. Employing this
correlation, a discovery limit of $m_{\phi^{++}} = 700$ GeV has been claimed
with only 30${\rm fb}^{-1}$ luminosity at the LHC, where the authors take into
consideration particle reconstruction efficiencies as well as Gaussian
distortion functions for the momenta and missing energy of final state
particles \cite{Hektor:2007uu}.

We conclude this section with the statement that little Higgs models with
$T$-parity and supersymmetry with $R$-parity would be hard to distinguish
apart at the LHC. Universal Extra Dimension (UED) with KK-parity would also
give similar signals. The best way to study them is to consider their
production via strong interaction and their decay via weak interaction. The
authors of \cite{Bhattacherjee:2009jh} have concentrated on final states
containing an unspecified number of jets, three or four leptons and missing
transverse momentum. They have asserted that the jet multiplicity
distributions are the crucial discriminating factors among the scenarios and
they have constructed several discriminating variables.  This is still an open
issue and constitutes a challenging {\em inverse problem}.

\section{Gauge-Higgs unification}
The basic idea of Gauge-Higgs unification (GHU) is that the Higgs boson would
arise from the internal components of a higher dimensional gauge field. As a
result, higher dimensional gauge invariance would protect the Higgs mass from
quadratic divergence.  When the extra space coordinate is not simply connected
(e.g. $S^1$), there are Wilson line phases associated with the extra
dimensional component of the gauge field (this is conceptually similar to
Aharonov-Bohm phase in quantum mechanics). Their 4d quantum fluctuation is
identified with the Higgs field. Higher dimensional gauge invariance does not
allow any scalar potential at the tree level. The scalar potential is
generated through radiative corrections.  The Higgs boson acquires a mass
through this radiatively generated potential.  One of the earliest
realizations of GHU was provided by Antoniadis in a work on extra dimension in
the supersymmetric context where the Higgs was coming from $N=4$
supermultiplet, i.e. from a higher dimensional gauge field
\cite{Antoniadis:1990ew}. But for the purpose of illustration we do not bring
in any supersymmetric aspect. We rather focus on the underlying dynamics of
GHU mechanism in non-supersymmetric extra dimensional context, for which we
proceed step by step \cite{Kubo:2001zc}.

\subsection{5d QED as an illustrative example}
The 5d Lagrangian, a function of the usual 4d coordinates ($x_\mu$) and the
5th space coordinate ($y$), is given by
\begin{equation}
\Lag(x,y) = -{1\over4} F_{MN}(x,y)F^{MN}(x,y) + \Lag_{\rm GF}(x,y) \, , 
~~~{\rm where} 
\end{equation}
$$
F_{MN}(x,y) = \p_MA_N(x,y)-\p_NA_M(x,y) \, . 
$$
The indices $M,N=(\mu, 5)$; with $\mu=0,1,2,3$. The symbol `GF' means
gauge-fixing.

The 5d gauge field $A_M$ transforms as a vector under the Lorentz group
SO(1,4). In the absence of gauge-fixing, the 5d QED Lagrangian is invariant
under a U(1) gauge transformation
$$
A_M(x,y) \to A_M(x,y)+\p_M\T(x,y) \, . 
$$
The compactification is on an orbifold $S^1/Z_2$, i.e., with $y \to (-y)$
identification. In order not to spoil gauge symmetry the following conditions
need to be satisfied, which allows a massless photon in 4d:
\begin{align}
A_M(x,y) &= A_M(x,y+2\pi R) \, , ~~
A_\mu(x,y) = A_\mu(x,-y) \, , ~~
A_5(x,y) = -A_5(x,-y) \, , \nonumber \\
\T(x,y) &= \T(x,y+2\pi R) \, , ~~~~
\T(x,y) =\T(x,-y) \, . 
\end{align}
The above conditions guarantee that the theory remains gauge invariant even
after compactification.  The Fourier mode expansions of different 5d fields 
are given by ($R$ is the radius of compactification). 
\begin{align}
A_\mu(x,y) & = {1\over\sqrt{2\pi R}} A_\mu^{(0)}(x) +
{1\over\sqrt{\pi R}} \sum_{n=1}^\infty A_\mu^{(n)}(x)
\cos\left({ny\over R}\right) \, , \nonumber \\
A_5(x,y) & = {1\over\sqrt{\pi R}} \sum_{n=1}^\infty
A_5^{(n)}(x) \sin\left({ny\over R}\right) \, , \\
\T(x,y) & = {1\over\sqrt{2\pi R}} \T^{(0)}(x) + {1\over\sqrt{\pi
R}} \sum_{n=1}^\infty \T^{(n)}(x)\cos\left({ny\over
R}\right) \, . \nonumber 
\end{align}
Above, $A_\mu^{(0)}(x)$ and $\T^{(0)}(x)$ are zero modes, which are the
relevant fields for ordinary 4d QED. As expected, there is no zero mode for
$A_5$.

The 4d effective Lagrangian is obtained by integrating out the fifth
coordinate, and is given by
$$
\Lag(x) = \int_0^{2\pi R} dy \Lag(x,y) \, . 
$$
The higher dimensional physics is reflected by the infinite tower of
Kaluza-Klein modes. A simple calculation yields the following 4d Lagrangian
\begin{eqnarray}
\label{4dL}
\Lag(x)&=& -{1\over4} F_{\mu\nu}^{(0)} F^{\mu\nu(0)} + \sum_{n=1}^\infty
\left[-{1\over4}F_{\mu\nu}^{(n)} F^{\mu\nu(n)} 
+{1\over2}\left({n\over R} A_\mu^{(n)}+\p_\mu
A_5^{(n)}\right)^2 \right] 
+ \Lag_{\rm GF}(x) \, . 
\end{eqnarray}
\begin{bquote}
\noindent\underline{\bf Steps leading to Eq.~(\ref{4dL})}
\begin{align*}
F_{MN}F^{MN} &=
F_{\mu\nu}F^{\mu\nu}+F_{\mu5}F^{\mu5}+F_{5\mu}F^{5\mu}
=F_{\mu\nu}F^{\mu\nu}+2F_{\mu5}F^{\mu5} \, ;\\
F_{\mu5}F^{\mu5}&=(\p_5A_\mu-\p_\mu A_5)^2=
(\p_5A_\mu)^2+(\p_\mu A_5)^2-2(\p_5A_\mu)(\p_\mu A_5) \, ;\\
\int_0^{2\pi R}d y (\p_5 A_\mu)^2&={n^2\over R^2} {1\over\pi R}
\left(A^{(n)}_\mu(x)\right)^2 \int_0^{2\pi R}dy\sin^2{ny\over R} = {n^2\over
R^2}(A_\mu^{(n)}(x))^2 \, ;\\
\int_0^{2\pi R} dy (-)\p_5A_\mu\p_\mu A_5&=\int_0^{2\pi R} dy \left({n\over
R}\right) {1\over\sqrt{\pi R}} A_\mu^{(n)}(x) \sin{ny\over R}
{1\over\sqrt{\pi R}} \p_\mu A_5^{(n)}(x)\sin {ny\over R}  \\
&=\left({n\over R}\right)A_\mu^{(n)}(x) \p_\mu A_5^{(n)}(x) \, .
\end{align*}
\end{bquote}
Now we shall show that the modes $A_5^{(n)}$, which are scalars with respect
to 4d Lorentz group, play the r\^ole of `would-be' Goldstone modes to be
`eaten up' by the massive $A_\mu^{(n)}$. In fact, in a sense, the mass
generation of heavy KK gauge modes by compactification can be viewed as a kind
of geometric Higgs mechanism.

We should keep in mind that the Lagrangian $\Lag(x)$ is still manifestly
gauge invariant by the joint actions of two transformations at each KK level:
\begin{align}
A_\mu^{(n)}(x) \to A_\mu^{(n)}(x) + \p_\mu\T^{(n)}(x) \, , ~~~~
A_5^{(n)}(x) \to A_5^{(n)}(x) - {n\over R}\T^{(n)}(x) \, . 
\end{align}

Now we use 't Hooft's gauge fixing condition by which the terms that mix
$A_\mu^{(n)}$ and $A_5^{(n)}$ are removed from the 4d effective Lagrangian.
We write
\begin{equation}
\Lag_{\rm GF}(x,y) = -{1\over 2\xi} \left(\p_\mu A_\mu(x,y) - \xi \p_5
A_5(x,y)\right)^2 \, . 
\end{equation}
Note that in the last equation the requirement of covariance of the gauge
fixing Lagrangian with respect to the $y$-direction has been sacrificed, which
is nothing serious as compactification $(S^1/Z_2)$ breaks SO(1,4) invariance
under ordinary 4d Lorentz transformation any way.

Now we calculate $\Lag(x)=\int_0^{2\pi R} dy \Lag (x,y)$ where $\Lag(x,y)$
contains the above $\Lag_{\rm GF}(x,y)$. All mixing terms involving
$A_\mu^{(n)}$ and $A_5^{(n)}$ are now reduced to total derivatives which are
irrelevant. Then the gauge-fixed 4d Lagrangian looks like
\begin{align}
\Lag(x)&= - {1\over4}
F_{\mu\nu}^{(0)}F^{\mu\nu(0)}-{1\over2\xi}(\p_\mu A_\mu^{(0)})^2 \nonumber \\
&+\sum_{n=1}^\infty\left[-{1\over4}F_{\mu\nu}^{(n)}F^{\mu\nu(n)} -
{1\over2\xi}\left(\p_\mu A_\mu^{(n)}\right)^2+{1\over2}\left({n\over
R}\right)^2 A_\mu^{(n)}A_\mu^{(n)}\right]\\
&+\sum_{n=1}^\infty\left[{1\over2}\left(\p_\mu A_5^{(n)}\right)^2 
 - {1\over2}\xi\left({n\over R}\right)^2 \left(A_5^{(n)}\right)^2\right] \, .
\nonumber 
\end{align}
The scalars $A_5^{(n)}$ with `gauge dependent masses' resemble the
would-be Goldstone bosons of an ordinary 4d Abelian theory in
$R_\xi$ gauge.
\begin{align}
  A_\mu^{(n)} ~\text{propagator}& ~~~~~\Rightarrow {1\over k^2-{n^2\over
      R^2}}\left[-g^{\mu\nu}+{(1-\xi)k^\mu
      k^\nu\over k^2-\xi\left({n\over R}\right)^2}\right] \, , \nonumber \\
  A_5^{(n)} ~\text{propagator}  & ~~~~~\Rightarrow
{1\over k^2-\xi\left({n\over R}\right)^2} \, . 
\end{align}
Clearly, the $A_5^{(n)}$ modes are unphysical, and they provide the
  longitudinal components of the massive $A_\mu^{(n)}$ states.

\subsection{5d SU(2) model as an illustration}
The gauge group is SU(2), the compactification is on $S^1/Z_2$, and we impose
a non-trivial $Z_2$ parity:
\begin{eqnarray}
P = \left(\begin{array}{cc}1&0\\
0&-1\end{array}\right)~,~\begin{array}{c}A_\mu \xrightarrow{Z_2} PA_\mu
P^\dagger \\ A_5\xrightarrow{Z_2} -PA_5P^\dagger \end{array} \, , 
\end{eqnarray}
where $A_\mu = A_\mu^a \tau_a$ is the Lie-algebra valued 5d gauge field. In
component form 
\begin{align}
A_\mu=A_\mu^a\tau_a&=\left(\begin{array}{cc} A_\mu^3&
A^1_\mu- i A^2_\mu\\ A^1_\mu+iA_\mu^2 & -A^3_\mu\end{array}\right) \, .\\ 
\nonumber \\
\therefore P A_\mu P^\dagger &= \left(\begin{array}{cc} 1 & 0\\ 0 &
-1\end{array}\right) \left(\begin{array}{cc} A_\mu^3&
A^1_\mu-iA^2_\mu\\ A^1_\mu+iA_\mu^2 & -A^3_\mu\end{array}\right)
\left(\begin{array}{cc} 1 & 0\\ 0 & -1\end{array}\right) \nonumber \\
\nonumber \\
& =
\left(\begin{array}{cc} A_\mu^3& (-)(A^1_\mu-iA^2_\mu)\\
(-)(A^1_\mu+iA_\mu^2) & -A^3_\mu\end{array}\right) \, . 
\end{align}
Clearly,
\begin{eqnarray}
A_\mu^3 & \xrightarrow{Z_2} A_\mu^3 \, , ~~~
(A_\mu^1, A_\mu^2) &\xrightarrow{Z_2} (-)(A_\mu^1, A_\mu^2) \, . 
\end{eqnarray}
Hence, $A_\mu^3(x,y)$ has zero mode $A_\mu^{3(0)}(x)$, but $A_\mu^1(x,y)$ and
$A_\mu^2(x,y)$ do not have zero modes.  Since $A_5\xrightarrow{Z_2}
-PA_5P^\dagger$, it is easy to show that $A_5^1(x,y)$ and $A_5^2(x,y)$ (and
not $A_5^3(x,y)$) have zero modes which can acquire vevs.  Thus we witness an
{\em explicit} breaking
$$
\begin{array}{c} G  \\ {\rm SU(2)} \end{array} \xrightarrow{P}  
\begin{array}{c} H \\ {\rm U(1)} \end{array} \, .
$$
We can therefore write
$$
\left<A_5^a\right> = \left(\left<A_5^{1(0)}\right>, 
\left<A_5^{2(0)}\right>, 0\right) \, . 
$$
Using the unbroken U(1) symmetry, we can assign the entire vev in one
component and hence without any loss of generality we can write
$\left<A_5^a\right>=(B,0,0)$ where $B$ is the vev.

The gauge boson masses originate from $F_{\mu5}^a F_a^{\mu5} = \left(\p_\mu
  A_5^a-\p_5 A_\mu^a + g\epsilon_{abc}A_\mu^b A_5^c\right)^2$.
The relevant term of the Lagrangian leading to the mass matrix is
$A_\mu^a(D_5D_5)_{ab}A_\mu^b$, where 
$$
(D_5D_5)_{ab} = \left(\begin{array}{ccc} \p_5\p_5 & 0 & 0\\
0 & \p_5\p_5-g^2B^2 & -2gB\p_5\\
0& 2gB\p_5 &\p_5\p_5-g^2B^2\end{array}\right) \, ,  
$$
with $a,b=1,2,3$ as adjoint representation indices.  There is no KK-number
mixing and this mass matrix holds for each $n$.  The derivatives in the above
matrix would act on the KK states.  For $n\ne0$, $A_\mu^{3(n)} \sim
{1\over\sqrt{\pi R}} \cos{ny\over R}$, $A_\mu^{1,2(n)} \sim {1\over\sqrt{\pi
    R}} \sin {ny\over R}$, which is a consequence of our choice of $P={\rm
  diag} (1,-1)$. Each derivative then picks up a factor $n/R$.  The KK gauge
boson mass-squared matrix turns out to be (for $n\ne 0$)
\begin{align}
\left(\begin{array}{ccc} {n^2\over R^2} & 0 & 0\\
0 & {n^2\over R^2}+{\a^2\over R^2} & {2\a n\over R^2}\\
0 & {2\a n\over R^2} & {n^2\over R^2}+{\a^2\over
R^2}\end{array}\right) \, , 
\end{align}
where $\a\equiv gBR$.  The eigenvalues are ${n^2\over R^2}$, ${(n+\a)^2\over
  R^2}$, ${(n-\a)^2\over R^2}$.  We have thus seen a two-stage symmetry
breaking: ($i$) SU(2) breaks to U(1) explicitly by the action of $P$, as a
result only $A_\mu^3$ has zero mode, and then ($ii$) U(1) breaks to {\em
  nothing} by the vev $B$, when $A_\mu^{3(0)}$ picks up a mass ${\a\over R}$.

Why the example of SU(2) is better than U(1)? In the U(1) example, the scalar
turned out to be unphysical. From SU(2) we got a physical scalar, which can
acquire a nonzero vev. However, we want a scalar which is a doublet under
SU(2), and the scalar we got in the above example is not a doublet of
SU(2). To achieve this, we move to SU(3).

\subsection{5d SU(3) as a toy model}
Now consider that the 5d gauge group is SU(3), which is compactified on
$S^1/Z_2$. The Lie algebra valued gauge fields are $A_M=A_M^a {\l^a\over
  2}$. Here, $\l^a$ are Gell-Mann matrices, given by
\begin{align*}
\l^1&=\left(\begin{array}{ccc}0&1&0\\1&0&0\\0&0&0\end{array}\right),~
 \l^2=\left(\begin{array}{ccc}0&-i&0\\i&0&0\\0&0&0\end{array}\right),~
 \l^3=\left(\begin{array}{ccc}1&0&0\\0&-1&0\\0&0&0\end{array}\right),~
 \l^4=\left(\begin{array}{ccc}0 &0&1\\0&0&0\\1&0&0\end{array}\right)\\
\l^5&=\left(\begin{array}{ccc}0&0&-i\\0&0&0\\i&0&0\end{array}\right),~
 \l^6=\left(\begin{array}{ccc}0&0&0\\0&0&1\\0&1&0\end{array}\right),~
 \l^7=\left(\begin{array}{ccc}0&0&0\\0&0&-i\\0&i&0\end{array}\right),~
 \l^8={1\over\sqrt{3}}\left(\begin{array}{ccc}1&0&0\\0&1&0\\0&0&-2
\end{array}\right) \, . 
\end{align*}
We impose $Z_2$-projection by requiring
\begin{eqnarray}
PA_\mu P^\dagger = A_\mu \quad \text{and}\quad PA_5P^\dagger=-A_5 \, ,
\quad\text{where}\quad P = \left(\begin{array}{ccc}-1&0&0\\
0&-1&0\\ 0&0&1\end{array}\right) = e^{i\pi\l_3} \, . 
\end{eqnarray}
The explicit transformations of the gauge boson fields are 
\begin{align}
\label{su3trans}
\left(\begin{array}{ccc}
A_\mu^3+{1\over\sqrt{3}}A_\mu^8 & A^1_\mu-iA_\mu^2 & A_\mu^4-iA_\mu^5\\
A_\mu^1+iA_\mu^2 &-A^3_\mu+{1\over\sqrt{3}}A_\mu^8 & A_\mu^6-iA_\mu^7\\
A_\mu^4+iA_\mu^5 &A^6_\mu+iA_\mu^7 &
-{2\over\sqrt{3}}A_\mu^8\end{array}\right) \xrightarrow{P}
\left(\begin{array}{ccc} \oplus & \oplus & \ominus\\
\oplus & \oplus & \ominus\\
\ominus & \ominus & \oplus\end{array}\right) \, , 
\end{align}
where $\oplus$ and $\ominus$ represent the relative signs upon transformation
under the given projection.  {\em For the $A_5$ scalars, $\oplus$ and
  $\ominus$ should be replaced by $\ominus$ and $\oplus$, respectively}. The
fields which are projected with $\oplus$ sign contain zero modes, but those
with the $\ominus$ sign do not have zero modes. 

As a consequence of the above projection, 
$$
\begin{array}{c} G\\ {\rm SU(3)}  \end{array}
\stackrel{P}{\longrightarrow}
\begin{array}{c} H \\ {\rm SU(2)} \times {\rm U(1)} 
\end{array} \, \, .
$$
Now, the ${\bf 8}$ generators of SU(3) are decomposed as {\bf 3 + 2 + 2 +1}
under the unbroken SU(2).  From Eq.~(\ref{su3trans}), it is clear that only
the triplet {\bf 3} $\left(A_\mu^1, A_\mu^2, A_\mu^3\right)$ and the singlet
{\bf 1} $\left(A_\mu^8\right)$ gauge bosons have zero modes. Also, the
components of the doublet {\bf 2} scalar $\left(A_5^4 - iA_5^5, ~A_5^6 -
  iA_5^7\right)^T$ have zero modes. We identify the {\em zero mode doublet
  scalar} with our Higgs doublet, which is expressed as
\begin{equation}
H_5^{(0)} = \left(\begin{array}{c} A_5^{4(0)}-iA_5^{5(0)}\\
A_5^{6(0)}-iA_5^{7(0)}\end{array}\right) \, . 
\end{equation} 
In other words, when $G\xrightarrow{P} H$, the generators of the {\em massless
  gauge bosons} belong to $H$, while those of the {\em massless scalars}
belong to the coset $G/H$.

We now turn our attention to the gauge transformations in bulk:
\begin{align*}
A_\mu &\rightarrow A_\mu + \p_\mu \T (x,y) + i[\T(x,y), A_\mu] \, , \\
A_5 &\rightarrow A_5 + \p_5 \T (x,y) + i[\T(x,y), A_5] \, . 
\end{align*}
For the scalars $A_5$, which correspond to the broken generators,
$\T(x,0)=\T(x,\pi R)=0$, but still $A_5\to A_5 + \p_5\T$. Because of this
shift symmetry, there cannot be any tree level potential for $A_5$.  Just like
gauge invariance forbids $A_\mu A_\mu$ term in the ordinary 4d QED Lagrangian,
the higher dimensional gauge invariance forbids $A_5 A_5$ term in the 5d
Lagrangian as well. But this is true only at tree level, as quantum
corrections generate the potential.

The quadratic $(A_5)^2$ and the quartic $(A_5)^4$ terms are generated at
one-loop level via two- and four-point diagrams with $A_5$ in external lines
and with KK fermions and bosons in internal lines. Such loops generate the
effective potential whose minimization yields the vev of $A_5$.  The gauge
loops tend to push $\langle A_5^0 \rangle$ to zero while minimizing the
potential, while the fermionic loops tends to shift $\langle A_5^0 \rangle$
away from zero in the minimum of the potential. In fact, the KK fermions are
instrumental for generating the correct vev. This way of breaking SU(2)
$\times $ U(1) symmetry to U(1)$_{\rm em}$ is called the {\em Hosotani
  mechanism} \cite{Hosotani:1983xw}. The one-loop generated Higgs mass is
given by
\begin{eqnarray} 
m_h^2 \simeq \frac{g^4}{128\pi^6} \frac{1}{R^2} \sum_{\rm KK} V''(\alpha) \, ,
\end{eqnarray}
where $\alpha$ is a dimensionless parameter arising from bulk interactions,
which corresponds to the minimum of the potential where the double-derivative
is calculated.  The summation is over all KK particles. Clearly, 5d gauge
symmetry is recovered in the limit $1/R \to 0$.

In fact, this $A_5$ is a symbolic representation of $H_5^{(0)}$.  A vev in
$H_5^{(0)}$ induces SSB of H $=$ SU(2)$\times$ U(1) to E $={\rm U(1)}_Q$.  The
composition of photon in this scenario is 
$\g_\mu \propto (A_\mu^3+{1\over\sqrt{3}}A_\mu^8)$.
Recalling that the composition of photon in the SM as given 
in Eq.~(\ref{photonZ}) is,
$$
\g_\mu = \sin \t_W W_\mu^3 + \cos \t_W B_\mu \, , 
$$
we obtain the following relations for the GHU scenario under consideration:
$$
\cot \t_W = {1\over\sqrt{3}} = \cot {\pi\over 3} \Rightarrow \t_W
= \pi/3 \Rightarrow \sin^2\t_W = {3\over4} \quad{\rm and}\quad
{M_W^2\over M_Z^2} = \cos^2\t_W={1\over4} ~~\therefore 
M_Z = 2M_W. 
$$
This is clearly experimentally ruled out! But this scenario provides the basic
intuitive picture of how a GHU scenario works through a simple illustration.
In this scenario,
$$
M_W^{(n)} = {n+\a\over R}~,~M_Z^{(n)} = {n+2\a\over
R}~,~m_\g^{(n)}={n\over R} \, . 
$$
The periodicity property demands that the spectrum will remain invariant under
$\alpha \to \alpha+1$. This restricts $\alpha$ in the range
$[0,1]$. Orbifolding further reduces it to $\alpha = [0,{1\over 2}]$. In
principle, $\alpha$ can be fixed from the $W$ mass.

\subsection{Realistic gauge-Higgs unification scenarios - 
a brief description}
There are quite a few obstacles that one faces in constructing a realistic
scenario. Since the Yukawa coupling arises from higher dimensional gauge
coupling, it turns out to be too small to produce the correct top quark
mass. Besides, one has also to worry about generating hierarchical Yukawa
interaction starting from higher dimensional gauge interaction which is after
all universal. The scalar potential is generated at one loop, which tends to
yield rather low Higgs boson mass. The compactification scale ($R^{-1}$)
required for this purpose turns out to be smaller than its experimental lower
limit. We briefly describe below some of the attempts made in removing these
obstacles.

\noindent $(i)$~~ It has been argued in \cite{Scrucca:2003ra} in the context
of a 5d $S^1/Z_2$ scenario that large brane localized kinetic term can help
jack up the Higgs mass to an acceptable range.  Another option is to break the
5d Lorentz symmetry in the bulk \cite{Panico:2005dh,Panico:2006fv}. The key
observation is that the stability of the loop generated scalar potential
relies essentially on 5d gauge symmetry and not so much on the SO(1,4) Lorentz
symmetry. If one breaks either explicitly or by some dynamics this Lorentz
symmetry keeping the SO(1,3) Lorentz symmetry in the ordinary space-time
dimension intact, then one can enhance the Higgs coupling to fermions. Such
breaking can be parametrized by the following pieces of the Lagrangian:
  \begin{eqnarray} 
{\Lag}_{gauge} &=& -\frac{1}{4} F_{\mu\nu} F^{\mu\nu}
    -\frac{\mathbf {a}}{4} F_{\mu 5} F^{\mu 5}~~;~~ \Lag_{Yuk} = \bar{\Psi}
    \left(i \gamma_\mu D^\mu - {\mathbf{k}} D_5 \gamma^5 \right)\Psi \, ,
\end{eqnarray} 
where the prefactors ${\mathbf {a}}$ and ${\mathbf{k}}$ need to be
phenomenologically tuned to match the data.

\noindent $(ii)$~~ If one goes to an even higher dimensional model, e.g. a 6d
GHU scenario, the gauge kinetic term contains a quartic interaction for the
internal components of the gauge fields, i.e. it yields a quartic term in the
Higgs potential at the tree level. Its strength of course depends on the gauge
coupling. The appearance of this tree level quartic coupling can solve the
`low Higgs mass' problem. But, in these scenarios, gauge symmetry allows some
orbifold localized operator which gives Higgs mass terms at the tree level,
and this brings back the quadratic cutoff sensitivity as encountered in the
SM. The question is, therefore, as how to tame this quadratic cutoff
sensitivity.  This was pursued in \cite{Scrucca:2003ut} with SU(3) gauge group
on $T^2/Z_N$ orbifold (with $N=2,3,4,6$). It was shown that only for $N=2$,
under the assumption of successful electroweak symmetry breaking, a condition
$m_h = 2 M_W$ has to be satisfied to keep the scalar potential free from
quadratic divergence.

\noindent $(iii)$~~ If one goes to the warped scenario \cite{Randall:1999ee},
additional features emerge \cite{Contino:2003ve}. AdS/CFT correspondence
\cite{Maldacena:1997re} tells us that a weakly coupled theory in 5d AdS is
equivalent to a strongly coupled 4d theory. In this case, the Higgs is a
composite particle, a pseudo-Goldstone boson of the strongly coupled CFT
sector. There is a global symmetry in the CFT sector that protects the Higgs
mass. Gauge and Yukawa interactions are introduced in the dual 5d AdS theory,
which explicitly break the global symmetry but do not induce quadratic
divergence to the Higgs mass at any loop. The Higgs mass can be large enough
thanks to the quartic interaction which can be generated dynamically at tree
level.  The quadratic term is, as expected, loop generated.  The all order
finiteness of the Higgs mass can be intuitively understood as follows.  The
Higgs is at the TeV brane and a scalar which breaks the gauge symmetry is at
the Planck brane and the information of this breaking reaches from Planck to
TeV brane by bulk propagators. This is a non-local effect which is the reason
behind the finiteness of the Higgs mass.  This type of model was further
consolidated in \cite{Agashe:2004rs} by considering a ${\rm SO(5)} \times {\rm
  U(1)}_{\rm B-L}$ symmetry in the bulk, which eventually gives SO(3)
custodial symmetry that prohibits large correction to the oblique $T$
parameter. The electroweak symmetry is dynamically broken by the top quark.

One distinct advantage of working in the warped space over the flat space is
noteworthy. Recall that in the GHU context the Yukawa coupling of the Higgs 
arises from higher dimensional gauge coupling.  In the context of Hosotani
mechanism \cite{Hosotani:1983xw} in flat space without any large brane kinetic
term, the 5d gauge coupling $g_5 \sqrt{R^{-1}} = g_4 \equiv g \sim 0.65$ is
rather small to yield the Higgs quartic coupling. On the other hand, in the
warped case the AdS dynamics gives a rather large 5d gauge coupling $g_5
\sqrt{k} \geq 4$ \cite{Agashe:2004rs}, which is why the Higgs quartic coupling
can be sufficiently large to yield the correct Higgs mass.

There are other GHU constructions in flat and warped space with different
features, which we are not going to cover here. We refer the readers to the
articles in Refs.~\cite{ghu} and to two excellent reviews on composite Higgs
scenarios \cite{Contino:2010rs,Serone:2009kf}.

\subsection{Comparison between gauge-Higgs/composite scenario 
and little Higgs models}
Conceptually, Gauge-Higgs models and little Higgs models are related
\cite{Contino:2003ve,Agashe:2004rs}. More precisely, through the AdS/CFT
correspondence gauge-Higgs unification in a 5d warped scenario
(Randall-Sundrum model) replicates a little Higgs model in 4d.  In the
conventional ({\em i.e. the way we developed the idea in this review}) little
Higgs models the sensitivity of the Higgs mass to the UV cutoff is logarithmic
at one-loop and quadratic at two-loop. In the composite picture, which is dual
to 5d gauge theory where the 5th component of the gauge boson makes the Higgs
boson, the Higgs mass is finite at all order. The little Higgs models are
calculable below the cutoff scale ($\sim 10$ TeV), while the QCD-like
composite models are calculable in the large $N$ limit allowing
$1/N$-expansion. There is another difference between the composite models and
the little Higgs models. The global symmetry that protects the Higgs mass in a
composite model is a symmetry of the strong CFT sector and not of the
SM. Hence the new TeV scale resonances form a complete multiplet of the global
group of the strong sector, unlike in the little Higgs models where the new
states are the partners of the SM particles. Another distinguishing feature is
the presence of a KK gluon in the extra dimensional models that is absent in
the conventional little Higgs constructions.

\subsection{Collider signals of gauge-Higgs unification models} 
Are there {\em smoking gun} signals of the GHU models?  These models generally
contain fermions with exotic electric charge, e.g. $(5/3)$. But the exact
value of the charge is a model-dependent question.  In most cases, the
lightest nonstandard particle turns out to be a colored fermion and not any
exotic (KK) gauge boson. This has got something to do with the fact that large
contribution from the exotic fermions are crucial in triggering correct amount
of electroweak symmetry breaking. Also, the gauge boson coupling to the
right-handed top quark in such scenarios is about (10-15)\% different from its
SM value. In a study \cite{Maru:2007xn}, the indirect effects of the KK
particles on the Higgs production via gluon fusion and Higgs decay to two
photons were analyzed in the context of a toy 5d scenario with SU(3) gauge
group on an $S^1/Z_2$ orbifold. If the KK states weigh around 1 TeV, the loop
effects provide about 10\% deviation from the SM results. Moreover, the
overall sign of the gluon-gluon-Higgs coupling was claimed to be opposite to
the one in the SM or the UED model, but consistent with the corresponding sign
in the little Higgs or the supersymmetric models. In a warped scenario with
${\rm SO(5)} \times {\rm U(1)}_X$ gauge symmetry in the bulk (chosen for
preserving custodial symmetry), the authors of \cite{Carena:2007tn} have
studied the LHC detection of a KK top quark which is strongly coupled to a KK
gluon.  In the composite Higgs context, one of the crucial tests is to measure
the scattering of the longitudinal gauge bosons ($V_L V_L \to V_L V_L$) and
find an excess event (see Ref.~\cite{Contino:2010rs} for a pedagogical
illustration).

\section{Higgsless scenarios}
The idea is to trigger electroweak symmetry breaking without actually having a
physical Higgs. The mechanism relies on imposing different boundary conditions
(BC's) on gauge fields in an extra-dimensional set-up.  The BC's can be
carefully chosen such that the rank of a gauge group can be lowered.  For the
purpose of illustration outlined in this review, we heavily rely on the
discussions given in \cite{Csaki:2005vy,Csaki:2003dt}.  
To start with, we consider a 5d gauge theory. The extra dimension is
compactified on a circle of radius $R$ with a $y \leftrightarrow (-y)$
identification, i.e. on a $S^1/Z_2$ orbifold. The fixed points are $y=0, \pi
R$. We can use different BC's at the two fixed points.

\subsection{Types of Boundary Conditions} 

Let us consider a 5d scalar field $\phi(x,y)$ in the interval $[0,\pi R]$. The
minimization of action requires {\em either} or {\em both} of the following:
\begin{itemize} 
\item $\phi|_{y=0,\pi R} =$ Constant. When the Constant $=0$, it is called the
  {\em Dirichlet BC}.
\item $\left(\p_5 \phi + V \phi\right)|_{y=0,\pi R} = 0$, where $V$ is some
  boundary mass parameter. When $V=0$, $\p_5 \phi = 0$, which is called {\em
    Neumann BC}. When $V\ne 0$, it corresponds to a mixed BC.
\end{itemize} 
Although we took a scalar field for demonstration, the BC's can be applied to
any other field as well.  We now perform some warm-up exercises to appreciate
the essential features of Higgsless scenarios.

\subsection{Breaking SU(2) $\to$ U(1)  by BC's}
This is a simple example to demonstrate that by appropriate choices of BC's we
can indeed get a massless gauge boson state (to be identified with the photon)
and massive states (to be identified with the $W$ and $Z$ boson). Consider a
SU(2) gauge symmetry in 5d. The gauge bosons are $A_M^a(x,y)$, where $a=1,2,3$
and $M = \mu,5$. Now we apply the BC's at the two fixed points. 
\begin{itemize} 
\item $\p_5 A_\mu^a|_{y=0} = 0$ for $a=1,2,3$, i.e. at the $y=0$ fixed point,
  we apply Neumann BC for all the three gauge bosons.
\item $A_\mu^{1,2}|_{y=\pi R} = 0$, $\p_5 A_\mu^3|_{y=\pi R} = 0$, i.e.  at
  the $y=\pi R$ fixed point, we apply Dirichlet BC for the first two
  components of the gauge bosons and Neumann BC for the third component.
\end{itemize} 
The $y$-dependent parts of the various KK mode gauge fields are then 
\begin{eqnarray} 
A_\mu^3 (y) &\Longrightarrow & \cos\left(\frac{ny}{R}\right)
~~(n=0,1,2, \cdot\cdot\cdot) \, , \nonumber \\
A_\mu^{1,2} (y) & \Longrightarrow & 
\cos\left(\frac{(2m+1)y}{2R}\right)
~~(m=0,1,2, \cdot\cdot\cdot) \, .
\end{eqnarray}
Their mass spectra are therefore given by 
\begin{eqnarray}
A_\mu^3  &\Longrightarrow & M_n = 0, \frac{1}{R}, \frac{2}{R}, \cdot\cdot\cdot
\, , \nonumber \\
A_\mu^{1,2}  & \Longrightarrow & M_m = \frac{1}{2R}, \frac{3}{2R},
\frac{5}{2R}, \cdot\cdot\cdot \, . 
\end{eqnarray}
Thus we identify 
\begin{equation} 
M_\g = 0, ~~M_W = \frac{1}{2R}, ~~ M_Z = \frac{1}{R} \, . 
\end{equation} 
Clearly, this is not a phenomenologically acceptable situation as $M_Z = 2
M_W$. The main problem here is that the gauge boson masses are independent
of the gauge couplings. Somehow, we have to bring that dependence in.

\subsection{Breaking SU(2) $\to$ `Nothing' by BC's}
Let us impose the following BC's. 
\begin{itemize} 
\item $\p_5 A_\mu^a|_{y=0} = 0$ for $a=1,2,3$. This is just like the previous
  example. 
\item $\p_5 A_\mu^a|_{y=\pi R} = V A_\mu^a|_{y=\pi R}$. This is a mixed
  BC. The $V \to 0$ limit corresponds to the Neumann BC and the $V \to \infty$
  limit corresponds to the Dirichlet BC. Notice that in the previous example,
  we took $V \to \infty$ limit for $a=1,2$ and $V \to 0$ limit for $a=3$.
\end{itemize} 
A general solution that satisfies the above BC's is 
\begin{equation}
\label{gensol}
A_\mu^a (x,y) = \sum_{n=1}^{\infty} A_\mu^{a(n)}(x) f_n(y), ~~{\rm with}~~
f_n(y) = \a_n \frac{\cos(M_n y)}{\sin(M_n \pi R)} \, .
\end{equation}
Since $f_n(y)$ is a cosine expansion, the BC at $y=0$ is trivially
satisfied. The BC at $y=\pi R$ leads to 
\begin{equation}
  \sum_{n=1}^{\infty} A_\mu^{(n)}(x) 
  \frac{(-)\a_n M_n \sin(M_n y)}{\sin(M_n \pi R)}\Big|_{y=\pi R}
  = V \sum_{n=1}^{\infty} A_\mu^{(n)}(x) 
      \frac{\a_n \cos(M_n y)}{\sin(M_n \pi R)}\Big|_{y=\pi R} \, , 
\end{equation}
which leads to the following transcendental equation from where the mass
spectrum is obtained: 
\begin{equation} 
\label{trans}
M_n \tan\left(M_n \pi R\right) = - V \, . 
\end{equation}
Now observe the following: 
\begin{enumerate} 
\item[($i$)] When $V=0$, which corresponds to Neumann BC for all $a=1,2,3$,
  gauge symmetry is unbroken.
\item[($ii$)] When $V\ne 0$, the SU(2) gauge symmetry is fully broken. The
  amount of breaking is controlled by $V$. The mass spectrum is given by the
  solution of the above transcendental equation.
\end{enumerate}
The normalization factor $\a_n$ is determined by requiring that the KK modes
are canonically normalized, i.e. 
\begin{equation} 
\int_0^{\pi R} dy f_n^2 (y)  = 1  \, .
\end{equation}
\begin{eqnarray}
\text{Therefore, using~ Eq.~(\ref{gensol}),}~~ \a_n = \frac{\sqrt{2}}
{\sqrt{\pi R ~{\rm cosec}^2 (M_n \pi R) + \frac{\cot(M_n \pi R)}{M_n}}} \, . 
\end{eqnarray}
Now using the transcendental equation (\ref{trans}), one can express 
\begin{equation} 
\a_n = \frac{\sqrt{2}}{\sqrt{\pi R \left(1+\frac{M_n^2}{V^2} \right)
- \frac{1}{V}}} \, .
\end{equation} 
Now we are all set to calculate the mass spectrum. Let us consider the
following two cases:
\begin{enumerate} 
\item[($i$)] $V=0$:~~ No breaking of gauge symmetry. All $A_\mu^a$ ($a=1,2,3$)
  have a cosine expansion.
\item[($ii$)] $V\ne 0$:~~ We assume $V \gg \frac{1}{R}$. Then the
  transcendental equation (\ref{trans}) implies that to the zeroth
  approximation $\cot(M_n \pi R) = 0$. This means $M_n \pi R =
  (2n+1)\frac{\pi}{2}$, i.e.
\begin{equation} 
M_n = \frac{2n+1}{2R}   ~~(n=0,1,2, \cdot\cdot\cdot) \, .
\end{equation}
Then to the next level of approximation, we take 
$M_n \pi R = (2n+1) \frac{\pi}{2} + \e$, where $\e$ is a small number. Then 
$
\cot(M_n \pi R) = \cot\left\{(2n+1)\frac{\pi}{2} + \e \right\}
= \cot(2n+1)\frac{\pi}{2} + 
\e \left\{-~{\rm cosec}^2 (2n+1)\frac{\pi}{2}\right\} = -\e \, .$ 
Putting back the above relation into the transcendental equation, we obtain
$\e = \frac{M_n}{V}$.  Therefore, $M_n \pi R = (2n+1) \frac{\pi}{2} + 
\frac{M_n}{V}$, i.e. 
\begin{equation}
M_n \simeq \frac{2n+1}{2R} \left(1+ \frac{1}{\pi RV} + 
\cdot\cdot\cdot\right)  ~~~~(n=0,1,2, \cdot\cdot\cdot) \, . 
\end{equation}
Clearly, there is no zero mode. SU(2) gauge symmetry is thus completely broken.
\end{enumerate}

\subsection{A model of EWSB by BC's: Higgsless scenario in flat space}
Right at the beginning, we set two goals: 
\begin{enumerate} 
\item[($i$)] The gauge boson masses have to be related to the gauge couplings.

\item[($ii$)] There should be a custodial symmetry in the bulk so as to be
  consistent with EWPT.
\end{enumerate} 

We therefore start with the gauge symmetry ${\rm SU(2)}_L \times {\rm SU(2)}_R
\times {\rm U(1)}_{B-L}$ in the bulk. The notations of gauge bosons and gauge
couplings are as follows ({\em Dimension of a 5d gauge coupling is
  $M^{-1/2}$}):
\begin{itemize} 
\item Group: ${\rm SU(2)}_L$, ~ gauge coupling: $g$, ~ gauge bosons:
  $A_M^{La}$ where $a=1,2,3$.  
\item Group: ${\rm SU(2)}_R$, ~ gauge coupling: $g$, ~ gauge bosons:
  $A_M^{Ra}$ where $a=1,2,3$.  
\item Group: ${\rm U(1)}_{B-L}$, ~ gauge coupling: $g'$, ~ gauge bosons:
  $B_M$. 
\end{itemize}
We denote the gauge bosons of the ${\rm SU(2)}_D$ group, which is the diagonal
subgroup of ${\rm SU(2)}_L \times {\rm SU(2)}_R$, as $A_M^{+a}$, where
$A_M^{\pm a} = \frac{1}{\sqrt{2}} \left(A_M^{La} \pm A_M^{Ra} \right)$. We
should remember that in order to have a zero mode of a generic gauge boson
$A_\mu$, i.e. to preserve the gauge symmetry, one should use Neumann BC: $\p_5
A_\mu = 0$. Although we display below the BC's of the gauge fields $A_\mu$ and
$B_\mu$ only, the conditions for $A_5$ and $B_5$ are not hard to obtain. We
just have to remember that the conditions have to be swapped between the $\mu$
and $y$ components. In other words, the Dirichlet BC's for gauge bosons mean
Neumann BC's for the corresponding scalars and vice versa. We now apply the
following BC's at the two fixed points:
\begin{itemize} 
\item \underline{$y=0$ fixed point}: 
 \begin{enumerate} 
 \item[($i$)] $\p_5 A_\mu^{+a} = 0$ and $\p_5 B_\mu = 0$ (i.e. ${\rm SU(2)}_L
   \times {\rm SU(2)}_R$ broken down to ${\rm SU(2)}_D$, also ${\rm
     U(1)}_{B-L}$ unbroken).
 \item[($ii$)] $A_\mu^{-a} = 0$ (i.e. the SU(2) orthogonal to ${\rm SU(2)}_D$ is
   broken).
\end{enumerate}

\item \underline{$y=\pi R$ fixed point}: 
 \begin{enumerate} 
 \item[($i$)] $\p_5 A_\mu^{La} = 0$ (i.e. ${\rm SU(2)}_L$ unbroken).
 \item[($ii$)] $\p_5 A_\mu^{R~1,2} = V A_\mu^{R~1,2}$, where $V=-\frac{1}{4}
   g^2v_R^2$. At the $y=\pi R$ brane we localize a scalar doublet under ${\rm
     SU(2)}_R$, which acquires a vev $v_R$ leading to ${\rm SU(2)}_R \times
   {\rm U(1)}_{B-L}$ breaking down to ${\rm U(1)}_Y$. Eventually, we take the
   $v_R \to \infty$ limit and the scalar will decouple without spoiling
   unitarity.
 \item[(iii)] $\p_5 A_\mu^{R3} = \frac{V}{g} \left(g A_\mu^{R3} - g'
     B_\mu\right)$.
 \item[(iv)] $\p_5 B_\mu = -\frac{Vg'}{g^2} \left(g A_\mu^{R3} - g'
     B_\mu\right)$.
 \end{enumerate}
 The last three BC's ensure that both ${\rm SU(2)}_R$ and ${\rm U(1)}_{B-L}$
 are broken when $V \ne 0$. Note additionally that $\p_5 \left(g'A_\mu^{R3} +
   gB_\mu\right) = 0$.  Finally, the only symmetry left unbroken is ${\rm
   U(1)}_Q$.
\end{itemize}

The BC's originate from the following consideration: The orbifold
  projection around $y=0$ fixed point has a ${\rm SU(2)}_L \rightleftharpoons
{\rm SU(2)}_R$ outer automorphism, while around $y=\pi R$ fixed point the
orbifold projections are ${\rm SU(2)}_L \leftrightarrow {\rm SU(2)}_L$ and 
${\rm SU(2)}_R \leftrightarrow {\rm SU(2)}_R$.  Define $\hat{y} = y + \pi
R$. The BC's can be derived from 
\begin{eqnarray} 
A_\mu^{La} (x,-y) &=& A_\mu^{Ra} (x,y) \, ; ~~  
B_\mu (x,-y) = B_\mu (x,y) \, ; \nonumber \\
A_\mu^{La} (x,-\hat{y}) &=& A_\mu^{La} (x,\hat{y}) \, ; ~~  
A_\mu^{Ra} (x,-\hat{y}) = A_\mu^{Ra} (x,\hat{y}) \, ; ~~  
B_\mu (x,-\hat{y}) = B_\mu (x,\hat{y}) \, .
\end{eqnarray}

Once the BC's are enforced, a given 4d gauge field is shared among many 5d
fields. We now take the $V \to \infty$ limit. Then the 5d gauge fields in the
($A_\mu^L, A_\mu^R, B_\mu$) basis can be expressed in terms of the 4d fields,
namely $\g_\mu$ (photon), $Z_\mu^{(n)}$ and $W_\mu^{\pm(n)}$, in the following
way $\left(A_\mu^{L,R\pm} = \left(A_\mu^{L,R1} \mp
    iA_\mu^{L,R2}\right)/\sqrt{2}\right)$:
\begin{eqnarray} 
\label{expansion_fields}
B_\mu (x,y) &=& \frac{1}{\sqrt{\pi R (g^2 + 2g'^2)}} 
\left[g\g_\mu(x) + \sqrt{2} g' \sum_{n=1}^\infty Z_\mu^{(n)}(x)
\cos\left(M_Z^{(n)}y\right)  \right] \, , \nonumber \\
A_\mu^{L3} (x,y) &=& \frac{1}{\sqrt{\pi R (g^2 + 2g'^2)}} 
\left[g'\g_\mu(x) - \sqrt{2} g \sum_{n=1}^\infty Z_\mu^{(n)}(x)
\frac{\cos\left(M_Z^{(n)}(y-\pi R)\right)}{2\cos\left(M_Z^{(n)}\pi R\right)}  
\right] \, , \nonumber \\
A_\mu^{R3} (x,y) &=& \frac{1}{\sqrt{\pi R (g^2 + 2g'^2)}} 
\left[g'\g_\mu(x) - \sqrt{2} g \sum_{n=1}^\infty Z_\mu^{(n)}(x)
\frac{\cos\left(M_Z^{(n)}(y+\pi R)\right)}{2\cos\left(M_Z^{(n)}\pi R\right)}  
\right] \, , \\
A_\mu^{L\pm} (x,y) &=& \frac{1}{\sqrt{\pi R}} \sum_{n=1}^\infty W_\mu^{n\pm}(x)
\cos\left(M_W^{(n)} (y-\pi R) \right) \, , ~~
A_\mu^{R\pm} (x,y) = \frac{1}{\sqrt{\pi R}} \sum_{n=1}^\infty W_\mu^{n\pm}(x)
\cos\left(M_W^{(n)} (y+\pi R) \right) \, . \nonumber 
\end{eqnarray} 
Thus, we obtain the massless photon $\g$, corresponding to the unbroken ${\rm
  U(1)}_Q$, and some KK towers of massive $W^{(n)\pm}$ and $Z^{(n)}$ gauge
bosons. The $Z^{(1)}$ and $W^{(1)\pm}$ are to be identified with the observed
$Z$ and $W^\pm$ bosons, respectively.

\subsubsection{The charged $\mathbf{W^{(n)\pm}}$ tower}
The solutions would be similar to the one as obtained from the transcendental
equation for the SU(2) $\to$ `nothing' case.
\begin{equation}
M_W^{(n)} \tan\left(2M_W^{(n)}\pi R \right) = - V = \frac{1}{4} g^2 v_R^2 \, ,
\end{equation}
which leads to the solution
\begin{equation}
\label{soln_w}
  M_W^{(n)} = \left(\frac{2n-1}{4R}\right) 
\left(1-\frac{2}{\pi R g^2 v_R^2} + \cdot\cdot\cdot\right) 
~~~ {\rm for}~~ n=1,2,\cdot\cdot\cdot 
\end{equation}

\subsubsection{The neutral $\mathbf{Z^{(n)}}$ tower}
We enforce the BC at $y=\pi R$: 
\begin{eqnarray}
\label{zn_1}
\p_5 A_\mu^{R3} = \frac{V}{g} \left(g A_\mu^{R3} - g' B_\mu\right) \, . 
\end{eqnarray}
The left-hand side (LHS) of Eq.~(\ref{zn_1}) is  
$$\p_5 A_\mu^{R3}\Big|_{y=\pi R} = \frac{\sqrt{2}g}{\sqrt{\pi R(g^2+2g'^2)}}
\sum_{n=1}^\infty M_Z^{(n)} \sin\left(M_Z^{(n)}\pi R\right) Z_\mu^{(n)}(x) \, .$$
The right-hand side (RHS) of  Eq.~(\ref{zn_1}) can be written as 
$$
\frac{\sqrt{2}g}{\sqrt{\pi R(g^2+2g'^2)}} \frac{v_R^2}{4}\sum_{n=1}^\infty
  \left[g'^2 \cos(M_Z^{(n)}\pi R) + g^2
   \frac{\cos\left(2M_Z^{(n)}\pi R\right)}{2\cos\left(M_Z^{(n)}\pi R\right)}
      \right] Z_\mu^{(n)}(x) \, . 
$$

Therefore,
$$
M_Z^{(n)} \sin\left(M_Z^{(n)}\pi R\right) = \frac{v_R^2}{4} 
\left[g'^2 \cos\left(M_Z^{(n)}\pi R\right) + g^2
    \frac{\cos\left(2M_Z^{(n)}\pi R\right)}{2\cos\left(M_Z^{(n)}\pi R\right)}
     \right] \, ,  
$$
which leads to the simplified form of the eigenvalue equation as
\begin{equation} 
\label{eqn_z}
M_Z^{(n)} \tan\left(M_Z^{(n)}\pi R\right) = \frac{v_R^2}{8} \left(g^2+2g'^2\right)
- \frac{v_R^2g^2}{8} \tan^2\left(M_Z^{(n)}\pi R\right) \, . 
\end{equation}

\subsubsection{Solution of Eq.~(\ref{eqn_z})}
We rewrite the equation as  
\begin{eqnarray*}
M_Z^{(n)} \pi R \tan\left(M_Z^{(n)}\pi R\right) = \frac{\pi R g^2 v_R^2}{8}
\left[\tan^2\left(M_0\pi R\right)- 
\tan^2\left(M_Z^{(n)}\pi R\right) \right] , 
{\rm where}~ 
\tan^2\left(M_0\pi R\right) = \left(1+ \frac{2g'^2}{g^2} \right)\, .
\end{eqnarray*}
Now take the limit $v_R \to \infty$. Then, $\left[\tan^2\left(M_0\pi R\right)-
  \tan^2\left(M_Z^{(n)}\pi R\right)\right] = 0$ is our zeroth approximation,
so that the LHS of Eq.~(\ref{eqn_z}) is finite. The solution is
$$
\tan\left(M_Z^{(n)}\pi R\right) = \pm \tan\left(M_0\pi R\right) \, .  
$$
Let us first take the $(+)$ sign solution and proceed. Then 
$$
\tan\left(M_Z^{(n)}\pi R\right) = + \tan\left(M_0\pi R\right)
= \tan\left(M_0\pi R +(n-1)\pi\right) \, , ~~~(n=1,2,\cdot\cdot\cdot)\, ,
$$
which means 
$$
M_Z^{(n)} = M_0 + \frac{n-1}{R} \, . 
$$
Now, instead of taking $v_R \to \infty$, if we take $v_R$ to be large and
expand in its inverse powers, we obtain
\begin{equation} 
\label{soln_z}
M_Z^{(n)} = \left( M_0 + \frac{n-1}{R}\right) 
\left[1-\frac{2}{(g^2+g'^2)v_R^2\pi R} + \cdot\cdot\cdot \right]  
~~(n=1,2,\cdot\cdot\cdot)
\end{equation}
If we take the $(-)$ sign solution in the zeroth order approximation, then
through similar steps, we obtain
\begin{equation} 
\label{soln_zp}
M_{Z'}^{(n)} = \left(-M_0 + \frac{n}{R}\right) 
\left[1-\frac{2}{(g^2+g'^2)v_R^2\pi R} + \cdot\cdot\cdot \right]
~~(n=1,2,\cdot\cdot\cdot)
\end{equation}
Thus we see there are two towers of neutral bosons: the $Z$ tower has a
spectrum given by Eq.~(\ref{soln_z}) and the $Z'$ tower spectrum is given by 
Eq.~(\ref{soln_zp}). It is not unexpected to have two towers, as the solutions
come from a quadratic equation.

\subsubsection{Range of $\mathbf{M_0}$}
Let us recall that $M_0 = \frac{1}{\pi R} \tan^{-1}
\sqrt{1+\frac{2g'^2}{g^2}}$. The maximum value of any $\tan^{-1}$ is
$\frac{\pi}{2}$. The minimum value of $\tan^{-1} \sqrt{1+\frac{2g'^2}{g^2}}$
is $\tan^{-1}(1) = \frac{\pi}{4}$.  These limits set the range of $M_0$: ~
\begin{equation}
\frac{1}{4R} < M_0 < \frac{1}{2R} \, . 
\end{equation}
For $v_R \to \infty$, we get the following range of the masses of the lightest
($n=1$) KK state of the $Z$ and $Z'$ towers
\begin{equation}
M_Z \equiv M_Z^{(1)} = M_0 = \left[\frac{1}{4R}, \frac{1}{2R} \right] \, , ~~~
M_{Z'}^{(1)} = -M_0+\frac{1}{R} = \left[\frac{1}{2R}, \frac{3}{4R} \right]
\, .
\end{equation}
In fact, the $Z'$ boson is heavier than the $Z$ boson level by level, i.e.
$M_{Z'}^{(n)} > M_Z^{(n)}$. The mass of the $W^{(1)}$ boson (which is in fact
the $W$ boson of the SM) is given by, putting $n=1$ in Eq.~(\ref{soln_w}) and
letting $v_R \to \infty$, 
\begin{equation} 
M_W \equiv M_W^{(1)} = \frac{1}{4R}  \, , ~~{\rm i.e.}~~  M_Z> M_W 
~~~{\rm as ~ expected} \, . 
\end{equation} 

\subsubsection{The 4d gauge couplings ($g_4$, $g_4^\prime$) and the $\rho$
  parameter}
From Eq.~(\ref{expansion_fields}) we take the expression for $B_\mu$ and look
at its expansion for $y=0$: 
\begin{equation}
\label{bmu}
B_\mu (x,0) = \frac{1}{\sqrt{\pi R (g^2 + 2g'^2)}} 
\left[g\g_\mu(x) + \sqrt{2} g'  Z_\mu^{(1)}(x) + ~~
{\rm higher} ~ Z_\mu^{(n)} ~ {\rm terms} \right] \, . 
\end{equation}
Note that the mass dimension of the 5d $B_\mu$ is $\frac{3}{2}$ while that of
the 4d $B_\mu$ is $1$.  Now we compare Eq.~(\ref{bmu}) with the SM expression
of $B_\mu$ in terms of the photon and $Z$ boson fields, namely,
\begin{equation} 
  B_\mu = \frac{1}{\sqrt{g_4^2+g_4^{\prime 2}}} 
\left[g_4 \g_\mu + g'_4 Z_\mu\right] \, .
\end{equation}
 It immediately follows that
$({g'_4}/{g_4}) = ({\sqrt{2}g'}/{g})$. We are now all set to calculate the
$\rho$-parameter in this scenario:
\begin{eqnarray} 
\frac{M_W^2}{M_Z^2} \equiv 
\frac{\left(M_W^{(1)}\right)^2}{\left(M_Z^{(1)}\right)^2}
= \frac{1}{16 R^2 M_0^2} = \frac{\pi^2}{16} 
\left(\tan^{-1}\sqrt{1+\frac{2g'^2}{g^2}} \right)^{-2}  
= \frac{\pi^2}{16} 
\left(\tan^{-1}\sqrt{1+\frac{g_4^{\prime 2}}{g_4^2}} \right)^{-2} \sim 0.85 \, .
\end{eqnarray}
Hence, 
\begin{equation} 
\rho \equiv \frac{M_W^2}{M_Z^2 \cos^2 \t_W} \sim 1.10 \, . 
\end{equation}
We summarize now what we learned from this scenario: 

\noindent ($i$) A big achievement is that the $W$ and $Z$ boson masses depend
on the gauge couplings. Without actually having a Higgs boson, just by
applying BC's on the boundaries, one can get the correct $W$ and $Z$ masses.

\noindent ($ii$) In this scenario $\Delta \rho \sim 10\%$ is far too large.
This scenario is thus disfavored by EWPT.  A slightly more acceptable value of
$\rho$ can be obtained by keeping a finite $v_R$ at the $y=\pi R$ fixed
point. Then unitarizing the theory would be a problem. The reason for such a
large $\Delta \rho$ is the following. Although the bulk and the $y = 0$ brane
respect custodial SU(2), the $y=\pi R$ brane does not. Since the KK wave
functions have significant presence around the $y=\pi R$ brane, a large
$\Delta \rho$ results. The remedy lies in expelling the higher ($n > 1$) KK
modes from the custodial symmetry breaking brane.

\subsection{Features of realistic Higgsless scenarios}
\subsubsection{Warped models and oblique parameters}
One of the advantages of going to the warped extra dimension is that the
contributions to the $S$ and $T$ parameters can be kept under
control. Following Ref.~\cite{Csaki:2003zu}, we consider a conformally flat
metric
\begin{eqnarray} 
\label{warped_metric}
ds^2 = h(y)^2 \left(\eta_{\mu\nu} dx^\mu dx^\nu - dy^2 \right) \, , 
\end{eqnarray}
where the extra spatial dimension is in the interval $[R,R^\prime]$. A flat
extra dimension scenario can be recovered if $h(y) =$ constant, while the AdS
limit is obtained when $h(y) = R/y$. Typically, $R^{-1} \approx M_{\rm Pl}$
and $(R^\prime)^{-1} \approx$ TeV scale. The gauge symmetry in the bulk
corresponds to ${\rm SU(2)}_L \times {\rm SU(2)}_R \times {\rm U(1)}_{B-L}$,
and the choice of the left-right gauge symmetry is motivated from the
requirement of a custodial symmetry for EWPT consistency.  The $W$ and $Z$
boson masses in this scenario are given by (with the approximation $R^\prime
\gg R$)
\begin{eqnarray}
M_W^2 \approx \frac{1}{R^{\prime 2} \ln\left(\frac{R^\prime}{R} \right)} \, , ~~~
M_Z^2 \approx \frac{g^2 + 2g^{\prime 2}}{g^2 + g^{\prime 2}}
\frac{1}{R^{\prime 2} \ln\left(\frac{R^\prime}{R} \right)} \, ,
\end{eqnarray}
where $g (= g_L = g_R)$ and $g^\prime$ are 5d SU(2) and U(1) gauge
couplings, respectively.  In the leading order $T$ (or equivalently $\Delta
\rho$) and $S$ are both vanishing -- this is the limit when the warp factor is
infinitely large, i.e. when the Planck brane is moved to the AdS boundary. For
a finite warped factor, $T$ and $S$ will be suppressed by
$\ln(R^\prime/R)$. Since a 5d warped model is dual, in the AdS/CFT sense, to a
4d theory involving a strongly coupled sector which is conformally invariant
between the Planck scale and the TeV scale, a lot of insight can be achieved
about the origin of $T$ and $S$ suppression from this correspondence. Weakly
charged left-right gauged symmetry in the 5d bulk ensures a global custodial
symmetry in the strongly coupled CFT side which keeps $T$ and $S$ under
control \cite{Agashe:2003zs}.

An important question that naturally arises in the Higgsless context is how to
generate the fermion masses. In the absence of a Higgs, one cannot write a
Yukawa coupling. But just like in the case of gauge bosons, appropriate
boundary conditions for fermions would generate their masses. But where to
localize the fermions? They cannot be localized at the UV (Planck) brane where
the gauge symmetry is that of the SM. The reason is that the theory at the UV
brane is chiral and there is no way a zero mode chiral fermion mass can be
generated. On the other hand in the IR (TeV) brane the unbroken gauge symmetry
corresponds to ${\rm SU(2)}_D$ which preserves isospin and would yield equal
up- and down-type masses. So the SM fermions have to be placed inside
vector-like multiplets residing in the 5d bulk which should feel different
gauge symmetry breaking at the two boundaries \cite{Csaki:2003sh}.  Needless
to mention that orbifold projection, or equivalently a set of appropriate
BC's, removes half of the vector-like spectrum yielding a chiral fermion
structure at the lowest KK level.  Also, when the fermions are delocalized
from the boundaries and judiciously placed at different locations in the bulk,
their couplings with the KK gauge bosons can be made to vanish, which
minimizes the $S$ parameter \cite{Agashe:2003zs}.

The third family continues to give some headache. The requirement of a large
top quark mass necessitates the localization of the $t_L$ (and hence $b_L$)
field(s) near the TeV brane. At this brane, because the unbroken gauge
symmetry is ${\rm SU(2)}_D \times {\rm U(1)}_{B-L}$, the $Zb_Lb_L$ coupling is
different from its SM value, which leads to a contradiction with the precision
measurement of the $Zbb$ vertex through $R_b$. This problem can be solved, but
at the price of making the model more complicated, e.g. by invoking a separate
mechanism of the top quark mass generation (analogous to the concept of
topcolor in technicolor models).  To sum up, the localization of the third
family is a major thorn in the construction of realistic Higgsless models.

\noindent {\em Moose models}:~ Several features of Higgsless models have been
investisgated in the context of {\em deconstructed} gauge theories by
discretizing the extra dimension. By doing it we get a finite set of 4d gauge
theories, each corresponding to a particular lattice site
\cite{ArkaniHamed:2001ca}. The fifth component of the gauge field, $A_5$,
which is the connection field, goes into the definition of the `link variable'
$\Sigma_i \equiv {\rm exp} \left(-ia A_5^{(i-1)} \right)$ realizing the
parallel transport between two lattice sites, where $a$ is the lattice
spacing. The link variables can be identified with `chiral fields' which
satisfy the condition $\Sigma \Sigma^\dagger = 1$ \cite{Accomando:2006ga}. In
this way, the 5d gauge theory is replaced by a collection of 4d gauge theories
with chiral fields $\Sigma_i$ having gauge interactions -- this is described
by `moose diagram'. A moose diagram is like a Feynman diagram where {\em
  lines} correspond to {\em links} and {\em vertices} to {\em gauge groups}.
If there is no loop, then one can show that $G = E - 1$, where $G$ is the
number of remaining Goldstone multiplets and $E$ is the number of external
links. Clearly, we need at least two external links to construct a minimal
model (which has only one Goldstone multiplet). It has been shown that in this
scenario the $S$ parameter can be made vanishingly small either by ideal
fermion delocalization \cite{Chivukula:2005xm}, or by introducing a dynamical
nonlocal field connecting the two ends of a moose \cite{Casalbuoni:2004id}.

\subsubsection{Tension between unitarity and EWPT}
This is a major issue that decides the fate of a Higgsless model. We
follow the discussions in Ref.~\cite{Cacciapaglia:2004rb}.  First, we ask the
obvious question: what unitarizes the theory in the absence of the Higgs? If
we consider the elastic scattering process $W_L^\pm Z_L \to W_L^\pm Z_L$, then
in the absence of the Higgs boson the amplitude will go like
\begin{eqnarray}
\left(g_{WWZZ} - g_{WWZ}^2\right) \left[a E^4 + b E^2 + \cdots  \right] \, ,
\end{eqnarray}
where the notations for the three- and four-point gauge couplings are
self-explanatory. In the Higgsless models, there are additional KK gauge
bosons.  The charged vector boson KK states $V_i^\pm$, which the same $V_i WZ$
Lorentz structure as the SM $WWZ$, would contribute to the above
amplitude. However, once we take into account the contributions from all the
states $i = 1,2,\cdots,\infty$ and two sum rules involving the trilinear gauge
couplings are satisfied, the new contributions completely cancel the $E^4$ and
$E^2$ growths. This is a consequence of higher dimensional gauge
symmetry. However, the residual growth would make the Higgsless theories break
down at a few TeV scale. In fact, at higher energies increasing number of
inelastic channels lead to unitarity violation by inducing a linear
growth. This is not unexpected from a 5d point of view, as the dimensionless
5d loop factor grows with energy as $g^2 E/24\pi^3$, where $g$ is the 5d gauge
coupling. In the warped Higgsless scenarios, the naive dimensional analysis
(NDA) cutoff would boil down to:
\begin{eqnarray}
\Lambda_{\rm NDA} \sim \frac{12\pi^4 M_W^2}{g^2 M_{W^{(1)}}} \, , 
\end{eqnarray}
which is around 12 TeV, putting $M_{W^{(1)}} \sim 1.2$ TeV. Explicit
calculation shows that this simple estimate is valid up to a factor of $1/4$
\cite{Papucci:2004ip}. What we thus learned is that in the Higgsless scenario,
because of the appearance of the new weakly coupled states in the TeV scale,
the unitarity saturation is postponed by roughly a factor of 10 beyond the SM
NDA cutoff scale $\Lambda_{\rm NDA}^{\rm SM} \sim 4\pi M_W/g \sim 1.8$
TeV. Clearly, heavier the KK $W$ boson the lower is the scale at which
perturbative unitarity is lost. Now, $M_{W^{(1)}}^2/M_W^2 = {\O}
\left(\ln(R^\prime/R)\right)$. If we increase $R$, i.e. lower the UV cutoff
from the Planck scale, then the first KK $W$ boson mass decreases from 1.2 TeV
to sub-TeV and the NDA cutoff scale goes up.  But one cannot arbitrarily
increase $R$, as this would increase the $T$ parameter which varies as
$1/\ln(R^\prime/ R)$ -- see Fig.~4 of \cite{Cacciapaglia:2004rb} -- even
though $S$ can be kept under control via fermion delocalization. The tension
between extending the domain of perturbative unitarity and at the same time
fitting precision electroweak data have also been discussed in scenarios
\cite{tension_higgsless}.

\noindent {\em Three- and four-site Higgsless models}:~ In the language of
deconstruction, delocalization of fermions correspond to allowing them to
derive their electroweak properties from more than one lattice sites or gauge
groups. It has been shown in Ref.~\cite{Casalbuoni:2005rs} that a linear moose
model, with several SU(2) gauge fields along the string and ${\rm SU(2)}_L$
and ${\rm U(1)}_Y$ as the two end-points, can reconcile EWPT constraints and
increased unitarity bound at the expense of some fine tuning. It has been
demonstrated in Ref.~\cite{Chivukula:2006cg} that several properties of the
Higgsless models, like ideal fermion delocalization, EWPT consistency, fermion
masses, etc., can be illustrated in a highly deconstructed model with only
three sites. The electroweak part of the gauge group corresponds to ${\rm
  SU(2)} \times {\rm SU(2)} \times {\rm U(1)}$, i.e. it contains only one
`interior' SU(2) group. It therefore contains only one set of ($W^\prime$,
$Z^\prime$) states, which can be arranged to be fermiophobic to minimize
precision electroweak corrections. If one extends this three-site model by one
more site, i.e. with one more interior SU(2) gauge group making it a four-site
Higgsless model, the gauge boson resonances need not be fermiophobic to
satisfy EWPT constraints \cite{Accomando:2008dm}.

\subsubsection{Collider signatures of Higgsless models}
The strongly coupled physics in the Higgsless scenario at a scale which is
roughly 10 times $\L^{\rm SM}_{\rm NDA}$ as a result of delayed unitarity
violation, is too large to be observed at the LHC. But it will be possible to
pin down those weakly coupled states which are responsible for unitarity
postponement. In this context we follow the analysis in
Ref.~\cite{Birkedal:2004au}. Different Higgsless models vary in different
aspects, like fermion placements and how the SM particles interact with the KK
states, but the mechanism by which $\L$ is raised is common to all. Weakly
coupled TeV-size new massive vector bosons $V_i$ (where $i$ is the KK label),
whose couplings to the SM gauge bosons are dictated by the {\em unitarity sum
  rules}, enforce the cancellation of the $E^2$ and $E^4$ terms in the
amplitudes of longitudinal gauge boson scattering thereby postponing unitarity
violation. What are the experimental signatures of $V_i$ bosons? It is
advantageous to consider the production of these vector bosons by the SM gauge
boson fusions as the couplings between the $V_i$ and the SM $W/Z$ bosons are
pretty model independent dictated by the unitarity sum rules. The sum rule
implies the following inequality
\begin{eqnarray} 
\label{g1}
g^{(1)}_{WZV} \leq \frac{g_{WWZ} M_Z^2}{\sqrt{3} M_1^\pm M_W} \, . 
\end{eqnarray}
Putting $M_1^\pm = 700$ GeV gives $g^{(1)}_{WZV} \leq 0.04$, which means that
the heavier the mass of $V_1$ the less is the chance to produce it at the
LHC. If we study the scattering channel $W^\pm Z \to W^\pm Z$, a process which
can be mediated by $V_1^\pm$ in the $s$-channel, there will be a sharp
resonance as soon as the $V_1$ threshold is crossed. Recall that a $t$-channel
Higgs exchange unitarizes this amplitude in the SM which therefore does not
give any resonance. Conventional theories of strong EWSB dynamics may give a
somewhat heavier ($\sim$ 2 TeV) resonance but that would be broad due to
strong coupling. But in the Higgsless theories, $V_1$ can be as light as 700
GeV, and the resonance will be narrow because the $V_1$ decay width is very
small. The reason is the following: the decay of $V_1^\pm$ takes place only in
a single channel, and the width is given by
\begin{eqnarray} 
\Gamma (V_1^\pm \to W^\pm Z) \approx \frac{\a (M_1)^3}{144 \sin^2\theta_W
  M_W^2} \, , 
\end{eqnarray} 
under the assumption that the unitarity sum rule is saturated by the first set
of KK vector boson states (i.e. with just $i=1$ of $V_i$). Putting $M_1 = 700$
GeV, the width turns out to be only about 13 GeV. We must remember that
$V_1^\pm$ do not have any significant fermionic couplings as otherwise EWPT
consistency will be jeopardized. Further details of $V_1^\pm$ search
strategies are beyond the scope of this review.

\section{Conclusions and Outlook}
\noindent {\bf 1.}~ We take a snapshot of all the limits on the SM Higgs mass:
\begin{itemize} 
\item Direct search: $m_h > 114.4$ GeV (at LEP-2, from
  non-observation in the $e^+e^- \to Zh$ channel).
\item EWPT: $m_h < 186$ GeV (at 95\% C.L., with direct search non-observation
  as a constraint in the fit).
\item Perturbative unitarity: $m_h < 780$ GeV (in the $2 W_L^+ W_L^- + Z_L
  Z_L$ channel).
\item Triviality: $m_h < 170$ GeV for $\Lambda = 10^{16}$ GeV (scalar quartic
  coupling should not hit the Landau pole).
\item Vacuum stability: $m_h > 134$ GeV for $\Lambda = 10^{16}$ GeV (quartic
  coupling should always stay positive).
\end{itemize} 
In the MSSM, there is a firm prediction on the upper limit of the lightest
Higgs mass. If the top squarks weigh around a TeV, then $m_h ~\ltap~ 135$ GeV.

\noindent {\bf 2.}~ All the BSM models we have considered are based on {\em
  calculability} and {\em symmetry} arguments. In all cases, the electroweak
scale $M_Z$ can be expressed in terms of some high scale parameters $a_i$,
i.e., $M_Z = \Lambda_{\rm NP} f(a_i)$, where $\L_{\rm NP}$ is the new physics
scale and $f(a_i)$ are calculable functions of physical parameters. In all
these models
\begin{itemize} 
\item The new physics scales originate from different dynamics:
  $\Lambda_{\rm SUSY} \sim M_S$ (the supersymmetry breaking scale);
  $\Lambda_{\rm LH} \sim f \sim F$ (the vev associated with $G \to H$ breaking);
  $\Lambda_{\rm GHU} \sim R^{-1}$ (the inverse radius of compactification).

\item The dynamical sign-flip of a scalar mass-square happens not only in
  supersymmetry, but also in little Higgs models and in Gauge-Higgs
  unification scenarios. In all cases, the large top quark Yukawa coupling
  plays a crucial r\^ole. A positive scalar quartic coupling can be arranged
  in all these scenarios.

\end{itemize}

\noindent {\bf 3.}~ In supersymmetry, the cutoff can be as high as the GUT or
the Planck scale. In little Higgs as well as in many variants of extra
dimensional scenarios the cutoff is significantly lower. The ultraviolet
completion in little Higgs models is an open question, though some attempts
have already been made in this direction.

\noindent {\bf 4.}~ In supersymmetry the cancellation of quadratic divergence
takes place between a particle loop and a sparticle loop. Since a particle
{\em cannot} mix with a sparticle, the oblique electroweak corrections and the
$Zb\bar{b}$ vertex correction can be kept under control. In the
non-supersymmetric scenarios (recall what happens in little Higgs models), the
cancellation occurs between loops with the same spin states. Such states {\em
  can} mix among themselves, leading to dangerous tree level contributions to
electroweak observables. This is the reason why a decoupling theory like
supersymmetry is comfortable with EWPT, while a technicolor-like
non-decoupling theory faces a stiff challenge from EWPT.

\vspace{0.1cm}

\noindent {\bf 5.}~ How to distinguish between the different models in
colliders? We have already discussed some of the smoking gun signals of
different scenarios. Here we highlight a few features that are the trademark
signals of some specific models. We first compare supersymmetry with little
Higgs models.  The chances are very high that one may mistake supersymmetry
with $R$-parity for little Higgs with $T$-parity or {\em vice versa} in the
LHC environment, since the pattern of cascade decays in the two models are
very similar and particle spin measurements are, in general, difficult. A
dictionary between superparticles and little Higgs heavy states is the
following: ($i$) Electroweak gauginos $\leftrightarrow$ $T$-odd gauge bosons,
($ii$) Sfermions $\leftrightarrow$ $T$-odd fermion doublets, ($iii$) Second
Higgs doublet $\leftrightarrow$ Scalar triplet, ($iv$) Higgsinos and gluino
$\leftrightarrow$ None, ($v$) None $\leftrightarrow$ $T$-even top
partner. What is interesting to observe is that there is no analog of gluino
in little Higgs models, and no analog of $t_+$ in supersymmetry.

In a general class of composite Higgs models (e.g. gauge-Higgs unification or
little Higgs), the strengths of $VVh$ and $VVhh$ couplings are different from
their SM predictions. A recent work suggests that double Higgs production via
$W_L W_L \to hh$ can be an interesting probe for verifying the compositeness
of the Higgs since the rate of this process is much larger (than in the SM) if
$h$ is a pseudo-Goldstone boson \cite{Contino:2010mh}.

The presence of a KK gluon in a gauge-Higgs unification model differentiates
it from a little Higgs model.  Moreover, the gauge-Higgs models have a special
feature that their lightest nonstandard particle is a colored fermion and not
a KK gauge boson. Such models also contain fermions with exotic electric
charge, whose value is different in different models.  The Higgsless models
are characterized by the presence of the $V_i$ vector boson states that delay
the unitarity saturation. The lightest of such states may pop up in the
scattering of $W^\pm Z \to W^\pm Z$ as an $s$-channel narrow resonance.

\vspace{0.1cm}

\noindent {\bf 6.}~ The model-builders have three-fold goals: (i) unitarize
the theory, (ii) successfully confront the EWPT, and (iii) maintain naturalness
to the extent possible. The tension arises as naturalness criteria requires the
spectrum to be compressed, while {\em EWPT compatibility} pushes the new
states away from the SM states.

\vspace{0.1cm}

\noindent {\bf 7.}~ A dark matter candidate is badly needed to justify
observational evidence. Besides the neutrino mass, dark matter provides the
only other concrete experimental motivation to go beyond the SM. The SM fails
to provide it. A favorite supersymmetric candidate is the lightest neutralino
if $R$-parity is conserved. The Little Higgs models provide a heavy stable
gauge boson if $T$-parity (which can be defined in `littlest' Higgs model) is
conserved. In extra dimensional models, the lightest Kaluza-Klein particle is
a stable dark matter candidate if KK-parity is conserved.

\vspace{0.1cm}

\noindent {\bf 8.}~ After all is said and done, the LHC is a {\em win-win}
machine in terms of discovery. If we discover the Higgs, we would expect to
discover also the new states that tame the unruly quantum correction to its
mass. If the Higgs is not there, the new resonances which would restore
unitarity in gauge boson scattering would be crying out for verification. In
order to identify the latter, we need the super-LHC (the high luminosity
option) to cover the entire spectrum. However, once we observe some new states
at the LHC, we definitely need a linear collider to know what these states
actually are.

\vspace{0.1cm}

\noindent{\bf Acknowledgments:}~ I am indebted to Romesh Kaul for sharing his
insights, particularly in little Higgs models, and for valuable comments on
the manuscript. I thank Steve King, Per Osland, Amitava Raychaudhuri, and
especially, Palash B.~Pal for reading the manuscript and suggesting
improvements. I also acknowledge several useful conversations with Avinash
Dhar, Marco Serone, Christophe Grojean, Probir Roy and Rohini Godbole. I am
thankful to the organizers of ($i$) the Advanced SERC School in High Energy
Physics in Hyderabad (2007), ($ii$) the RECAPP Workshop at HRI, Allahabad
(2008), and ($iii$) the 15th Vietnam School of Physics (2009) for invitation
to lecture on Standard Model and beyond highlighting electroweak symmetry
breaking, based on which this review is written. I acknowledge hospitality at
the Physics Department of T.U.~Dortmund, CERN Theory Division and the HEP
section of ICTP, during the course of writing this review.  This work is
partially supported by the project No.~2007/37/9/BRNS of BRNS (DAE), India,
and DST/DAAD project No.~INT/DAAD/P-181.

\small{

}
\end{document}